\documentclass[twocolumn]{aastex631}

\usepackage{amsfonts}
\usepackage{amsmath}
\usepackage{amssymb}

\usepackage{booktabs}
\usepackage{color}
\usepackage{comment}
\usepackage{CJK}
\usepackage{float}
\usepackage{gensymb}
\usepackage{graphicx}
\usepackage{hyperref}
\usepackage{makecell}
\usepackage{mathrsfs}
\usepackage{multirow}
\usepackage{newtxtext,newtxmath}
\usepackage{orcidlink}
\usepackage{overpic}
\usepackage{rotating}
\usepackage{subfigure}
\usepackage{threeparttable}
\usepackage{ulem}
\usepackage{xcolor}

\makeatletter
\renewcommand\normalsize{%
   \@setfontsize\normalsize\@xpt{10.56}
   \abovedisplayskip 2.2mm \@plus2\p@ \@minus1\p@
   \abovedisplayshortskip \z@ \@plus3\p@
   \belowdisplayshortskip 2.2mm \@plus2\p@ \@minus1\p@
   \belowdisplayskip \abovedisplayskip
   \let\@listi\@listI}
\normalsize
\renewcommand\small{%
   \@setfontsize\small\@ixpt{9.68}%
   \abovedisplayskip 8.5\p@ \@plus3\p@ \@minus4\p@
   \abovedisplayshortskip \z@ \@plus2\p@
   \belowdisplayshortskip 4\p@ \@plus2\p@ \@minus2\p@
   \def\@listi{\leftmargin\leftmargini
               \topsep 4\p@ \@plus2\p@ \@minus2\p@
               \parsep 2\p@ \@plus\p@ \@minus\p@
               \itemsep \parsep}%
   \belowdisplayskip \abovedisplayskip
}
\renewcommand\footnotesize{%
   \@setfontsize\footnotesize\@viiipt{8.36}%
   \abovedisplayskip 6\p@ \@plus2\p@ \@minus4\p@
   \abovedisplayshortskip \z@ \@plus\p@
   \belowdisplayshortskip 3\p@ \@plus\p@ \@minus2\p@
   \def\@listi{\leftmargin\leftmargini
               \topsep 3\p@ \@plus\p@ \@minus\p@
               \parsep 2\p@ \@plus\p@ \@minus\p@
               \itemsep \parsep}%
   \belowdisplayskip \abovedisplayskip
}
\renewcommand\scriptsize{\@setfontsize\scriptsize\@viipt\@viiipt}
\renewcommand\tiny{\@setfontsize\tiny\@vpt\@vipt}
\renewcommand\large{\@setfontsize\large\@xiipt{14}}
\renewcommand\Large{\@setfontsize\Large\@xivpt{18}}
\renewcommand\LARGE{\@setfontsize\LARGE\@xviipt{22}}
\renewcommand\huge{\@setfontsize\huge\@xxpt{25}}
\renewcommand\Huge{\@setfontsize\Huge\@xxvpt{30}}
\def\@listi{\leftmargin\leftmargini
            \parsep 4\p@ \@plus2\p@ \@minus\p@
            \topsep 8\p@ \@plus2\p@ \@minus4\p@
            \itemsep4\p@ \@plus2\p@ \@minus\p@}
\let\@listI\@listi
\@listi
\def\@listii {\leftmargin\leftmarginii
              \labelwidth\leftmarginii
              \advance\labelwidth-\labelsep
              \topsep    4\p@ \@plus2\p@ \@minus\p@
              \parsep    2\p@ \@plus\p@  \@minus\p@
              \itemsep   \parsep}
\def\@listiii{\leftmargin\leftmarginiii
              \labelwidth\leftmarginiii
              \advance\labelwidth-\labelsep
              \topsep    2\p@ \@plus\p@\@minus\p@
              \parsep    \z@
              \partopsep \p@ \@plus\z@ \@minus\p@
              \itemsep   \topsep}
\def\@listiv {\leftmargin\leftmarginiv
              \labelwidth\leftmarginiv
              \advance\labelwidth-\labelsep}
\def\@listv  {\leftmargin\leftmarginv
              \labelwidth\leftmarginv
              \advance\labelwidth-\labelsep}
\def\@listvi {\leftmargin\leftmarginvi
              \labelwidth\leftmarginvi
              \advance\labelwidth-\labelsep}
\makeatother

\newcommand{\kms}{{\rm km}\,{\rm s}^{-1}}

\begin{document}
\begin{CJK*}{UTF8}{gbsn}

\title{Almost Optically Dark Galaxies in DECaLS (I): Detection, Optical Properties and Possible Origins}


\author[0000-0002-4915-4137]{Lin Du (杜林)}
\affiliation{Key Laboratory of Optical Astronomy, National Astronomical Observatories, Chinese Academy of Sciences, 20A Datun Road, Chaoyang District, Beijing 100012, People's Republic of China}
\affiliation{School of Astronomy and Space Science, University of Chinese Academy of Sciences, Beijing 100049, China}

\author[0000-0003-4546-8216]{Wei Du (杜薇)}
\altaffiliation{wdu@nao.cas.cn}
\affiliation{Key Laboratory of Optical Astronomy, National Astronomical Observatories, Chinese Academy of Sciences, 20A Datun Road, Chaoyang District, Beijing 100012, People's Republic of China}

\author[0000-0003-0202-0534]{Cheng Cheng (程诚)}
\affiliation{Key Laboratory of Optical Astronomy, National Astronomical Observatories, Chinese Academy of Sciences, 20A Datun Road, Chaoyang District, Beijing 100012, People's Republic of China}
\affiliation{Chinese Academy of Sciences South America Center for Astronomy, National Astronomical Observatories, Chinese Academy of Sciences, Beijing 100101, People's Republic of China}

\author[0000-0001-6083-956X]{Ming Zhu (朱明)}
\affiliation{CAS Key Laboratory of FAST, National Astronomical Observatories, Chinese Academy of Sciences, Beijing 100101, People's Republic of China}

\author[0000-0001-9838-7159]{Haiyang Yu (于海洋)}
\affiliation{CAS Key Laboratory of FAST, National Astronomical Observatories, Chinese Academy of Sciences, Beijing 100101, People's Republic of China}
\affiliation{School of Astronomy and Space Science, University of Chinese Academy of Sciences, Beijing 100049, China}

\author[0000-0002-4333-3994]{Hong Wu (吴宏)}
\affiliation{Key Laboratory of Optical Astronomy, National Astronomical Observatories, Chinese Academy of Sciences, 20A Datun Road, Chaoyang District, Beijing 100012, People's Republic of China}
\affiliation{School of Astronomy and Space Science, University of Chinese Academy of Sciences, Beijing 100049, China}

\begin{abstract}
  We report the discovery of eight optical counterparts of ALFALFA extragalactic objects from DECaLS, five of which are discovered for the first time. These objects were flagged as H\textsc{i} emission sources with no optical counterparts in SDSS before. Multi-band data reveal their unusual physical properties. They are faint and blue ($g-r=-0.35\sim0.55$), with quite low surface brightness ($\mu_{\rm g,peak}=24.88\sim26.41\,{\rm mag}/{\rm arcsec}^2$), irregular morphologies, low stellar masses ($\log_{10}(M_{*}/M_\odot)=5.27\sim7.15$), low star formation rates ($SFR=0.21\sim9.24\times10^{-3}\,{M_\odot}\,{\rm yr}^{-1}$), and remarkably high H\textsc{i}-to-stellar mass ratios ($\log_{10}(M_{\rm H\textsc{i}}/M_{*}) = 1.72\sim3.22$, except AGC\,215415). They deviate from the scaling relations between H\textsc{i} and optical properties defined by the ALFALFA sample and the baryonic Tully-Fisher relation. They agree well with the main sequence of star-forming galaxies but exhibit low star-forming efficiency. Based on their physical properties and environments, we speculate that six of these objects may have originated from tidal processes, while the remaining two appear to have isolated origins. They may have had a relatively calm evolutionary history and only begun to form stars recently.

\end{abstract}



\section{Introduction}
\label{sect:intro}

The Arecibo Legacy Fast ALFA H\textsc{i} (ALFALFA) Survey is a blind H\textsc{i} survey which has detected $\sim31500$ extragalactic H\textsc{i} sources across a sky area of more than $7000\,{\rm deg}^2$. Due to improved sensitivity and angular resolution, it provides us with a confident map of H\textsc{i} density distribution in the nearby universe at redshift $z<0.06$. When the H\textsc{i} sources in the 40$\%$ sky area of ALFALFA ($\alpha$.40 catalog) were first published, an attempt was made to align them with the optical counterparts from the seventh data release of the Sloan Digital Sky Survey \citep[SDSS DR7,][]{2009ApJS..182..543A}. However, $\sim1.26\%$ of them, identified by their velocities as extragalactic H\textsc{i} sources, were found to fail to match with the optical counterparts \citep{2011AJ....142..170H}.

With great attention to the origin and physical properties of these \textit{failed} sources, much research has been carried out to investigate them. It has been revealed in recent years that some of these sources are tidal debris from interacting galaxies, OH megamaser impostors \citep{2016MNRaS.459..220S}, ultra-compact high-velocity clouds \citep[UCHVCs,][]{2019AJ....157..183J}, and H\textsc{i}-bearing ultra-diffuse galaxies \citep[HUDs,][]{2017ApJ...842..133L}. However, out of these extragalactic H\textsc{i} sources without an optical counterpart in SDSS, there remain $\sim50$ sources that have not yet found their optical counterparts even in deeper optical observations, and it is difficult to interpret or classify them based on H\textsc{i} data alone.  

One perception is that they are probably the \textit{dark galaxies}. Unlike other optical faint objects that are difficult to detect, dark galaxies are believed to be dominated by dark matter and have no stellar components. Previous observations have identified several potential dark galaxy candidates, such as VIRGOH\textsc{i}21 \citep{2005ApJ...619L...1M} and the optically invisible H\textsc{i} clouds \citep{2007ApJ...665L..15K}. Simulations have also confirmed that dark galaxies can form under small-scale dark matter halos \citep{2006MNRAS.368.1479D}. Despite this, none of the dark galaxy candidates have been confirmed as genuine dark galaxies yet. The existence of \textit{totally dark} galaxies is still a matter of debate. The other perception is more conservative, arguing that they are \textit{almost dark} sources, which have not been detected due to the optical detection limitations. However, as these sources lack clear optical counterparts in current surveys, their physical properties and origins remain largely unexplored.

To better understand the characteristics of these rare sources, \citet{2015AJ....149...72C} launched a project called ``Almost Dark Campaign''. This project aimed to conduct follow-up multi-band observations for the isolated \textit{almost dark} sources. To ensure high precision and sensitivity, many advanced instruments were employed, such as the Hubble Space Telescope (HST) and the Wisconsin-Indiana-Yale-NOAO Telescope (WIYN) for deep optical imaging, as well as the Very Large Array (VLA) and Westerbork Synthesis Radio Telescope (WSRT) for H\textsc{i} synthesis mapping. Several discoveries were made through this project. Some focused on identifying large samples and large-scale structures, such as the identification of a sample of 115 isolated HUDs with unique properties \citep{2017ApJ...842..133L}, and the detection of extended optically dark H\textsc{i} features in Leo Cloud \citep{2016MNRAS.463.1692L}. Others focused on individual sources, also revealing many distinctive characteristics.

H\textsc{i}\,1232+20 and AGC\,229101 are two peculiar systems with extreme nature that have been found among the studies. Unlike the properties previously found in H\textsc{i} sources in ALFALFA, H\textsc{i}\,1232+20 and AGC\,229101 have extremely high H\textsc{i}-to-stellar mass ratios (81 and 49, respectively), blue colors and remarkably low star formation rates ($6.9\times10^{-4}\,{M_\odot}\,{\rm yr}^{-1}$ for AGC\,229385), which deviate from the scaling relations of the $\alpha.40$ sample \citep{2012ApJ...756..113H}. They are located at the faint end of the baryonic Tully-Fisher relation, deviating from the empirical relation \citep{2018AJ....155...65B}. There have been very few such objects in the ALFALFA survey, which are difficult to classify based on existing galaxy classifications. The lack of samples and research makes the physical properties and origins of such objects still a mystery. There is a hypothesis that these sources may imply a class of objects just around the threshold for star formation. The gas densities on such a source lie just above and below the threshold of the gas density required for star formation, which could inhibit the star-forming process for an extended period \citep{2015ApJ...801...96J}.

In this work, we studied eight sources detected by the Dark Energy Camera Legacy Survey \citep[DECaLS,][]{2019AJ....157..168D}, which are H\textsc{i} sources in the 100$\%$ sky area of ALFALFA ($\alpha.100$ catalog) that failed to find an optical counterpart in the crossover with SDSS data \citep{2020AJ....160..271D}. Among the eight optical counterparts, five of them are discovered for the first time. Similar to AGC\,229385 and AGC\,229101, these sources exhibit low luminosities with surface brightness at least one magnitude fainter than the definition of low surface brightness galaxies (LSBGs). Most of them also show extremely high H\textsc{i}-to-stellar mass ratios, which deviate significantly from the distribution of the H\textsc{i} population in the ALFALFA survey. Meanwhile, these sources may expand the limited sample of systems with the extreme nature described above, such as H\textsc{i}\,1232+20, where AGC\,229385 is located. 

We organize this paper as follows: in Section \ref{sec:02}, we describe the process of discovering the optical counterparts of the eight objects, and introduce the multi-band images and data used in this work, including H\textsc{i}, optical, and ultraviolet (UV); in Section \ref{sec:03}, we present the details of data processing and the physical parameters of the stellar and H\textsc{i} gas components; in Section \ref{sec:04}, we show the scaling relations of these sources, and discuss their physical properties; in Section \ref{sec:05}, we analyze the environment of these galaxies, speculate on their origins, and attempt to explain their peculiar physical properties; and in Section \ref{sec:06}, we summarize the work, and look forward to the direction of the following research. Throughout this work, Hubble constant is assumed as $H_0=70\,{\rm km}\,{\rm s}^{-1}{\rm Mpc}^{-1}$, and the distances are sourced from \citet{2018ApJ...861...49H}. All magnitudes are in the $AB$ magnitude system \citep{1983ApJ...266..713O}. We assumed the stellar initial mass function (IMF) of \citet{2003PASP..115..763C}, and converted all reference data accordingly with conversion factors, unless otherwise specified.

\section{Data}\label{sec:02}
\subsection{Parent sample and sources}

To investigate the properties of more \textit{almost dark} H\textsc{i} sources in ALFALFA, we utilized the following method to select \textit{almost dark} sources. 

Firstly, we used two catalogs to construct our parent sample. The $\alpha.100$-SDSS catalog from \citet{2020AJ....160..271D} provides essential optical and H\textsc{i} properties for objects cross-matched in the SDSS and ALFALFA surveys. It includes information such as the right ascension, declination, optical band magnitude, as well as the velocity and distance determined by H\textsc{i} emission. However, some H\textsc{i} sources within the SDSS footprint failed to match an optical counterpart. These sources are marked as ``3'' in the ``photometry flag'' column, amounting to a total of 567 sources. These sources fall into two scenarios: they either have multiple possible optical counterparts in the large H\textsc{i} beam region, making it difficult to determine the most likely optical counterpart, or fail to detect any optical emissions in SDSS. We cross-matched these 567 sources with the $\alpha.100$ catalog \citep{2018ApJ...861...49H}, which provides detailed H\textsc{i} parameters about the ALFALFA H\textsc{i} sources, complementing the information of velocity width of the H\textsc{i} line profile ($W_{50}$, $W_{20}$), integrated H\textsc{i} line flux density, and H\textsc{i} mass for these sources. The catalog of these 567 sources with detailed H\textsc{i} parameters served as our parent sample.

Secondly, based on the H\textsc{i} centroid of the sample, we conducted a thorough check of each source by eye in DECaLS, using a circle with a diameter of $3.3^{\prime}$, which matches the resolution of a single beam in ALFALFA. In the process, we discovered that out of 567 sources failing to match an optical counterpart in SDSS, 21 are either not covered by DECaLS or only with single-band observations; $\sim130$ sources are in the regions exhibiting severe spurious signals, such as pupil ghosts and significant contamination from bright sources. We initially excluded these sources due to poor imaging quality. Among the remaining sources, we found over 50 bright, large-angular-diameter galaxies, some of which even fill the entire ALFALFA beam size. These sources may plausibly belong to the first case of the parent sample described above, exhibiting dust lanes along with complex stacking of stars and H\textsc{ii} regions, which makes it impractical to assign a single optical counterpart to the H\textsc{i} sources under the matching algorithm. Besides, around 50 sources exhibit morphology of small-scale blue compact dwarfs; 7 sources were found to be positioned on large-scale, Galactic cirrus cloud, some position of which can match with that of the sources in the Galactic Arecibo L-Band Feed Array H\textsc{i} (GALFA-H\textsc{i}) Compact Cloud Catalog \citep{2012ApJ...758...44S}. These sources were also excluded from further identification. We continued to search for \textit{almost dark} galaxies from the remaining 243 sources, based on the criteria that the sources should fail to get detected or identified in SDSS images, but should be detected in DECaLS images. We ultimately identified 24 sources that meet the criteria, all of which exhibit faint and relatively diffuse features. After a thorough evaluation, we excluded sources with red color that are unlikely to exhibit strong H\textsc{i} emission; sources contaminated by foreground pollution that are difficult to correct; sources located near large-scale Galactic cirrus where the membership is uncertain; and AGC\,229385, an already extensively studied \textit{almost dark} galaxy. Finally, there were only eight sources left, composing the sample for further research. 

\begin{figure*}
    \centering
    \includegraphics[width=0.24\textwidth]{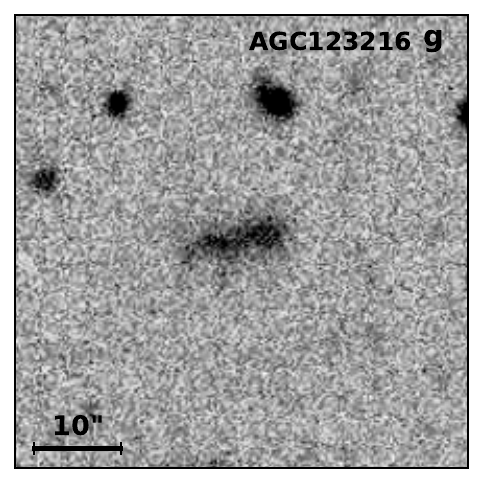}
    \includegraphics[width=0.24\textwidth]{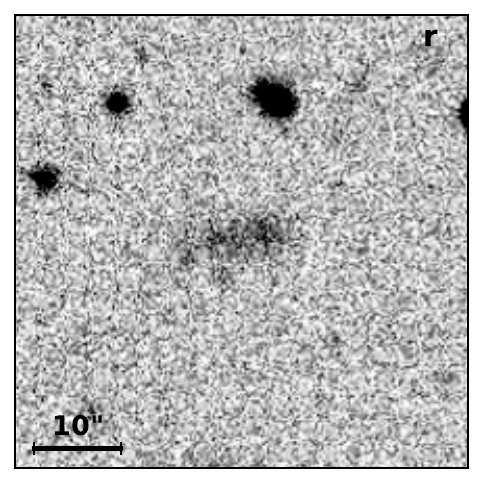}
    \includegraphics[width=0.24\textwidth]{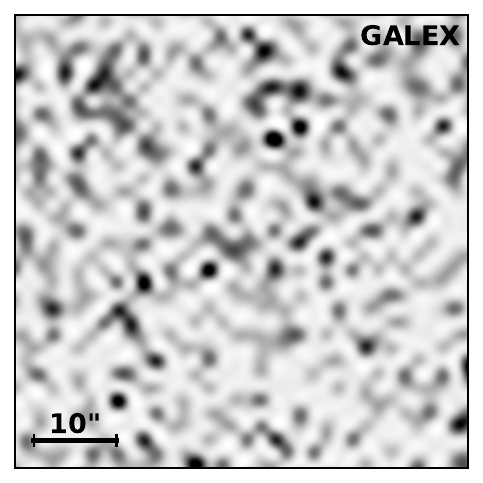} 
    \\
    \includegraphics[width=0.24\textwidth]{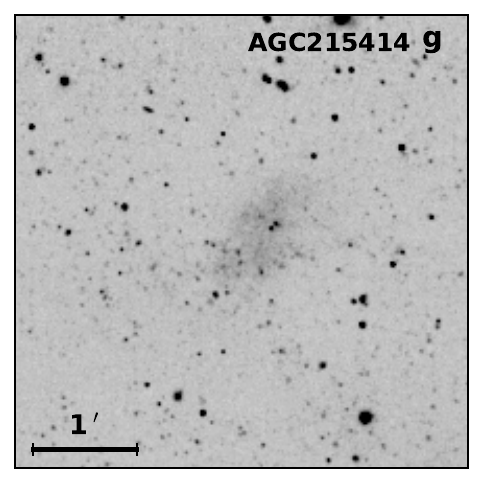}
    \includegraphics[width=0.24\textwidth]{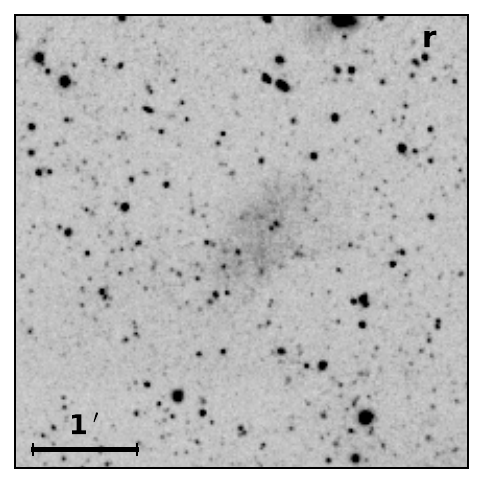}
    \includegraphics[width=0.24\textwidth]{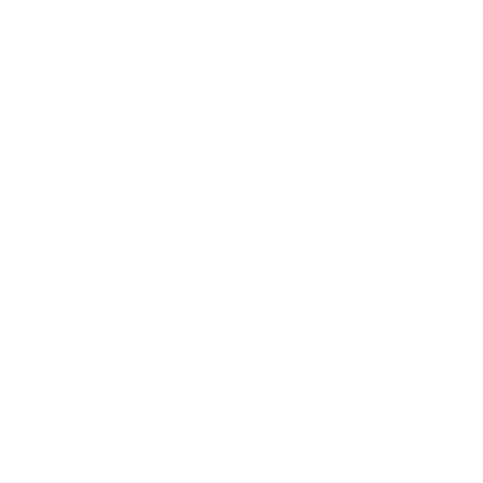} 
    \\
    \includegraphics[width=0.24\textwidth]{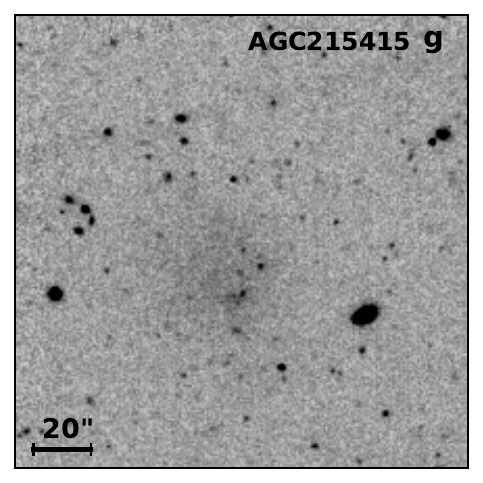}
    \includegraphics[width=0.24\textwidth]{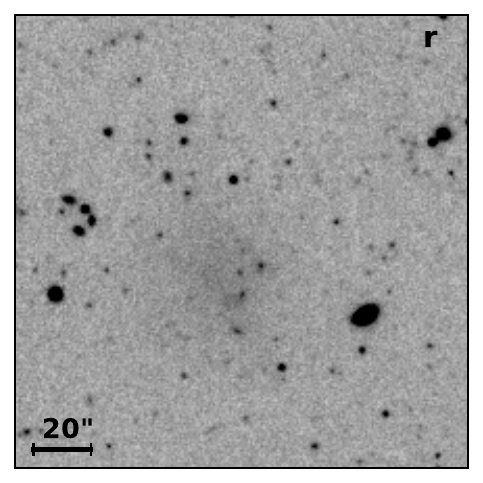}
    \includegraphics[width=0.24\textwidth]{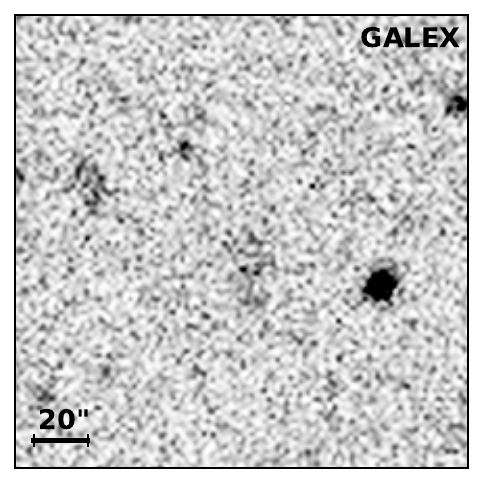} 
    \\
    \includegraphics[width=0.24\textwidth]{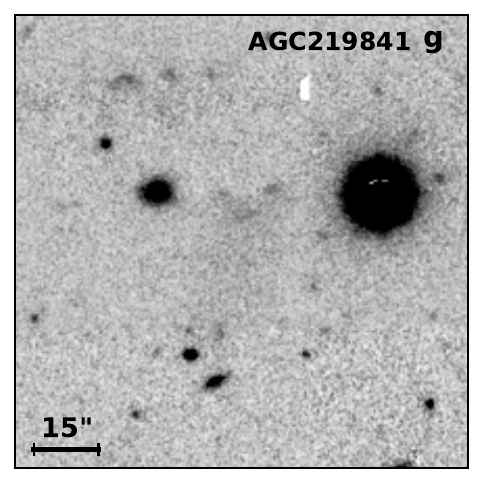}
    \includegraphics[width=0.24\textwidth]{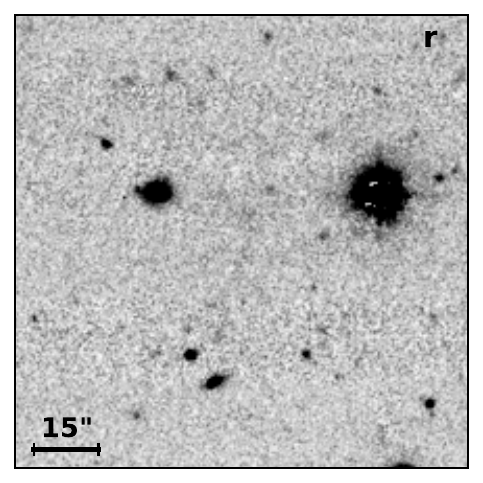}
    \includegraphics[width=0.24\textwidth]{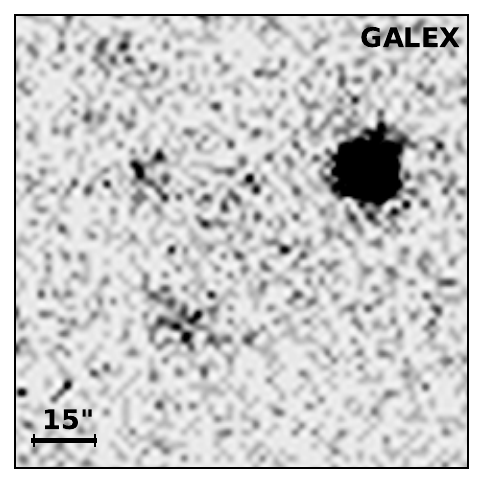} 
    \\
    \includegraphics[width=0.24\textwidth]{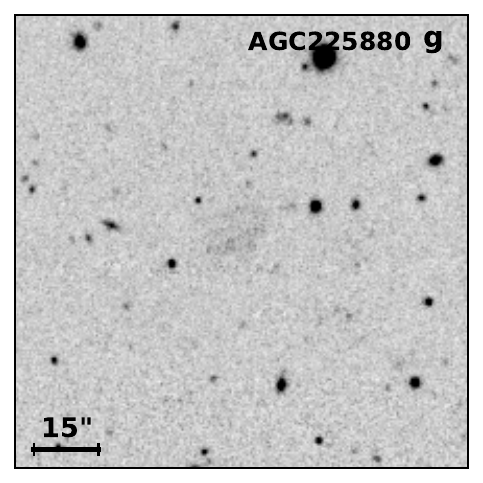}
    \includegraphics[width=0.24\textwidth]{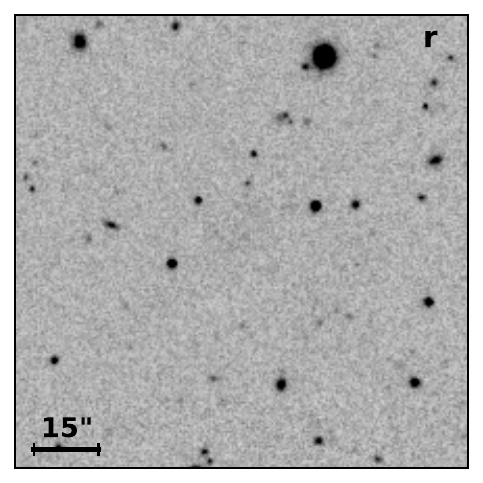}
    \includegraphics[width=0.24\textwidth]{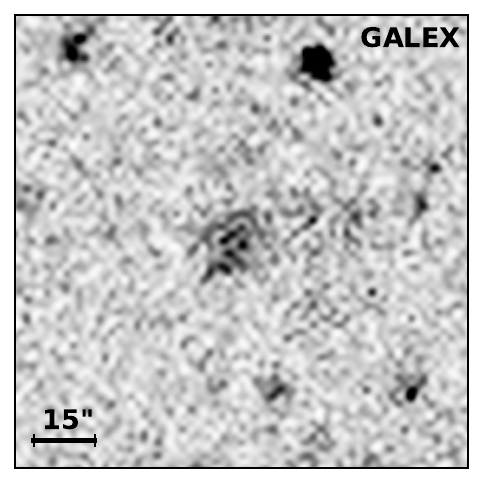} 
    \caption{Optical and UV images of our sources, with the same order as in Table \ref{tab:01}. The left and middle columns correspond to the $g$- and $r$-band images from DECaLS, respectively, and the right column is the NUV images from GALEX. The images are oriented N-up and E-left. Our sources are in the center of the images. All the sources have a faint but clear optical counterpart in the $g$ band and a fainter one in the $r$ band. AGC\,225880 and AGC\,229197 have obvious NUV counterparts, while the counterparts of the other targets are a bit ambiguous. AGC\,215414 is not covered by the footprint of GALEX.}
    \label{fig:01}
\end{figure*}

The eight sources, which were eventually selected out of 567 sources, included AGC\,123216, an \textit{almost dark} dwarf galaxy with a suspected tidal origin \citep{2023AJ....165..197G}; AGC\,215414, the tail of the NGC\,3628 plume \citep{2014ApJ...786..144N} in Leo Triplet; AGC\,215415, considered as an ultra-diffuse galaxy (UDG) candidate in \citet{2018A&A...615A.105M}, and remaining five newly discovered targets\footnote{It is mentioned that ground-based WIYN had revealed a very faint stellar component associated with AGC\,229361 according to private communication with H. Pagel in 2018 in Section 4.6 of \citet{2019AJ....157...76B}. However, before this paper, the emission of the stellar component in AGC\,229361 had not been demonstrated or studied.}. The archival parameters and multi-band images of the eight sources are shown in Table \ref{tab:01} and Fig. \ref{fig:01}, respectively.

\subsection{H\textsc{i} Data}
The H\textsc{i} data of the eight sources are mainly from archival ALFALFA data. ALFALFA is a wide area extragalactic H\textsc{i} line survey, which measures two regions of the sky, covering ${7{\rm h}30{\rm m}<{\rm R.A.}<16{\rm h}30{\rm m}}$, $0\degree<{\rm Dec.}<+36\degree$ in the northern Galactic hemisphere and ${22{\rm h}<{\rm R.A.}<03{\rm h}}$, $0\degree<{\rm Dec.}<+36\degree$ in the southern hemisphere, and extends outward to 250\,Mpc. It uses a 7-beam Arecibo L-band feed array, with beam sizes of $3.3^{\prime}$ along the azimuth direction and $3.8^{\prime}$ along the zenith angle direction. It follows a two-pass drift scan mode and tiles sky along the declination direction \citep{2005AJ....130.2598G}.

We acquired the fully processed H\textsc{i} data\footnote{The data can be accessed at \url{http://egg.astro.cornell.edu/alfalfa/data/index.php}.}  from the $\alpha.100$ catalog \citep{2018ApJ...861...49H}. We obtained the H\textsc{i} parameters of our sources, including right ascension and declination, velocity width, heliocentric velocity, distance, and H\textsc{i} mass. Table \ref{tab:01} shows a detailed account of these parameters.

Additionally, observations from VLA have covered three of the sources, which are AGC\,215414, AGC\,225880, and AGC\,229197, under D, C, and D configurations, with resolutions of about $80^{\prime\prime}$, $25^{\prime\prime}$, and $75^{\prime\prime}$, respectively \citep{2014ApJ...786..144N,2010ApJ...725.2333K,1995AJ....109.2415C}. Compared with ALFALFA, VLA has advantages in pointing accuracy, frequency coverage, and imaging capabilities, with an angular resolution at least five times better than that of ALFALFA for H\textsc{i} observations in the L band. We acquired more precise H\textsc{i} centroid and H\textsc{i} mass for these three sources from VLA\footnote{For AGC\,215414, the H\textsc{i} centroid is not provided in \citet{2014ApJ...786..144N}, so we used the H\textsc{i} centroid given by AFLAFLA; the H\textsc{i} mass is given in the form of a range, and we took the median value.}, which are derived from the H\textsc{i} intensity or velocity maps, and the integrated H\textsc{i} flux, respectively.

\subsection{Optical Data}\label{sec:02-03}

The optical data used in this study are from the seventh public data release of DECaLS, corresponding to the final, full sky coverage of the survey program from August 2014 to March 2019. The data set includes optical data at $g$, $r$, and $z$ bands captured by DECam on the Blanco 4-m telescope at Cerro Tololo Inter-American Observatory, along with processed images and catalogs, where the $5\sigma$ point sources depths of $g$, $r$, and $z$ bands can reach 24.7, 23.9, and 23.0\,mag, respectively.

We downloaded the coadded images and inverse variance images of $g$, $r$, and $z$ bands of the eight sources. The coadded images are reduced through the NOIRLab Community Pipeline and processed by the Tractor\footnote{see \url{https://github.com/legacysurvey/legacypipe} for more details.}, including astrometric calibration, photometric characterization, inverse-variance weighted coaddition of images, background correction, and removal of false signals \citep{2019AJ....157..168D}. The inverse variance images corresponding to the coadded images are a measure of the accuracy of the data in each pixel on the coadded images, with the value of each pixel on the former being the inverse variance calculated from the corresponding pixel on the latter. They are constructed based on the sum of the inverse variances of the individual input images in units of ${\rm nanomaggies}^{-2}$ per pixel, using the same weighted overlay as from single-epoch images to coadded images. The processed images are science-ready images that can be used directly for photometry and error estimation, with good quality and proper background subtraction \citep{2022MNRAS.515.5335L}.

\subsection{UV and IR Data}\label{sec:02-04}
The UV data come from the Galaxy Evolution Explorer \citep[GALEX,][]{2005ApJ...619L...1M}, which conducted a series of imaging and spectroscopic surveys in the near-ultraviolet (NUV, $1770-2730\Angstrom$) and far-ultraviolet (FUV, $1350-1780\Angstrom$) from April 2003 to June 2013. As parts of the imaging survey, All-Sky Imaging Survey (AIS), Medium Imaging Survey (MIS), and Deep Imaging Survey (DIS) cover areas of 27000, 5000, and $365\,{\rm deg}^2$, with depths of 20.5, 23.5, and 25.0\,mag, respectively\footnote{The three values correspond to the limit magnitudes required for the scientific objectives of GALEX surveys, derived from Table 1 of \citet{2005ApJ...619L...1M}. Specific bands for these magnitudes are not specified. The detailed limited magnitudes of FUV and NUV bands under different exposure time are shown in Table \ref{tab:04} and section \ref{sec:03-04}.}. GALEX has a relatively poor image resolution (4.2/$5.3^{\prime\prime}$ FWHM, FUV/NUV), and the field of view (FOV) is about $1.2^\circ$ diameter ($1.28/1.24^\circ$). However, photometric accuracy beyond the range of $1.1^\circ$ will be severely disturbed by edge artifacts \citep{2011MNRAS.411.2770B}.

After a search on MAST, seven sources are in the coverage of GALEX, except AGC\,215414. We checked the UV images of the sources and confirmed that they are more than $3^{\prime}$ away from the edge of the FOV to ensure photometric accuracy. We selected the NUV and FUV images with the longest exposure time and accessed the data using \texttt{gPhoton} \citep{2016ApJ...833..292M}. \texttt{gPhoton} is a database product that is associated with software packages and provides a user-friendly way to obtain and process photon-level images of GALEX General Release (GR) 6/7. It contains several modules, one of which is \texttt{gAperture}. \texttt{gAperture} can perform aperture photometry on GALEX UV images by doing a cone search on the sky positions of the individual photon event at the spatial resolution of the data.

In addition, we also downloaded the infrared images of these sources in the Wide-field Infrared Survey Explorer (WISE). However, our samples are almost invisible in infrared images. This indicates that these sources are not only inherently lacking in hot and warm dust, but also less affected by dust shielding.

\section{Analysis and Results}\label{sec:03}
In this section, we examine the characteristics of the stellar and H\textsc{i} gas components. We provide the physical parameters including magnitude, color, surface brightness, stellar mass, H\textsc{i} mass, and star formation rate of these sources. Detailed information about our data processing methods and techniques is also present.

\subsection{Photometry}\label{sec:03-01}
Magnitude and color are important indicators to characterize the global optical properties of the sources. We performed aperture photometry on the coadded images in DECaLS to obtain these two parameters. Before conducting photometry, we took the following steps to ensure proper image processing.

Firstly, we examined images in the $g$, $r$, and $z$ bands and found that the optical counterparts are only visible in the $g$ and $r$ bands, and barely visible in the $z$ band. So we discarded the $z$-band images and processed only the $g$- and $r$-band images.

Secondly, we preliminarily determined the optical center and size of the sources using $g$-band images, which are with the highest signal-to-noise ratio (S/N). We cropped the $g$- and $r$-band images to ensure that sources are centered and frames are 5-10 times the size of the sources.

Thirdly, we examined the local background of the sources to determine whether the coadded images could be used directly for photometry. We used a series of methods to make this determination, as described below: \romannumeral1) we conducted aperture photometry with a series of increasing radii, and evaluated the quality of the background subtraction by observing whether the flux curve in the apertures rises with the radius and then flattens; \romannumeral2) we performed aperture photometry on the dark background at different locations and checked whether the flux approaches zero; \romannumeral3) we estimated and extracted background using \texttt{Background2D} in \texttt{Photutils} of \texttt{Astropy}, and checked whether there are large-scale gradients in the background; \romannumeral4) we conducted sigma clipping manually on the background value and found all images showing the absolute value of less than 0.0003. After thorough analysis, we confirmed that the background had been properly processed, without over or under-subtraction and without large gradients.

After examination, we began to carry out the photometry. We first used software \texttt{SExtractor} \citep{1996A&AS..117..393B} to detect sources on the smoothed $g$-band images. Photometry was performed using \texttt{APER} and \texttt{AUTO} modes on the original coadded images. However, due to the faintness and irregular morphology, we encountered difficulties in detecting our sources completely on some images. 

To address the issue, we adopted an innovative approach that combines the utilization of \texttt{MTObjects} \citep{10.1007/978-3-319-18720-4_14} for target detection and the application of \texttt{SExtractor} for source photometry. \texttt{MTObjects} detects 2D image sources using statistical tests with the ``Max-Tree'' algorithm. It performs exceptionally well in detecting faint extended sources, surpassing traditional tools like \texttt{SExtractor}. We performed the source extraction by \texttt{MTObjects} on the smoothed and bright-sources-masked images, using default parameters but reducing the \texttt{move\_factor} to enlarge the spread of faint extended sources. The segmentation images are shown in the right panel of Fig. \ref{fig:02}. Then we masked all the other sources on the coadded images based on the segmentation images and then used \texttt{SExtractor} for aperture photometry on the masked images. To determine the source center and aperture size, we used the detection results of the $g$-band images, proceeded with a manual visual inspection, and removed a portion of unreliable pixels at the edge of the segmentation. This process aims to minimize the potential over-detection by \texttt{MTObjects} of extended pixels beyond the actual periphery of the sources \citep{10.1007/978-3-319-18720-4_14}, and create a more accurate \textit{true} source detection segmentation. The basis for the inspection is that excessive, small ``dendritic'' structures at the edges of galaxies and extended, filamentous structures without environmental disturbance will be inconsistent with the peripheral gravitational balance and the dynamic equilibrium of galaxies, so they are most likely the results of over-detection. We carefully removed part of the ``dendritic'' portions from galaxy peripheries by empirical judgment, which are structures that are highly unlikely to reflect the true morphology of a galaxy. We used the center detected by \texttt{MTObjects} as the initial input for \texttt{SExtractor} and set the initial photometric aperture as the radius covering the ``true'' sources in the detection of \texttt{MTObjects}. With reliable parameters, \texttt{SExtractor} successfully performed aperture photometry on our sources. Some minor adjustments have also been made to the apertures and centers to fully cover sources while minimizing the influence of background noise.

We obtained the photometric flux of each source in the $g$ band and calculated the $g$-band magnitude, using the flux-magnitude conversion relation provided in the header file of the images. Similarly, we performed photometry on the $r$-band images using the same center, aperture, and approach as the $g$-band images. Then we obtained the $g-r$ color (without extinction correction), ranging from -0.098 to 0.55 among the eight sources.

Subsequently, we performed extinction corrections on the $g$- and $r$-band magnitudes and the color. The range of $g-r$ color (without extinction correction) suggests that the sources are quite blue, indicating minimal impact from the internal dust and negligible obscuration by foreground dust. Intrinsic dust extinction can thus be disregarded, with only Galactic extinction requiring consideration. We utilized a Python module \texttt{dustmaps} to access the two-dimensional (2D) ``SFD'' dust map \citep{1998ApJ...500..525S} of interstellar dust reddening and extinction, and applied the relation from Table 6 of \citet{2011ApJ...737..103S} based on \citet{1999PASP..111...63F}, with $A(\lambda)/E(B-V)=3.237$ for the $g$ band and 2.176 for the $r$ band. The magnitudes with and without Galactic extinction correction of the $g$ and $r$ bands are listed in Col.(1) and Col.(2) of Table \ref{tab:02}, respectively, and both colors before and after the extinction correction can be found in Col.(5) of Table \ref{tab:02}. We evaluated the impact of the extinction correction on stellar mass, a physical parameter closely related to the magnitude and color, the calculation of which will be described in detail in Section \ref{sec:03-03}, and found that the stellar mass of these sources remained consistent within 2\% regardless of whether the correction was applied or not.

Furthermore, given that our method involved subjective decisions about the size of the sources, we took additional measurements to ensure the reliability of our photometric results. We performed another photometry on the $g$-band images employing the $R_{e\!f\!f}$ and center given by \texttt{MTObjects} and then compared the results. The flux of our sources, measured within the $R_{e\!f\!f}$ of \texttt{MTObjects}, exhibits a minimal difference, within $12\%$, except for AGC\,258609 (but also $<30\%$). The $g$-band photometric results based on $R_{e\!f\!f}$ from \texttt{MTObjects} and the $R_{e\!f\!f}$ given by \texttt{MTObjects} are listed in Col.(6) and Col.(11) of Table \ref{tab:02}, respectively.  

Additionally, we compared our photometric results of AGC\,215414, AGC\,215415, and AGC\,123216 with those reported in three different studies \citep{2014ApJ...786..144N,2018A&A...615A.105M,2023AJ....165..197G}. The comparison revealed discrepancies in the $g$-band magnitudes and $g-r$ colors. Specifically, the $g$-band magnitudes of AGC\,215414, AGC\,215415, and AGC\,123216 in the literature are found to be 17.1, 18.18 and 20.80\,mag, respectively, with the corresponding colors of 0.45, 0.881 and $-0.11$; while our photometry yielded $g$-band magnitudes of 17.63, 18.68, and 20.63 mag, respectively, with $g-r$ colors of 0.29, 0.55, and 0.046. It was found that, for AGC\,123216, our photometric results are in complete concordance with those reported by \citet{2023AJ....165..197G}, within the allowed margin of error. However, for the other two sources, our results are $\sim0.5$\,mag fainter in magnitude, and $\sim0.2$ bluer in color without considering the error. We analyzed the cause and found that it was possibly due to the choice of photometry methods. We used an aperture photometry method, while in \citet{2014ApJ...786..144N} and \citet{2018A&A...615A.105M}, they used an exponential or a S\'{e}rsic brightness profile to fit the surface brightness distribution. The discrepant results obtained from the aperture photometry and fitting photometry indicate that the sources being studied may not fit well with a standard exponential or S\'{e}rsic profile, or their edges may be too faint to determine an aperture that captures adequate optical emission through conventional methods. To derive measurements that more accurately reflect the ``real'' physical properties of these sources, it may be imperative to employ deeper images or more advanced methods of detection and photometry for faint objects.

After the above verification and comparison, we confirmed the reliability of our measurements. The flux of the sources under our photometry is neither overestimated nor underestimated.

\subsection{Surface Brightness}
As shown in Fig. \ref{fig:01}, the eight sources are faint, with low S/Ns and irregular morphologies. We attempted to fit the surface brightness profile of these sources using a classical method by \texttt{GALFIT} \citep{2011ascl.soft04010P} with the coadd point spread function (PSF) from DECaLS, which is an extended PSF model combined with a flexible fittable inner Moffat profile and a fixed outer power-law profile. However, the classical method failed to provide an accurate fit for six out of the eight sources, regardless of applying exponential, S\'{e}rsic, or a combination of both models. The small angular diameter of the sources resulted in insufficient sampling, making accurate fitting difficult. Therefore, we followed the method of surface brightness measurements on the \textit{almost dark} galaxy AGC\,229385 \citep{2015ApJ...801...96J}, and took a similar approach to measure the surface brightness.

We used the coadded images of our sources and masked the bright stars above the $5\sigma$ detection threshold. We determined the position angle, major axis, and minor axis by ellipse fitting of \texttt{SExtractor}. Based on this, we placed a series of boxes on the optical images to cover the sources, taking into account their morphology and size. The boxes are sized to both capture variations in brightness and minimize the influence of background noise. By tracing along the major axis as much as possible and considering the morphology of the sources, we positioned the boxes accordingly. Using this method, we could determine the surface brightness along the approximate direction of the major axis and the peak surface brightness of the sources. This enables us to avoid situations where the surface brightness fitting cannot be conducted due to the low S/N at the edge of the sources.

The sizes and positions of the boxes overlaid on the unmasked $g$-band images of the eight sources are shown in the left panel of Fig. \ref{fig:03}. We measured the total flux of all the pixels within each box, calculated the surface brightness, and plotted curves of the surface brightness varied by the different positions of the sources. The surface brightness curves can be found in the right panel of Fig. \ref{fig:03}. 

For noise estimation, we used the inverse variance images corresponding to the coadded images in $g$ and $r$ bands of the same region. As introduced in \ref{sec:02-03}, the value of each pixel in the inverse variance images is the inverse variance calculated from the coadded images, and we derived the variance of each pixel and calculated the error of the flux within each box by unbiased estimation, which follows
\begin{equation}
    {\rm error}=\sqrt{\frac{\sum{\rm var}}{(0.262l)^2\times(l^2-1)}},
\end{equation}
where $\sum{\rm var}$ is the sum of the variances of pixels in the boxes derived from the inverse variance images, 0.262 is the pixel scale of the coadded image of DECaLS in the unit of ${}^{\prime\prime}/{\rm pixel}$, and $l$ is the length of each box in the unit of pixels. With the error, we also gave the upper and lower limits of the surface brightness within each box.

Finally, for evaluating the background noise, we placed several boxes with a size of $20\times20$ pixels on the inverse variance images, where correspond to dark, starless background on the coadded images, and calculated the average error $\sigma$ of the pixels within the boxes. The confidence levels with $3\sigma$ as the detection threshold are drawn in the right panel of Fig. \ref{fig:03}, and images of both $g$ and $r$ bands show a fairly high S/N, with the surface brightness exceeding the $3\sigma$ sky level at almost every detection. Therefore, the detected signals could accurately reflect the emissions of sources, especially when measured at the peak. The peak surface brightness of the eight sources in the $g$ band is listed in Col.(10) of Table \ref{tab:02}.

\subsection{Stellar Mass and HI-to-stellar mass ratio}\label{sec:03-03}

Stellar mass $M_{*}$ is a critical indicator to characterize a galaxy, and it can be estimated through spectral energy distribution (SED) fitting based on multi-band luminosity. SED fitting involves fitting stellar population synthesis (SPS) models to multi-band luminosity, making key assumptions including the IMF, star formation history (SFH), nebular emission, and dust attenuation law. Several SED modeling codes have been developed to estimate the physical properties of galaxies from broadband data, such as MAGPHYS \citep{2012IAUS..284..292D}, FAST \citep{2018ascl.soft03008K}, CIGALE \citep{2019A&A...622A.103B}, and Prospector \citep{2021ApJS..254...22J}. The physical properties of a galaxy derived from SED fitting depend on the choice of SPS models and physical assumptions.

In cases where multi-band data are not available, color-stellar mass-to-light ratio relations (CMLRs) can be used to estimate $M_{*}$. CMLRs are the empirical relations between color, single-band luminosity, and $M_{*}$. They are derived from SED fitting methods applied to various samples of galaxies. They provide a convenient method that does not require multi-band data and works as long as the color and luminosity of one band are available. Multiple calibrations of CMLRs in multiple bands have been proposed: some are based on the observations of different types of galaxies, such as the CMLR calibrations for spiral galaxies in \citet[B03]{2003ApJS..149..289B}, for irregular dwarf galaxies in \citet[H16]{2016AJ....152..177H}, and for LSBGs in \citet[D20]{2020AJ....159..138D}; while others are based on model data of SPS, such as \citet[Z09]{2009MNRaS.400.1181Z}, and \citet[IP13]{2013MNRaS.430.2715I}.

Due to the lack of infrared data, we used the CMLR method to calculate the $M_{*}$ of our sources. When applying this method to a galaxy, both the priority among different CMLRs and the distinction of $M_{*}$ obtained from an individual CMLR in different bands need to be considered \citep{2020AJ....160..122D}. We determined the suitability of these calibrations and selected the calibration of H16 for our estimation, since the irregular morphology, small size, and blue color of our sources are similar to those of the sample of dwarf irregular (dIrr) galaxies used in their paper.

The calculation of the $M_{*}$ of eight sources is as follows. In Section \ref{sec:03-01}, we got the $g-r$ colors, which had been corrected for Galactic extinction. We brought them into the CMLR of H16, which is 
\begin{equation} \label{eq:02}
  \log_{10}(M_{*}/L_g)=-0.601+1.294\times(g-r).
\end{equation}
The $M_{*}$ can be estimated from the stellar mass-to-light ratio $M_{*}/L_g$ and the $g$-band luminosity $L_g$. We converted the apparent magnitude $m_g$ to absolute magnitude $M_g$ by 
\begin{equation}
  M_g=m_g-5\log_{10}{(\frac{d}{10\,\rm pc})},
 \label{eq:03}
\end{equation}
where $d$ is the distance given by ALFALFA, as shown in Col.(10) of Table \ref{tab:01}. We did not apply K-correction since all sources have low redshifts, with the highest being $z=0.017$. For the convenience of comparing the $M_{*}$ with the H\textsc{i} mass $M_{\rm H\textsc{i}}$, both $L_g$ and $M_{*}$ were converted into solar units, and the absolute magnitude of the sun $L_\odot$ in the $g$ band is $5.05$ mag \citep{2018ApJS..236...47W}. The $L_g\,[L_\odot]$  could be obtained from 
\begin{equation}
  L_g\,[L_\odot]=10^{-0.4(M_g-M_\odot)},
\end{equation}
then the $M_{*}$ could be calculated by Eq. (\ref{eq:02}). The error of the $M_{*}$ is calculated through the error transfer formula, based on the photometric error of $g$-band images given by \texttt{SExtractor}, the error of $d$, and the uncertainty of the CMLR. The estimated $\log_{10}M_{*}$ and the error of our sources are shown in Col.(7) of Table \ref{tab:02}.

Furthermore, it is important to make a comparison of the $M_{*}$ estimates obtained under different calibrations, as it is unclear which one is more applicable. One of the most significant factors that affect the uncertainty of the $M_{*}$ is the IMF \citep{2001ApJ...550..212B}. As the low-mass stars contribute significantly to the total mass of a galaxy but contribute little to the total luminosity, an IMF with more massive stars gives a smaller estimate of total stellar mass. The Salpeter IMF \citep{1955ApJ...121..161S} is based on disk stars and shows that the number of stars decreases rapidly with increasing mass in an exponential form. The Chabrier IMF \citep{2003PASP..115..763C} considers different stellar components and is in the form of a logarithm-normal distribution. Another common IMF, given by \citet{2001MNRAS.322..231K}, is based on field stars and is in the form of a broken power law. It has a similar shape to the Salpeter IMF at higher mass but flattens out at lower mass.

Different CMLRs adopt different IMFs. The CMLRs of B03 are based on a ``diet'' Salpeter IMF, which is a modified form of reducing the number of faint low-mass stars in Salpeter IMF; while most CMLRs, such as Z09, H16, and D20, are based on the Chabrier IMF; additionally, the relation of IP13 is based on the Kroupa IMF. This brings in zero-point deviations between the CMLRs. Here we thus specifically evaluate the variations in $M_{*}$ estimates produced by various IMF and CMLR assumptions, shifting our assumption of the IMF and CMLR temporarily, while the rest of the paper consistently adopts the assumption of the Chabrier IMF and H16 CMLR. We unified the CMLRs of B03, Z09, IP13, H16, and D20 to the same IMF of \citet{1955ApJ...121..161S}, \citet{2001MNRAS.322..231K}, and \citet{2003PASP..115..763C}, respectively. The conversions between IMFs in CMLRs are achieved by adding correction factors to $\log_{10}(M_{*}/L)$. We followed the conversion factor of +0.15 given by \citet{2003ApJS..149..289B} to convert ``diet'' Salpeter IMF to Salpter IMF, of -0.093 given by \citet{2008MNRAS.383.1439G} to convert ``diet'' Salpeter IMF to Chabrier IMF, and of +0.057 given by \citet{2016AJ....152..177H} to convert Kroupa IMF to Chabrier IMF. Based on the conversions between each IMF, we estimated the corresponding $M_{*}$ for each CMLR. Table \ref{tab:03} presents the converted $\log_{10}M_{*}$ under different IMFs of B03, Z09, IP13, H16, and D20. We found that under different assumptions of IMFs and CMLRs, the B03 CMLR with a Salpeter IMF would give the highest mass, and the Z09 CMLR with a Kroupa IMF would give the lowest mass, with a difference of $0.45\sim0.92$ in $\log_{10}(M_{*}/M_\odot)$ of our sources.

Subsequently, we calculated the H\textsc{i}-to-stellar mass ratio $M_{\rm H\textsc{i}}/M_{*}$ of the eight sources using the $M_{*}$ estimates in Table \ref{tab:02} and the archival $M_{\rm H\textsc{i}}$ from VLA for AGC\,215414, AGC\,225880, and AGC\,229197 \citep{2014ApJ...786..144N,2010ApJ...725.2333K,1995AJ....109.2415C} and from ALFALFA for the rest five sources \citep{2018ApJ...861...49H}. As shown in Table \ref{tab:01}, the $M_{\rm H\textsc{i}}$ from VLA agree well with those from ALFALFA for the three sources with both observations. Except for AGC\,215415, the derived $\log_{10}(M_{\rm H\textsc{i}}/M_{*})$ ratios for these sources are remarkably high, ranging from 1.69 to 2.77, even using the lowest estimates from various CMLRs and IMFs.

\subsection{Star Formation Rates}\label{sec:03-04}

The star formation rate (SFR) provides us with an insight into the evolution history and current state of galaxies. Various indicators, such as H$\alpha$ emission, radio continuum emission, total infrared (TIR) flux, UV flux, X-ray emission, and composite multi-wavelength luminosity, can be used to determine SFRs. 

Given the limited available data,  we relied on the UV flux indicator to estimate SFRs. This method is based on the observation that the youngest stellar populations, such as OB stars, will ionize the hydrogen surrounded and emit most of the energy in the UV band. If dust attenuation is negligible, the UV flux has a linear relation with SFRs.

We preliminarily examined GALEX FUV and NUV images for the eight sources with the longest exposure time. Only two sources (AGC\,225880 and AGC\,229197) show significant NUV emissions, and only AGC\,229197 has relatively good imaging quality in the FUV band. Nonetheless, for the other five sources, we can still at least provide the upper limits for the SFRs. 

We obtained the NUV and FUV magnitudes by \texttt{gAperture}, as introduced in Section \ref{sec:02-04}. We provided the coordinates, photometric apertures, background annulus sizes, and time ranges of the sources to \texttt{gAperture}, and then it could help carry out the aperture photometry and output the magnitude. For the consistency of photometry, we chose the same center and aperture as the $g$ and $r$ bands. The background annulus sizes depend on the radius within which there is no other emission, as such an area can represent the background value well\footnote{We did not perform PSF matching between the photometry apertures of UV and optical bands, considering the weak UV emission and the potential background issues with larger apertures. As seen in Section \ref{sec:04-03}, our conclusions will not change by a larger UV aperture.}. We corrected the Galactic extinction using values from \citet{2007ApJS..173..293W}, which adopted $A(\lambda)/E(B-V)=8.24$ for the FUV band and 8.2 for the NUV band, based on the extinction law of \citet{cardelli1989relationship}. The FUV and NUV magnitudes before and after extinction correction are listed in Col.(3) and Col.(4) of Table \ref{tab:02}, respectively. 

To calculate the SFR from UV luminosity $L_\nu$, a commonly used classical calibration was given by \citet{1998ARA&A..36..189K}, which assumed a Salpeter IMF \citep{1955ApJ...121..161S}, solar metallicity, a stellar mass interval of $0.1\sim100\,M_\odot$, and a continuous SFH over a past period of $0\sim100\,{\rm Myr}$. The conversion equation is given by \citep{1998ARA&A..36..189K}, which is
\begin{equation}
\vspace{-0.1cm}
    {SFR}\,(M_\odot\,{\rm yr}^{-1})=1.4\times10^{-28}L_\nu
    \,({\rm erg}\,{\rm s}^{-1}\,{\rm Hz}^{-1}).
\label{eq:05}
\end{equation}

However, there may be several problems when using the formula. The effect of dust attenuation is a concern since it is much larger at the UV band than at the other wavelengths. But given the faintness and blue color of our sources, the level of dust attenuation should be sufficiently low and the effect can be reasonably ignored. Furthermore, observations revealed that the impact of dust is negligible at low SFRs \citep{2010MNRAS.403.1894D}, while as the SFRs rise, the impact of dust becomes increasingly critical. Another problem is that Eq. (\ref{eq:05}) is given based on a Salpeter IMF, while the $M_{*}$ was calculated under the assumption of a Chabrier IMF in this work, and we converted the IMF of Eq. (\ref{eq:05}) to a Chabrier IMF using the conversion factor of 1.58, as recommended by \citet{2007ApJS..173..267S}.

Before calculating SFRs using UV magnitudes, we assessed the confidence that the measured UV emissions were from the sources rather than the background. A straightforward comparison between the measured magnitude and the detection limit of GALEX under a similar exposure time can make the judgment. If our measured magnitude is brighter than the corresponding detection limit, UV emissions from the sources may be detected. Table \ref{tab:04} presents the exposure time and measured magnitudes of the FUV and NUV images of our sources, as well as $5\sigma$ detection limits for various sky survey strategies of GALEX \citep{2005ApJ...619L...7M}. We found that the measured FUV magnitudes for six sources, except for AGC\,229197, are significantly deeper than the detection limits of GALEX at a comparable exposure time. Therefore, the flux we measured on the FUV images is more likely to come from the background than from the emission of the sources. For NUV images, all sources, except for AGC\,123216, exhibit magnitudes shallower than the 5$\sigma$ detection limit of GALEX at corresponding exposure time, indicating that the flux of NUV images is highly likely from the sources themselves. Therefore, we used NUV band data to calculate SFRs due to more reliable source emissions, despite FUV being a relatively more accurate diagnostic tool for measuring SFRs.

We obtained the absolute NUV magnitude ${M}_{\rm NUV}$ using the same way as Eq. (\ref{eq:03}), and deduced the luminosity $L_\nu$ by a basic transformation, given by 
\begin{equation}
L_{\rm NUV}({\rm erg}\,{\rm s}^{-1}\,{\rm Hz}^{-1})=
10^{-0.4({M}_{\rm NUV}-51.6)}.
\end{equation} 

Col.(9) of Table \ref{tab:02} presents the calculated SFRs of our sources before and after extinction correction. We found that whether to perform the extinction correction or not would result in a maximum difference of about 20$\%$ in SFRs. The SFRs are quite low, which could suggest that the star formation process has not yet commenced on the observed scale or has been somehow suppressed in recent years.

\section{Scaling Relationship and Discussion}\label{sec:04}
 
In Section \ref{sec:03}, we obtained a set of properties for the eight sources from multi-band data. In this section, we aim to investigate the scaling relations among these properties and compare them with those of the $\alpha.100$ sample, the \textit{almost dark} galaxy, LSBGs, and HUDs. In all relations, we uniformly used the physical parameters of our sources after the Galactic extinction correction.

\subsection{$H\textsc{i}$ Mass, Stellar Mass and (NUV-$r$) color}

As the reservoir from which all stars form, H\textsc{i} gas plays an indispensable role in the process of star formation. It is related to various physical parameters, such as $M_{*}$ and color. The relations among them can provide insights into past evolutionary processes, such as recent star formation \citep{2018MNRaS.476.4488H}, ram-pressure stripping \citep{2022arXiv220802634J}, and tidal interactions. Therefore, investigating the scaling relations is crucial to infer the current evolution stage of sources and create a coherent picture of galaxy evolution.

Here we selected four parameters, $\log_{10}M_{\rm H\textsc{i}}$, $\log_{10}M_{*}$,  $\log_{10}(M_{\rm H\textsc{i}}/M_{*})$ and the (NUV-$r$) color, which were confirmed to appear close relations with each other in $\alpha.40$ sample \citep{2012ApJ...756..113H}, reflecting the ratio of recent and past star formation.

Following the work of \citet{2012ApJ...756..113H}, we constructed the $\alpha.100$-GALEX-SDSS sample, and used it as a comparison of our sources. This sample was obtained by cross-matching the $\alpha.100$-SDSS catalog from \citet{2020AJ....160..271D} with the GALEX-SDSS sample from \citet{2017ApJS..230...24B}. The latter is a combination of AIS GUVcat and SDSS DR14, providing a set of unique UV sources and their parameters from the AIS of GALEX. We matched the sources by SDSS ID, selected the sources classified as galaxies, and then built the $\alpha.100$-GALEX-SDSS sample, which contains 11943 galaxies. Fig. \ref{fig:04} shows the scaling relations of our sources, the $\alpha.100$-GALEX-SDSS sample, and the \textit{almost dark} galaxy AGC\,229385. All of our sources fall outside the $2\sigma$ range ($95\%$) of the $\alpha.100$-GALEX-SDSS sample, and most of them also exhibit significant deviations from the scaling relation of the $\alpha.100$ sample. Additionally, at least six of our sources lie close to AGC\,229385 in each relation. 

Specifically, as demonstrated in Fig. \ref{fig:04}(a) and \ref{fig:04}(b), in the (NUV-$r$) versus $\log_{10}M_{\rm H\textsc{i}}$ and $\log_{10}M_{*}$ versus $\log_{10}M_{\rm H\textsc{i}}$ relations, most of our sources exhibit similar $\log_{10}M_{\rm H\textsc{i}}$ values to that of the $\alpha.100$ sample, while the (NUV-$r$) color and $\log_{10}M_{*}$ of our sources are lower than the overall typical values for the $\alpha.100$ sample, which range $-1.2\sim0.9$ and $5.2\sim7.1M_\odot$, respectively. For the (NUV-$r$) versus $\log_{10}(M_{\rm H\textsc{i}}/M_{*})$ and $\log_{10}M_{*}$ versus $\log_{10}(M_{\rm H\textsc{i}}/M_{*})$ relations, as shown in Fig. \ref{fig:04}(c) and \ref{fig:04}(d), our sources deviate from the trend of typical distribution of the $\alpha.100$ sample, with extremely high $\log_{10}(M_{\rm H\textsc{i}}/M_{*})$, ranging from 1.81 to 3.27, except for AGC\,215415. This suggests they may have unusual and extreme physical origins and evolution histories. Among these relations, our sources exhibit a significantly high $M_{\rm H\textsc{i}}/M_{*}$, low $M_{*}$, and distinct blue colors.

\subsection{Baryonic Tully-Fisher Relation}\label{sec:04-02}

The baryonic Tully-Fisher Relation (BTFr) is a strongly correlated relation between baryonic mass $M_{\rm bar}$ and rotation velocity. It was first proposed by \citet{2000ApJ...533L..99M} in the form of $M\propto v^4$, and has been confirmed through the application to various types of galaxies \citep{2021MNRAS.508.1195P, 2022ApJ...940....8M}. It provides a redshift-independent measurement of distance and a rigorous test for galaxy-forming models. 

We investigated the $M_{\rm bar}$ and rotation velocity of our sources by BTFrs. We followed the recommendation of \citet{2012AJ....143...40M} and represented the rotational velocity as $V=W _{\rm int}/2$, where $W_{\rm int}$ is the intrinsic linewidth of the H\textsc{i} line profile. Studies have shown that $W_{50}$ is more prone to the influence of noise and blending, making it a less reliable estimator of the intrinsic linewidth, and $W_{20}$ is less sensitive to these effects and is therefore considered a more reliable estimator of the intrinsic linewidth in most cases \citep{2016AJ....152..157L,2017MNRAS.469.2387P}. The simulation results also found that at lower masses, $W_{50}$ deviates from the expected trends of other velocity estimators and tends to underestimate velocities more severely \citep{2016MNRAS.459..638B}. Therefore, we used archival $W_{20}$ in ALFALFA as the estimators of the observing linewidth, which is measured at the $20\%$ level of the single peak or each of the two peaks of H\textsc{i} line profile. We obtained the uncertainty of $W_{50}$ from ALFALFA archival data, derived from the statistical error of S/N of H\textsc{i} \citep{2011AJ....142..170H,2018ApJ...861...49H}, and converted it to the uncertainty of $W_{20}$. 

As described in \citet{2018ApJ...861...49H}, it is difficult to provide a robust estimate of $W_{20}$ for a lower peak due to the low S/N. To ensure the reliability of the archival $W_{20}$ parameters, we checked the H\textsc{i} line profiles of our sources and found that all of the eight sources show relatively narrow single-horned velocity profiles instead of pronounced double-horned profiles. We refitted the lines using single and double Gaussian, Lorentz, and Voigt profiles, and obtained the best fitting parameters, including the FWHM and fitting error. We compared them with the archival parameters and found that seven of the eight sources matched within the margin of error, except for AGC\,219841, where a potentially double-horned profile significantly disturbs the linewidth estimation. The H\textsc{i} line profiles for the eight sources are shown in Appendix \ref{appe}.

To determine the intrinsic linewidth, we performed the $g$-band photometry in \texttt{AUTO} mode of \texttt{SExtractor} to obtain the semi-major axis (a) and semi-minor axis (b). We applied the relations $\cos{i}=a/b$ (hereafter ``ab-ratio'') and $W_{\rm int}=W_{\rm obs}/\sin{i}$ to perform inclination correction. It is worth noting that none of our sources show very low inclinations ($i<30^\circ$): AGC\,219841, AGC\,229197, and AGC\,258609 have inclinations lower than $45^\circ$, with values of $38.5^\circ$, $36.3^\circ$, $40.1^\circ$, respectively; and the other five sources show inclinations greater than $45^\circ$. This leads to a very puzzling situation: for most of the sources, the inclination angles characterized by the optical ab-ratio indicate that these sources are not face-on; however, the H\textsc{i} profiles are narrowly single-horned instead of double-horned, as shown in Appendix \ref{appe}. Possible explanations for this puzzling issue are as follows: \romannumeral1) the uncertainty of the measurement of inclination angles is significant due to the low S/N, which may result in incorrect results; \romannumeral2) their rotational velocities may be extremely low, or they are not rotation-support but pressure-support systems; and the velocity distribution map of AGC\,123216 corroborates this, revealing a notably low rotational velocity of about $15\,\kms$ \citep{2023AJ....165..197G}; \romannumeral3) the kinematics of the gas and stellar components may be different, leading to the optical ab-ratio no longer being a suitable probe for tracking the inclination of the overall sources. Meanwhile, these possibilities also imply that the $W_{20}$ parameters of these sources may not be ideal indicators of rotation velocity.

For the baryonic mass, we calculated it as $M_{\rm bar}=M_\ast+M_{\rm gas}$, where $M_\ast$ is the stellar mass calculated in Section \ref{sec:03-03}. For the estimation of $M_{\rm gas}$, we considered a simple form of $M_{\rm gas}=\gamma M_{\rm H\textsc{i}}$, where $\gamma$ is determined by assuming a cosmological hydrogen fraction of $75\%$ hydrogen and $25\%$ helium. Using the assumption of the primordial hydrogen fraction ($\gamma=1.33$) instead of solar value ($\gamma=1.4$) is more appropriate for our sources, as they are more similar to low-luminosity, gas-rich late-type galaxies \citep{2012AJ....143...40M}. The estimation of the baryonic mass error $\delta M_{\rm bar}$ followed the method of \citet{2016ApJ...816L..14L} in forms of 
\begin{small}
\begin{equation}
\vspace{-0.12cm}
    \delta M_{\rm bar}=\sqrt{(\delta M_g)^2+(\gamma_\ast\delta L_g)^2
    +(L_g\delta\gamma_\ast)^2+\left(2M_{\rm bar}\frac{\delta d}{d}\right)^2},
\end{equation}
\end{small}
where $\delta M_{\rm gas}$, $\delta L_g$, $\delta d$ are the errors of gas mass, $g$-band luminosity, and distance, respectively. $\delta M_{\rm gas}$ and $\delta d$ are obtained from the error provided by ALFALFA, while $\delta L_g$ is derived from the error of flux given by \texttt{SExtractor}, calculated through the error transfer formula. $\gamma\ast$ is the stellar mass-to-light ratio of galaxies, and the error $\delta\gamma\ast$ is derived from the CMLR coefficient error in \citet{2016AJ....152..177H}.

Fig. \ref{fig:05} presents the locations of our sources on the BTFr diagram. We compared our sources to a sample of 121 late-type disk galaxies from SPARC \citep{2016AJ....152..157L}, using $W_{20}$ as the linewidth estimator. The resulting BTFr is described as follows: 
\begin{equation}
\vspace{-0.12cm}
    \log_{10}\frac{M_{\rm bar}}{M_\odot} = (3.75\pm0.08)\log_{10}\frac{W_{2      
                                                        0}}{2}+(1.99\pm0.18).
    \vspace{-0.07cm}
\end{equation}
As shown in Fig. \ref{fig:05}, before the inclination correction, all of our sources deviate from the empirical BTFr towards the heavier baryonic mass end, which are donated by red pentagrams. After inclination correction, five of our sources (numbers 3, 4, 5, 6, and 8) agree with the BTFr comparatively, while the other three (numbers 1, 2, and 7) still deviate significantly from the BTFr, which are all shown in grey stars. The agreement of BTFr shows a well-defined distribution of baryonic and dark matter, implying that the five sources may have undergone a relatively quiescent evolutionary history, with limited merging and interactions with other galaxies. Their angular momentum may stay relatively conserved during the past evolving histories. The deviation towards the heavier baryonic mass end of the BTFr suggests that the sources may have undergone tidal interactions. Alternatively, it may require the introduction of special formation scenarios or new theories to explain the unknown cause, as discussed in \citet{2020NatAs...4..246G}.

Besides, it is essential to emphasize the significant effect of inclination correction. The inclination derived from \texttt{SExtractor} is highly uncertain due to the low surface brightness (LSB) and irregular morphology of these sources. Additionally, research has indicated that the inclination angle determined through the optical band may not accurately represent the inclination angle of H\textsc{i} gas \citep{2009A&A...505..577T}. A concerning scenario is that, when considering all factors, sources with inaccurately estimated inclinations could be located at any positions indicated by the arrows in Fig. \ref{fig:05}. However, although this possibility cannot be completely discounted, the overall trend of the sources towards the heavier baryonic mass end indicates that the impact of inclination uncertainty is probably not significant enough to invalidate this tendency.

\subsection{SFR and Mass Relation} \label{sec:04-03}

Based on a bimodal distribution in color, luminosity, and other properties found in a large sample analysis \citep{2001AJ....122.1861S, 2004ApJ...600..681B}, galaxies can be categorized into two groups, star-forming galaxies and quiescent galaxies \citep{2004MNRAS.351.1151B}. Blue galaxies like our sources are usually star-forming, and researches show that their SFRs increase with stellar mass up to a limit and then decline. It is known as ``the main sequence (MS)'', well established for star-forming galaxies across a wide range of redshifts \citep{2007ApJ...660L..43N, 2007ApJ...670..156D}.

The relation between SFR and $M_\ast$ has also been studied in the low mass galaxies and LSBGs. They both show a deviation from the MS determined by massive and bright galaxies, with their SFRs being lower than the expected SFRs of their mass, which may suggest a decrease of star-forming efficiency (SFE) at low mass and LSB end \citep{2010AJ....139..447H,2012ApJ...757...85G,2012AJ....143..133H,2018ApJS..235...18L}. 

We compared our sources with two MSs of the star-forming galaxy at $z=0$, which are derived from the blue star-forming galaxies without the contribution of active galactic nuclei (AGN) from SDSS \citep{2007A&A...468...33E}, and 22816 star-forming galaxies (both blue and red galaxies) in the NOAO Extremely Wide-Field Infrared Imager Medium-Band Survey \citep{2012ApJ...754L..29W}, respectively. We also compared our sources with the MS of LSBGs \citep{2017ApJ...851...22M}, which is based on 56 late-type LSBGs with the $M_{*}$ ranging from $5\times10^{6}$ to $7\times10^{9}M_\odot$. As shown in the left panel of Figure \ref{fig:06}, our sources are situated at the low-mass end and match well with the relations of the star-forming galaxies. However, despite having similar surface brightness and morphology to the LSBGs, they differ from the relation of LSBGs, exhibiting higher SFRs. 

We also compared our sources with the $\alpha.100$-GALEX-SDSS sample. To ensure the validity of the comparison, we estimated the SFRs of $\alpha.100$-GALEX-SDSS sample using the same method as our sources in Section \ref{sec:03-04}. The outermost contour represents $95\%$ of the $\alpha.100$ sample, and our sources are in the outliers. We included the HUD sample from \citet{2017ApJ...842..133L} as a comparison and found that these HUDs also agree well with the MSs of star-forming galaxies, but our sources exhibit lower masses than the HUDs. Moreover, the position of the \textit{almost dark} galaxy AGC\,229385 on the diagram is very close to ours, which is also consistent with the MS characteristics of star-forming galaxies, and it shows very low mass, with SFR higher than that of the galaxies with such surface brightness.

However, when we plotted our sources on the $\log_{10}M_{\rm H\textsc{i}}$ and SFR relation diagram, as shown in the right panel of Figure \ref{fig:06}, our sources show the SFR characteristics opposite to that shown in the left panel, with most of them exhibiting lower SFRs compared to the $\alpha.100$-GALEX-SDSS sample. Meanwhile, the positions of our samples are consistent with that of the LSBG and HUD samples on the diagram, and also conform to the previously observed SFR characteristics of other LSBG samples \citep{2019ApJS..242...11L}.

The higher SFR of our sources than that of the LSBGs with similar $M_{*}$ may indicate that our sources have more effective mechanisms for star formation, such as efficient gas supply or star formation feedback, possibly triggered by mergers or interactions with other galaxies. The environmental factors, such as the pressure of the surrounding intergalactic medium or ram-pressure stripping caused by motion through dense medium, could also be responsible for the higher SFR. 

On the other hand, the lower SFR compared to the $\alpha.100$-GALEX-SDSS sources with similar $M_{\rm H\textsc{i}}$ suggests a low efficiency of H\textsc{i} gas conversion into stars. This inefficiency might stem from a lower rate of atomic-to-molecular gas conversion, which limits the direct supply needed for star formation. Alternatively, the gas component of our sources is so extended that the surface density of cold gas fails to reach the threshold for star formation \citep[e.g.][]{1989ApJ...344..685K,2001ApJ...555..301M}. Other factors, such as the gas stability and specific characteristics of the interstellar medium (ISM), would also result in the potential inefficiency in converting gas into stars, impeding the star formation process. 

Considering both of the above, one plausible explanation for our findings is that our sources have undergone a period of suppressed star formation in the past, potentially due to a variety of factors. However, they have recently initiated the process of star formation, and are currently in the early stages. Consequently, they possess low $M_{*}$ and maintain significant reserves of H\textsc{i} gas, exhibiting slightly higher SFRs than the MS of LSBGs but lower SFRs than that of the $\alpha.100$ sample on the $M_{\rm H\textsc{i}}$ versus SFR diagram.

\section{Environments and Possible Origins}\label{sec:05}

The environment is a crucial factor when considering the origins of LSB objects. In low-density regions, these objects are thought to form in situ, yet the reasons why they have not evolved significantly for such a long time remain unclear. In galaxy groups or clusters, interactions between galaxies compress and stretch interstellar medium, creating structures such as tidal bridges and tails between galaxies, which can lead to star formation outside galaxies. In this section, we aim to determine the possible origins of these galaxies by analyzing their environments. To distinguish between H\textsc{i} and optical emissions, we appended the suffix ``$OC$'' to the source name for the optical counterpart, while the original name denotes the H\textsc{i} component of the source.

\subsection{Two Tidal Dwarfs\footnote{``Dwarfs'' here means dwarf galaxy candidates, not stars; it holds the same meaning in subsequent mentions.}: AGC\,215414 and AGC\,229197}\label{sec:05-01-01}
Among our sources, the origin of AGC\,215414 and AGC\,229197 appears to be the most identifiable. These two systems exhibit clear tidal features or tendencies towards tidal origins. Prior research has conducted multiple investigations on the two systems. 

\subsubsection{AGC\,215414}
Possibly due to the faintness, AGC\,215414\,OC was not initially detected in the cross-matching between SDSS and ALFALFA surveys. AGC\,215414\,OC is located at the tip of the plume extending eastward from NGC\,3628. Observations and simulations indicated that NGC\,3628 and NGC\,3627 (M66), members of the Leo Triplet galaxy system, experienced a tidal encounter 800\,Myr ago, resulting in the formation of the plume \citep{1978AJ.....83..219R, 1979ApJ...229...83H}.  AGC\,215414\,OC has been identified as a tidal dwarf galaxy in later studies and observations \citep{2014ApJ...786..144N}.
A H\textsc{i} velocity distribution map of the tidal tail from the VLA observations (Project code: AB1074, PI: A. Bolatto) has also been presented in Fig. 3 of \citet{2014ApJ...786..144N}, which revealed a distinct difference in velocities between the plume and AGC\,215414, accompanied by a noticeable bending of the plume towards AGC\,215414 at its far end. This suggests that AGC\,215414 may not be gravitationally bound to the plume and could potentially evolve into a self-gravitating system, indicating that AGC\,215414\,OC is a tidal-origin galaxy.

Optical images from DECaLS could further support the evidence. AGC\,215414\,OC exhibits the inward-bending characteristic in both the $g$ and $r$ bands, as demonstrated in the upper panel of Fig. \ref{fig:07}, which shows the $g$-band image overlaid by H\textsc{i} distribution contours from VLA. These images also reveal numerous star-forming regions within AGC\,215414\,OC, suggesting that a considerable number of stars have already started to form inside it.

\subsubsection{AGC\,229197}

AGC\,229197 is a part of the intergalactic gas system called H\textsc{i}\,1225+01, located on the outskirts of the Virgo Cluster. This system is composed of two prominent H\textsc{i} clumps, each with $\rm M_{H\textsc{i}}\sim10^9\rm M_\odot$, distributed along the northeast (NE) and southwest (SW) directions, and connected by an intermediate H\textsc{i} bridge, which has $\rm M_{H\textsc{i}}\sim10^8\rm M_\odot$. The coordinate and H\textsc{i} mass of AGC\,229197 by ALFALFA survey is consistent with those of the bridge from VLA observations. The H\textsc{i} intensity map and velocity field from VLA also revealed a disturbed morphology of the SW component, along with a consistent radial velocity gradient in the SW clump and the bridge between the two clumps. Numerical simulations suggested that the system is the result of an undergoing prograde encounter between the NE and SW clumps, with the bridge component being pulled out from the SW clump by the NE clump \citep{1995AJ....109.2415C}.

The lower panel of Fig. \ref{fig:07} displays the H\textsc{i} distribution and the optical counterparts of H\textsc{i}\,1225+01. The NE clump comprises a blue LSBG; while the SW clump does not show any detectable stellar emission, which is considered a promising \textit{dark galaxy} candidate. AGC\,229197\,OC is precisely situated at the center of the intermediate H\textsc{i} bridge. 

Whether the optical emission we found corresponds to H\textsc{i} gas is up for discussion. We estimated the physical size of AGC\,229197\,OC based on the effective radius given in Table \ref{tab:02}. Assuming the optical counterpart is at the same distance of the H\textsc{i} component, we calculated a physical scale of $\sim234$\,pc for AGC\,229197\,OC. This is significantly larger than the 10\,pc threshold typically considered for the upper limit size of globular clusters \citep{2007AJ....133.2764B}. This suggests that \mbox{AGC \,229197\,OC} might have been a galaxy in the early stage of star formation, and it is likely to originate from the tidal process between the encounter of the NE and SW clumps. The specific location and star formation may also hold significant implications for understanding the evolutionary stage of the entire H\textsc{i}\,1225+01 system.

Although the possibility that the optical counterpart could be a distant background source cannot be excluded based on the available data, it is noteworthy that AGC\,229197\,OC coincides with a peak of H\textsc{i} in the H\textsc{i}\,1225+01 system. Furthermore, the system possesses limited stellar components that correspond to the optical characteristics of galaxies.  Therefore, it is more plausible to consider AGC\,229197\,OC as part of the system rather than a background galaxy. To confirm the redshift and proceed with further investigation, it is imperative to obtain an optical spectrum of this source.

\subsection{Four Group Dwarfs: AGC\,123216, AGC\,215415, AGC\,219841, and AGC\,258609}\label{sec:05-01-02}

The four sources are situated in medium-density environments, which do not show significant tidal features. Whether they have tidal origins or are products of galaxy interactions like AGC\,215414 and AGC\,229197 requires further investigation. Therefore, we searched for galaxies near the four sources and selected those with the same redshift as them. We examined a region with a diameter of 0.6\,Mpc around each source, based on the findings of the Stephan's Quintet in \citet{2022Natur.610..461X}: the H\textsc{i} large-scale structure of Stephan's Quintet, which may be produced by the tidal interactions of the five galaxies in the center, is known to extend up to 0.6\,Mpc, and is one of the most extended H\textsc{i} structures that could result from galactic interactions. To ensure consistency in distance measurements, we used the Hubble flow distance given by NASA/IPAC Extragalactic Database (NED) as a redshift indicator, which does not account for corrections for the Local Group, Virgo cluster, Great Attractor, and Shapley, as their effects are nearly identical for objects at the same redshift in the same direction. Considering the allowed redshift error, we found a number of galaxies with optical emission in the vicinity of the four sources, all sharing the same redshift. This suggests a potential co-evolution among these galaxies. 

Additionally, it is important to note that a reliable determination of the distance of objects should also take the proper motion velocity into account, and the sample of galaxies with the same redshifts under our method may not be complete. Nonetheless, it is sufficient for conducting discussions on the environment and making reliable indications about the origins of our sources.

\subsubsection{AGC\,123216} 

AGC\,123216\,OC is a faint, peanut-shaped object surrounded by NGC\,807, NGC\,805, MRK\,365, and KUG\,0202+283 (numbers 3, 4, 5, and 2, respectively). Their relative positions, coordinates, and redshifts are shown in Fig. \ref{fig:08} and Table \ref{tab:05}. NGC\,807, an early-type galaxy rich in H\textsc{i} and CO, is located to the northeast of AGC\,123216\,OC. VLA observations of the H\textsc{i} distribution revealed that it consists of two nearly equal-mass components: a smooth central H\textsc{i} disk, and two tidal arms extending towards the northwest and southeast directions. The distribution of CO emission shows asymmetry, with 70$\%$ of it located towards the southeast. NGC\,807 also exhibits some optical tidal features \citep{2013AJ....145...56L}. NGC\,805 is an SB0 galaxy, located very close to AGC\,123216\,OC, only $2.9^{\prime}$ away towards the almost due west (slightly north) of AGC\,123216\,OC. MRK\,365 is an Sc galaxy with a double nucleus and exhibits significant spiral arm structures \citep{2004AJ....128...62G}. KUG\,0202+283 is located to the southeast of AGC\,123216\,OC. It appears as a late-type barred spiral galaxy with a blue color. Multiple star-forming regions are present along the spiral arms, which indicates that the galaxy is undergoing a violent star-formation process. Despite these observations, research on NGC\,805, MRK\,365, and KUG\,0202+283 remains limited. However, the spatial positions of these galaxies suggest that they are most likely related to each other, and the three galaxies NGC\,807, NGC\,805, and Mrk\,365 have been identified to compose a compact group of galaxies, USCG\,U098 \citep{2002A&amp;A...391...35F}. It is widely recognized that compact groups of galaxies typically contain a significant amount of diffuse gas dominated by dark matter. Subsystems are also prone to forming in more loosely associated areas and then evolving through gravitational processes \citep{1997ARa&A..35..357H}. 

The spatial attributes of AGC\,123216\,OC suggest that it could be tidal debris from the interaction of member galaxies in the compact group, where a recent collapse of molecular gas has triggered the star formation. However, the concentration and morphology of the three galaxies do not indicate signs of interaction, and their colors are also different from that of AGC\,123216\,OC. Therefore, it is also possible that AGC\,123216\,OC was formed in situ, isolated from other galaxies. Additionally, there is no evidence to definitively rule out the possibility that it originated from another galaxy, KUG\,0202+283. Nevertheless, the distinct peanut-shaped morphological profile of AGC\,123216\,OC suggests that it could likely have already evolved into a self-gravitational system over time under gravitational force. 

Recent research presented the H\textsc{i} map of AGC\,123216, obtained through observations from WSRT, which was shown in Fig. 1 and Fig. 7 of \citet{2023AJ....165..197G}. In their work, the origin of AGC\,123216 is claimed to be ambiguous, and two potential scenarios for the origin were prompted: the interaction process from NGC\,807, behind which is a merger event of gas-rich spiral galaxies; or a flyby interaction with MRK\,365. Notably, the size of H\textsc{i} component is significantly larger than that of the stellar component ($R_{\mathrm{eff,H\textsc{i}}}/R_{\mathrm{eff,*}}$\textgreater 4). Moreover, in contrast to the profile exhibited by AGC\,123216\,OC, the H\textsc{i} column density distribution forms an almost perfectly round shape, and the H\textsc{i} velocity map shows a rotational pattern that is almost aligned with the major axis of the stellar component. The rotational velocities are low, with $V_{r}<15\,\kms$. These features suggest AGC\,123216\,OC might have undergone a recent star formation and is currently undergoing a gas relaxation process.

\subsubsection{AGC\,215415}

AGC\,215415\,OC is a faint, diffuse galaxy located southeast of Leo Triplet, with consistent redshift. It has been identified as an H\textsc{i}-only (without optical emissions) member of the Leo Triplet Group (M\,66 Group) by velocity dispersions and spatial density distributions. With a Hubble flow distance of AGC\,215415 at  $\sim10$\,Mpc, a FOV of $3.4^\circ$ corresponds to the physical scale of 0.6\,Mpc, as illustrated in the lower left panel of Fig. \ref{fig:08}. Such a large FOV contains numerous galaxies with the same redshift, most of which are identified as member galaxies or probable member galaxies of M\,66 Group \citep{2009AJ....138..338S}. Table \ref{tab:06} lists the information of these galaxies.  

It is interesting to note that besides the three dominant galaxies that consist of the Leo Triplet (num.2-4\footnote{``num.2-4'' represents the sources numbered 2-4 in Table \ref{tab:06} and the lower left panel of Fig. \ref{fig:08}, hereafter adopting the same convention.}, marked by red circles), the rest of these galaxies either display very diffuse morphology and blue color (num.5-13, marked by orange circles), or are H\textsc{i} objects without a detected optical counterpart (num.14-22, marked by blue circles). Furthermore, among the parts of galaxies with optical counterparts, IC\,2715 (num.12), IC\,2767 (num.10), IC\,2782 (num.9), and IC\,2787 (num.7) have been identified as LSBGs in previous literature \citep{2004AJ....128...62G, 2019MNRAS.483.1754D, 2020ApJS..248...33H}; AGC\,215414\,OC (num.8) also meets the standard; for the rest AGC\,215286 (num.13) and AGC\,215303 (num.5), we did a simple photometry and found they can also be classified as LSBGs; and for the last galaxy, NGC\,3666 (num.6), we fitted an exponential disc model on it and its disc also shows the LSB characteristics.

The surroundings of AGC\,215415\,OC are characterized by a large number of LSB objects, which could be a result of interactions in the region. Given the abundant gas composition of the environment, it is possible that these interactions are dominated by the Leo Triplet. Such interactions between galaxies will lead to the diffusion and mixing of gas and stellar components, which gradually relax into new systems under the influence of gravity and viscosity. These interactions may also trigger and promote the star formation process, finally resulting in more galaxies within the environment with bluer colors, diffuse morphologies, and LSB features.

In the surrounding image of AGC\,215415 depicted in Fig. \ref{fig:08}, the positions of the sources lacking optical counterparts (marked by blue circles) indicate that they are likely to share a common origin, NGC\,3628 (num.2). However, it is challenging to make reliable deductions about the origin of AGC\,215415\,OC, as it is situated at the center of the FOV, at a distance of 1.39$\degree$ from NGC\,3628. Nevertheless, it is reasonable to consider the important role of the cluster environment, and the formation of AGC\,215415\,OC may be attributed to possible interaction processes. Another intriguing idea is that it may originate from the same encounter event that produced AGC\,215414 $\sim800\,{\rm Myr}$ ago. More further observations are required to advance these hypotheses.

\subsubsection{AGC\,219841}

The morphology of AGC\,219841\,OC appears diffuse and irregular, with additional brightness contamination from the nearby star and galaxy further complicating the discernibility. AGC\,219841 is located in a complex environment surrounded by galaxies with the same redshift, as listed in Table \ref{tab:07}. The galaxies are arranged in a northeast-southwest direction, as shown in the upper right panel of Fig. \ref{fig:08}. Referring to \citet{2008A&A...479..927T}, which identified numerous groups of galaxies in SDSS DR5 based on the Friends-Of-Friends (FOF) Algorithm, we found that these galaxies correspond to the galaxies in the southeastern part of the same galaxy group.

As depicted in the surrounding image of AGC\,219841, the three nearest galaxies in the northeast direction are NGC\,3991, NGC\,3994, and NGC\,3995 (numbers 5, 4, and 3, respectively). These galaxies consist of a relatively loose (with the farthest projected distance of $\rm \sim50\,kpc$) galaxy triplet named Arp\,313 \citep{1966ApJS...14....1A}. Among them, NGC\,3994 and NGC\,3995 constitute a close pair of interacting galaxies. Spectral observations of them revealed an enhanced surface density of SFR in their center and outer regions \citep{2019A&A...622A.180L}. NGC\,3991 is a peculiar galaxy with a warped disk and an unusual brightness distribution. Multi-band observations have revealed a global rotation feature, and deduced that it underwent a rare, non-continuous star formation produced by several successive starbursts \citep{2013MNRaS.435.1958B}. The VLA observations revealed the complex and internally turbulent H\textsc{i} gas distribution around the three galaxies and a small companion to the east of NGC\,3995, as shown in Fig. 3 of \citet{2004AJ....127.1900W}. This vast cloud extends towards the east, south, and west. However, many positions of the cloud were found to lack optical counterparts in previous studies. Particularly, the optical counterpart of H\textsc{i} emission is entirely absent in the western direction, where AGC\,219841\,OC is located. Therefore, we believe that AGC\,219841\,OC could have formed as a result of an intricate interaction process occurring within the same H\textsc{i} cloud associated with the Arp\,313 Triplet.

In addition, the high-sensitivity H\textsc{i} observations from the Five-hundred-meter Aperture Spherical Telescope (FAST) revealed a large-scale H\textsc{i} gas structure of more than $30^\prime$ in the northeast-southwest direction of AGC\,219841, with the H\textsc{i} column density contours overlaid on the surrounding image of AGC\,219841\,OC \citep{2024SCPMA..6719511Z}. This provides strong observational evidence for the potential co-evolution of these galaxies. Meanwhile, we are aware that the relatively broad H\textsc{i} spectral line width of AGC\,219841 compared with other sources may not only be attributed to the intrinsic rotational velocity or velocity dispersion of the gas, but also to the contribution of gas exchange within the group of galaxies. More specific details about the origin and evolution of this system will be discussed in later work.

\subsubsection{AGC\,258609}

AGC\,258609\,OC has the lowest peak-surface-brightness among the eight sources, and the morphology is difficult to discern by eye. Table \ref{tab:08} lists the galaxies with similar redshifts within 0.6\,Mpc, and the lower right panel of Fig. \ref{fig:08} illustrates their positions in the environment. Among them, NGC\,5990 (number 2) is a luminous infrared galaxy (LIRG) in the Great Observatories All-sky LIRG Survey \citep[GOALS,][]{2009PASP..121..559A}, classified as a \mbox{Seyfert\,II} galaxy. The optical Integral Field Unit (IFU) spectrum from the Wide Field Spectrograph (WiFeS) showed some post-starburst galaxy features and a lack of correlation between gas density and SFR, which suggests the possibility of significant dust shielding or gas outflows \citep{2017ApJS..232...11T, 2018A&A...618A...6K}. Visual inspection of the Spitzer Space Telescope Infrared Array Camera (IRAC) 3.6\,${\rm \mu}$m image revealed that it is a pre-merger \citep{2013ApJS..206....1S}. Among the five galaxies listed in Table \ref{tab:08}, CGCG\,050-098 (number 3) is acknowledged as the companion galaxy of NGC\,5990 \citep{2006AJ....132..197W}. Additionally, CGCG\,050-091 (number 6) and CGCG\,050-095 (number 4) are classified as satellite galaxies of NGC\,5990 with retrograde orbits, while CGCG\,050-093 (number 5) is another satellite galaxy with a prograde orbit \citep{2006ApJ...645..228A}. 

These observations indicate that the region around AGC\,258609\,OC has experienced or is currently undergoing a significant star formation process and is in a pre-merger stage. The environment consists of a primary galaxy and several companion galaxies with different orbits, which is likely to result in asymmetric tidal torques, triggering gas inflows and outflows \citep{1996ApJ...464..641M}. These findings strongly imply that tidal interactions may be responsible for the origin of AGC\,258609\,OC, whose progenitor is likely to be the gas ejected through these processes.

\subsection{Two isolated Dwarfs: AGC\,225880 and AGC\,229361}\label{sec:05-01-03}

In contrast to the sources discussed above, \mbox{AGC\,225880\,OC} and \mbox{AGC\,229361\,OC} appear to exist in isolation, without any nearby optically visible galaxies accounting for their origins or any potential interactions. The prolonged absence of star formation is confusing, which might be due to the inefficiency of converting H\textsc{i} gas to molecular gas or the inability of molecular gas to collapse and form stars. Additional observations are necessary to draw more definitive conclusions on the dynamics of these sources.

\subsubsection{AGC\,225880}
\mbox{AGC\,225880\,OC} is found in the outskirt of the Virgo cluster and is only $12.5^{\prime\prime}$ away from the H\textsc{i} emission centroid of Cloud 1 South (C1S) in \citet{2010ApJ...725.2333K}, which strongly suggests that they are related. H\textsc{i} observations from VLA revealed that Cloud 1 (C1) is an isolated cloud, with no optical counterpart identified in previous studies. According to the research conducted by \citet{2010ApJ...725.2333K}, the nearest optical structure of C1 is positioned $3.8^{\prime}$ towards the northeast direction, and the closest late-type galaxy with a matching redshift of the Virgo cluster is situated at a distance of $1\degree$ towards the northeast. Moreover, the study claimed that C1 is the only gas structure that is definitively extragalactic and unambiguous, and is not associated with another galaxy outside the Local Group. 

Although there are strong indications that AGC\,225880\,OC might correspond to the H\textsc{i} component of C1S, we also examined other galaxies within 0.6\,Mpc that share the same redshift. It was found that in addition to Cloud 1 North (CIN) and C1S, there are two other galaxies that share the same redshift, and the information and positions of all these galaxies are shown in Table \ref{tab:09} and the left panel of Fig. \ref{fig:09}. The two additional galaxies, IC\,3028 and LEDA\,1402698 (numbers 4 and 5, respectively), are located at a projection distance of 0.48 and $0.52\degree$ (approximately 0.21 and 0.19\,Mpc) from AGC\,225880, respectively. IC\,3028 (VCC\,24), a member in the Virgo Cluster Catalog, is a blue compact dwarf (BCD) with a star-forming nuclear and elliptical low-surface-brightness component \citep{2006AJ....132.2432L,2014A&A...562A..49M,2018ApJ...859....5L}; LEDA\,1402698 (EVCC\,101), a member in the Extended Virgo Cluster Catalog, is a poorly studied blue diffuse galaxy, and it does not show any clear center or nucleus in either of the optical of infrared images \citep{2020MNRAS.494.1784A}. Given the distance and morphology of these two sources, neither of them is likely to be related to AGC\,225880\,OC. Therefore, we tend to infer that AGC\,225880\,OC is associated with C1.

More details are revealed in the zoom-in area of the left panel Fig. \ref{fig:09}, which corresponds to the beam area of ALFALFA, covered C1N, C1S, and AGC\,225880\,OC. The center of AGC\,225880\,OC, coincides with a depression in the gas contour of C1S, which is marked as a light green cross. AGC\,225880\,OC is located north of C1S, between C1N and C1S, and closer to C1S. A bold conjecture is that C1N and C1S were once connected, given the isolated nature of C1 and their special position with AGC\,225880\,OC. The star-forming process of AGC\,225880\,OC may have ionized the gas in the corresponding region, leading to the separation of the two cloud structures that we currently observe.

\subsubsection{AGC\,229361}

AGC\,229361\,OC stands out as the galaxy candidate with the highest $M_{\rm H\textsc{i}}/M_{*}$ among the eight sources in this paper. It is located on the periphery of the Virgo Cluster. Two galaxies share the same redshift with it within a 0.6\,Mpc scale: VirgoH\textsc{i}\,27 and AGC\,229360. Table \ref{tab:10} and the right panel of Fig. \ref{fig:09} provide detailed information about their positions, distances, and separations.

VirgoH\textsc{i}\,27, a confirmed extended member of the Virgo Cluster, is situated $1.4^\prime$ southeast of AGC\,229361. Arecibo measurements show an intense peak flux of H\textsc{i} gas, with $W_{50}=52\,\kms$ and the heliocenter velocity of $1658\,\kms$. Although there are two 2MASS optical objects nearby, one lies outside the range of the H\textsc{i} beam, and the other has a small size and a red color, inconsistent with the H\textsc{i} component \citep{2004MNRAS.349..922D}. In the absence of a convincing optical counterpart, VirgoH\textsc{i}\,27 remains enigmatic.

AGC\,229360, an H\textsc{i} emission source that lacks an optical counterpart in both the ALFALFA-SDSS Galaxy Catalog and DECaLS, is located nearly $5^{\prime}$ due north of AGC\,229361. The heliocentric velocity of AGC\,229360 is $1573\,\kms$, almost identical to that of AGC\,229361, with the former being $88\,\kms$ lower. The $W_{50}$ and $W_{20}$ of AGC\,229360 are 61 and $89\,\kms$, respectively, which is approximately $2\sim3$ times greater than that of AGC\,229361.

According to \citet{2019AJ....157...76B}, the consistency of the spatial positions and velocities of AGC\,229360 and AGC\,229361 indicated that they might be a pair. Moreover, they could be associated with AGC\,229385\footnote{AGC\,229385 is an \textit{almost dark} galaxy that exhibits enigmatic properties \citep{2015ApJ...801...96J,2018AJ....155...65B,2019arXiv190107557B}, and we have compared our sources with it in various relations and found that their properties are very similar.} since all the three are isolated \textit{almost dark} sources, and their projected positions and velocities differ by only $1.4\degree$ and $313\,\kms$, respectively. The similarity suggested that they might have originated from the Local Void \citep{1987nga..book.....T} following a similar trajectory.

However, given the scale at which interactions may occur, a more plausible assumption based on available data is that the system constituted by AGC\,229361, AGC\,229360, and VirgoH\textsc{i}\,27 appears to be isolated. Besides, although the beam size of ALFALFA may cause some overestimation of $M_{\rm H\textsc{i}}$, it is insufficient to account for the abnormally high $M_{\rm H\textsc{i}}/M_{*}$ displayed by AGC\,229361. These findings provide valuable insights into the formation scenarios of AGC\,229361\,OC, which may have emerged under unique conditions in the early universe that hindered star formation for unclear reasons. All speculations mentioned above require further observations to verify.

\subsection{Space Distribution and Clustering Properties}

When examining the spatial distribution of our sources, a distinct preference in the distribution of the spatial positions appears. With the exceptions of AGC\,123216 and AGC\,258609, the right ascensions of our sources fall between $11\,{\rm h}\sim12.5\,{\rm h}$, which are within the Virgo cluster or the Leo Triplet.

Objects in or near the galaxy clusters or groups tend to have more opportunities to occur interactions than those in the field, so they tend to exhibit the characteristics of early-type galaxies, with greater mass, redder color, and lower SFRs \citep{2006PASP..118..517B,2010ApJ...721..193P}. Although the features displayed by our sources do not seem to match the typical characteristics of galaxies within the clusters, it could at least suggest that the environment of galaxy clusters is more likely to produce galaxies with such unique properties, and the six galaxies are likely the products of interactions in the very recent past.

\subsection{Deviation of optical centroid from the H\textsc{i} centroid}

Besides the specific environments and the global spatial distributions of these galaxy candidates, we observed a deviation of approximately $0.21^\prime\sim1.18^\prime$ between the optical counterpart centroid and the H\textsc{i} centroid of these sources, as shown in Fig. \ref{fig:10}. While the beam size and pointing accuracy of ALFALFA may affect the accurate determination of the H\textsc{i} centroid to $30^{\prime\prime}$ for low-S/N sources in regions of source confusion, as discussed in \citet{2011AJ....142..170H}, it is crucial to note that this offset phenomenon could provide us with additional clues to the origin and evolutionary history of these galaxies.

There are several reasons why the optical centroid and H\textsc{i} centroid of sources do not align. \citet{1986ApJ...306..466H} found that the H\textsc{i} deficiency in Virgo Cluster galaxies could result in a discrepancy between the optical center and the H\textsc{i} center, which may be due to the tidal stripping of gas from the galaxies as they move through the dense intracluster medium. Similarly, the minor mergers and tidal interactions could alter the distribution of the H\textsc{i} gas and cause such phenomenon \citep{2001ASPC..240..257H}. Another possibility is that the H\textsc{i} distribution is intrinsically asymmetric. The gravitational influence of dark matter can cause the offset, where the H\textsc{i} gas is pulled away from the optical center due to the gravitational force of the dark matter, resulting in an inconsistency between the distribution of the gas and stellar components \citep{2002A&A...390..829S,2005A&A...442..137N}.

\section{Summary}\label{sec:06}

In this work, we presented the discovery of eight optical counterparts of ALFALFA extragalactic objects, which were previously flagged as H\textsc{i} emission objects with no optical counterparts in SDSS. We did photometry for multi-band images, obtained physical parameters, and compared them with the empirical scaling relations of galaxies. We found that the physical properties of these galaxy candidates are distinct from the $\alpha.100$ sample and deviate from classical relations, but similar to that of the \textit{almost dark} galaxy AGC\,229385 and HUDs. Their properties can be summarized as follows:\\

1) Faint and blue objects: These objects have peak surface brightness in the range of $25.13\sim26.76\,{\rm mag}/{\rm arcsec}^2$ and $g-r$ color ranging from -0.35 to 0.55.

2) Irregular morphology: The morphology of these objects is irregular and hard to describe.

3)	Low $M_{*}$ and high $M_{\rm H\textsc{i}}/M_{*}$: The estimated $M_{*}$ are quite small (10$^{5.27\sim7.15}$\,$M_\odot$), while $M_{\rm H\textsc{i}}$ are relatively high (10$^{6.97\sim9.48}$\,$M_\odot$), resulting in extremely high $M_{\rm H\textsc{i}}/M_{*}$ (10$^{1.72\sim3.22}$, except AGC\,215415).

4) Low SFRs and SFE: The SFRs derived from NUV photometry are relatively low ($0.21\sim9.24\times10^{-3}{M_\odot}\,{\rm yr}^{-1}$). They are well consistent with the extension of the MS of galaxies at the low mass end, showing the characteristics of star-forming galaxies. However, they show low SFE, with high $M_{\rm H\textsc{i}}$ compared to the SFRs.

5) At the low mass end of the BTFr with potential deviation: These objects are at the low mass end of the BTFr, with three deviating from the relation and biased towards the heavier end of the baryonic mass.

6) Potential origins of interaction or isolation: six of the eight candidates have optically visible galaxies nearby with the same redshift, suggesting they may originate from tidal interactions associated with known structures; the remaining two isolated clouds fail to find associated optical structures, indicating that they may have formed in situ, fled from the Local Void, or formed under extreme formation conditions, which hindered their star formation for an extended period. The proportion of sources with these two types of origins happens to coincide with the prediction of ``roughly $3/4$ of the `dark' H\textsc{i} sources are located in fields'' in \citet{2011AJ....142..170H}.\\

To summarize, it is insufficient to conclusively determine the origin and evolution of these galaxy candidates from available observations. However, based on their blue colors, high $M_{\rm H\textsc{i}}/M_{*}$, low SFRs, and environment features, it is likely that star formation in them occurred only recently. This could have been caused by tidal interactions or other unclear reasons after a long period of stability. The presence of these galaxy candidates provides new low-mass samples for the BTFrs and galaxy MS, and might even lead to discussions about various cosmological paradigms.

Improving data quality is crucial for reliable future studies on these sources. High-precision multi-band observations, high-resolution H\textsc{i} synthesis maps, and deep optical observations will facilitate dynamical studies and provide accurate distances. These will aid in better interpretation of the physical properties of galaxy candidates and help enhance constraints on their origins and evolution scenarios.

\begin{acknowledgements}

This work is supported by the National Key R$\&$D Program of China (grant No. 2022YFA1602901). D.W. acknowledges the support from the Youth Innovation Promotion Association, Chinese Academy of Sciences (No. 2020057). We are also grateful for the support of the National Natural Science Foundation of China (NSFC; grant Nos.12090041 and 12090040). Additional support comes from the Strategic Priority Research Program of the Chinese Academy of Sciences (grant Nos. XDB0550100 and XDB0550102) and the Open Project Program of the Key Laboratory of Optical Astronomy, National Astronomical Observatories, Chinese Academy of Sciences. D.W. has also been supported by science research grants from the China Manned Space Project and the NSFC (grant Nos. U1931109 and 11733006).

\end{acknowledgements}

\appendix 
\section{H\textsc{i} Spectrum of the eight sources}\label{appe}
As a supplement to Section \ref{sec:04-02}, Fig. \ref{fig:overall} shows the H\textsc{i} spectral line profiles of the eight sources obtained from the observations of ALFALFA. These H\textsc{i} lines mainly exhibit single-horned profiles, with AGC\,215415 and AGC\,219841 also slightly displaying asymmetric double-horned features. Such line profiles may indicate that these sources are face-on or have very low rotational velocities, possibly with gas exchanges and asymmetrical H\textsc{i} distributions.



\begin{deluxetable*}{lllllllllllll}
  \tablenum{1}
  \tablecaption{Information Of Eight Sources From Archival Data}
  \label{tab:01}
  \tablewidth{0pt}
  \tablehead{\colhead{}&\colhead{AGCNr} & \colhead{Name} & \colhead{RA\_OC} & \colhead{DEC\_OC} & \colhead{RA\_H\textsc{i}} & \colhead{DEC\_H\textsc{i}} & 
    \colhead{$W_{50}$} & \colhead{$W_{20}$} & \colhead{Vhelio} & \colhead{Dist} & \colhead{$\log_{10}M_{\rm H\textsc{i}}$} \\
    \colhead{}&
    \colhead{} & \colhead{} & \colhead{(J2000)} & \colhead{(J2000)} & \colhead{(J2000)} & \colhead{(J2000)} & \colhead{(${\rm km}/{\rm s}$)} & 
    \colhead{(${\rm km}/{\rm s}$)} & \colhead{(${\rm km}/{\rm s}$)} & \colhead{(Mpc)} & \colhead{$(M_\odot)$}\\
    \colhead{} &\colhead{(1)}&
    \colhead{(2)} & \colhead{(3)} & \colhead{(4)} & \colhead{(5)} & \colhead{(6)} & \colhead{(7)} & \colhead{(8)} & 
    \colhead{(9)} & \colhead{(10)} & \colhead{(11)}}
  \startdata
  \multicolumn{11}{c}{Data Of Eight Sources From ALFALFA} \\
1 & 123216 &          & 31.1855046  & 28.8030914 & 31.17792  & 28.80583 & 28  & 51  & 5111 & 70.5 & 8.65 \\
2 & 215414 & LeoTrip. & 170.809     & 13.7139    & 170.79625 & 13.70555   & 27  & 47  & 878  & 10   & 8.6  \\
3 & 215415 & Leogrp.  & 171.144     & 12.675     & 171.14249 & 12.67778  & 19  & 36  & 1004 & 10   & 6.97 \\
4 & 219841 &          & 179.2788219 & 32.2646447 & 179.28708 & 32.25444  & 105 & 147 & 3133 & 48.4 & 9.48 \\
5 & 225880 &          & 182.2004    & 11.9113    & 182.18959 & 11.92139  & 29  & 42  & 1230 & 15.3 & 7.65 \\
6 & 229197 & H\textsc{i}\,1225c  & 186.8427    & 1.5231     & 186.83542 & 1.5225    & 39  & 57  & 1285 & 15.2 & 8.6  \\
7 & 229361 &          & 186.6694    & 19.7542    & 186.67084 & 19.76139  & 22  & 37  & 1661 & 27.9 & 8.49 \\
8 & 258609 &          & 236.4716    & 2.4995     & 236.49126 & 2.5      & 57  & 85  & 3842 & 57.3 & 8.88 \\
  \hline 
  \multicolumn{11}{c}{Higher Resolution Data: VLA H\textsc{i} Map} \\
 & 215414 & Leo TDG      & 170.809  & 13.7139  &           &             &           &   &           &  & 8.58 \\
 & 225880 & Cloud 1 South & 182.2004 & 11.9113  & 182.1975    & 11.9133    & $20\pm8$ &   & $1225\pm3$&  & 7.4   \\
 & 229197 & H\textsc{i}\,1225c      & 186.8427 & 1.5231  & 186.8397768 & 1.5234657 &    &  &$1288\pm5$ &  & 8.447 \\
\enddata
\tablecomments{Col.(1): Entry number in Arecibo General Catalog (AGC); \\Col.(2): Name of the potentially corresponded optical counterpart or H\textsc{i} cloud, based on ALFALFA and VLA observations; \\Col.(3) \& (4): Right ascension and declination of the centroid of the optical counterpart in DECaLS (J2000); \\Col.(5) \& (6): Right ascension and declination of the centroid of the H\textsc{i} source by ALFALFA (J2000), with a declination-dependent error in the order of $20^{\prime\prime}$ after correction for systematic telescope; \\Col.(7) \& (8): Velocity width of the H\textsc{i} line profile measured at the width of the $50\%$ and $20\%$ level of each of the two peaks (or of the single peak if only one is present); \\Col.(9): Heliocentric velocity in the observed frame, calculated from the midpoint of the channel of $W_{50}$; \\Col.(10): Adopted distance, determined by cosmological distance or the local universe peculiar velocity model respectively, with cz $\approx6000\,\kms$ as the bound; \\Col.(11): H\textsc{i} mass (in solar units), computed via the standard formula with the integrated H\textsc{i} line flux density and the assumed distance. \\To distinguish between H\textsc{i} and optical emissions, the optical counterparts are labeled with the suffix ``$OC$''. For example, \textit{AGC\,123216} represents the H\textsc{i} emission of the source, while \textit{AGC\,123216 OC} represents its optical counterpart.}
\end{deluxetable*}


\setcounter{table}{1} 
\setlength\tabcolsep{2.5pt}
\begin{table*}[!t]\footnotesize
  \centering
  \caption{Derived Properties Of Eight Sources}
  \label{tab:02}
  \begin{tabular*}{\hsize}{@{\ \ }@{\extracolsep{\fill}}cc|cccccc@{\ \ }}
    \hline\hline
    &AGC & $m_{\rm g}\,|\,m_{\rm g,corr}$ & $m_{\rm r}\,|\,m_{\rm r,corr}$  & $m_{\rm FUV}\,|\,m_{\rm FUV,corr}$ & $m_{\rm NUV}\,$$|$$\,m_{\rm NUV,corr}$  & $g-r\,|\,(g-r)_{\rm corr}$ & $m_{\rm g,appr,MTO}$  \\
     & &$(mag)$  &$(mag)$  &$(mag)$ &$(mag)$  & &$(mag)$ \\
        & & (1)  &(2)  &(3) &(4)  &(5) & (6)\\
    \hline 
   1 & 123216 & $20.83\,|\,20.63\pm0.028$  & $20.72\,|\,20.59\pm0.049$     & $21.66\,|\,21.16\pm0.43$   & $21.39\,|\,20.89\pm0.22\,\,$   &\,\,$0.11\,|\,0.046\pm0.056$     &20.91 \\
   2 & 215414 & $17.73\,|\,17.64\pm0.013$  & $17.41\,|\,17.35\pm0.013$     & ---                        & ---                            &\,$0.32\,|\,0.29\pm0.019\,$      &17.75 \\
   3 & 215415 & $18.80\,|\,18.68\pm0.020$  & $18.22\,|\,18.14\pm0.015$     & $20.25\,|\,19.94\pm0.20$   & $18.93\,|\,18.61\pm0.017$      &\,$0.59\,|\,0.55\pm0.025\,$      &18.84 \\
   4 & 219841 & $20.08\,|\,20.01\pm0.021$  & $19.78\,|\,19.73\pm0.037$     & $21.48\,|\,21.30\pm0.37$  & $19.98\,|\,19.80\pm0.10\,\,$   &\,$0.30\,|\,0.28\pm0.042\,$      &20.21 \\
   5 & 225880 & $20.71\,|\,20.63\pm0.042$  & $20.87\,|\,20.82\pm0.089$     & $21.56\,|\,21.36\pm0.37$   & $20.49\,|\,20.29\pm0.036$      &\,\,$-0.16\,|\,-0.19\pm0.098$     &20.80 \\
   6 & 229197 & $21.68\,|\,21.61\pm0.067$  & $21.37\,|\,21.33\pm0.060$     & $22.85\,|\,22.69\pm0.17$  & $21.58\,|\,21.42\pm0.056$      &\,$0.31\,|\,0.29\pm0.090\,$      &21.65 \\
   7 & 229361 & $21.59\,|\,21.47\pm0.098$  & $21.90\,|\,21.82\pm0.23\,\,$  & $21.26\,|\,20.96\pm0.34$   & $20.66\,|\,20.35\pm0.14\,\,$   &$-0.31\,|\,-0.35\pm0.25$  &21.48 \\
   8 & 258609 & $21.93\,|\,21.58\pm0.098$  & $21.37\,|\,21.13\pm0.080$     & $22.33\,|\,21.43\pm0.60$   & $\,20.76\,|\,19.86\pm0.049$      &\,$0.56\,|\,0.44\pm0.13\,\,\,$   &21.59 \\
    \hline\hline
   & AGC & $\log_{10}M_{*}$\,$|$\,$\log_{10}M_{\rm {*,corr}}$ & $\log_{10}(M_{\rm H\textsc{i}}/M_{*})$\,$|$\,$\log_{10}(M_{\rm H\textsc{i}}/M_{\rm *,corr})$ & $SFR\,|\,SFR_{\rm corr}$ & $\mu_{\rm g,peak}$\,$|$\,$\mu_{\rm g,peak,corr}$ & $R_{\rm g,eff,MTO}$\\
  & & $(/M_\odot)$  &$({\rm dex})$  &$(10^{-3}{M_\odot}\,{\rm yr}^{-1})$  &$({\rm mag}/{\rm arcsec}^2)$ & $({\rm kpc})$&\\
   & &(7)  &(8)  &(9) &(10)& (11) &\\
    \hline
   1 & 123216 &  $6.92\,|\,6.93\pm0.18$  & $1.73\,|\,1.72\pm0.20$  & $4.18\,|\,5.35\pm0.86\,\,$   & 25.13\,$\,|\,24.93$\,$_{+0.011}\atop^{-0.012}$     & 1.15 &\\
   2 & 215414 &  $6.74\,|\,6.74\pm0.29$  & $1.86\,|\,1.86\pm0.35$  & ---                          & \,\,25.68\,$\,|\,25.59$\,$_{+0.0054}\atop^{-0.0054}$   & 0.73 &\\ 
   3 & 215415 &  $6.65\,|\,6.66\pm0.39$  & $0.32\,|\,0.31\pm0.43$  & $\,0.80\,|\,1.04\pm0.034$  & 25.58\,$\,|\,25.46$\,$_{+0.011}\atop^{-0.011}$     & 0.54 &\\
   4 & 219841 &  $7.15\,|\,7.15\pm0.27$  & $2.33\,|\,2.33\pm0.28$  & $7.12\,|\,9.24\pm1.41\,\,$   & 26.15\,$\,|\,26.08$\,$_{+0.011}\atop^{-0.011}$     & 1.39 &\\
   5 & 225880 &  $5.29\,|\,5.30\pm0.20$  & $2.11\,|\,2.10\pm0.25$  & $0.44\,|\,0.58\pm0.14\,\,$   & 26.13\,$\,|\,26.05$\,$_{+0.026}\atop^{-0.025}$     & 0.32 & \\
   6 & 229197 &  $5.51\,|\,5.51\pm0.35$  & $2.94\,|\,2.93\pm0.38$  & $0.16\,|\,0.21\pm0.054$      & 24.95\,$\,|\,24.88$\,$_{+0.014}\atop^{-0.013}$     & 0.23 &\\
   7 & 229361 &  $5.27\,|\,5.27\pm0.28$  & $3.22\,|\,3.22\pm0.30$  & $1.27\,|\,1.64\pm0.27\,\,$   & 25.88\,$\,|\,25.76$\,$_{+0.022}\atop^{-0.022}$     & 0.45 &\\
   8 & 258609 &  $6.88\,|\,6.89\pm0.44$  & $2.00\,|\,1.99\pm0.44$  & $4.88\,|\,6.33\pm0.45\,\,$   & 26.76\,$\,|\,26.41$\,$_{+0.035}\atop^{-0.034}$     & 1.21 &\\
    \hline\hline
    \end{tabular*}
    \tablecomments{These parameters are obtained based on data from DECaLS, ALFALFA, and GALEX. All parameters listed here (except Col.(6) and Col.(11)) contain the results before and after Galactic extinction correction, where we use the subscript suffix ``$corr$'' to present the names of the parameters after extinction correction and separate the results with the symbol ``$|$''. The errors are based on the results after extinction correction. Col.(6) lists the apparent magnitude of the $g$ band measured by \texttt{MTObjects} without correcting extinction.}
 
\end{table*}    

\begin{table*}[!t]
  \centering
  \caption{The Stellar Mass Of Eight Sources Under Different CMLRs And IMFs}
  \label{tab:03}
  \begin{tabular*}{\hsize}{@{\ \ }@{\extracolsep{\fill}}cccccccccc@{\ \ }}
    \hline\hline
CMLR  & IMF      &123216 &215414 &215415 &219841 &225880 &229197 &229361 &258609  \\
\hline
      & ``diet'' Salpeter   & 7.03  & 6.91  & 6.88  & 7.31  & 5.35  & 5.68  & 5.29  & 7.08  \\
      & Salpeter            & 7.18  & 7.06  & 7.03  & 7.46  & 5.50  & 5.83  & 5.44  & 7.23  \\
B03   & Kroupa              & 6.88  & 6.76  & 6.73  & 7.16  & 5.20  & 5.53  & 5.14  & 6.93  \\
      & Chabrier            & 6.94  & 6.82  & 6.79  & 7.22  & 5.26  & 5.59  & 5.20  & 6.99  \\
  \hline
      & Salpeter            & 6.77  & 6.78  & 6.88  & 7.18  & 4.97  & 5.54  & 4.82  & 7.03  \\
Z09   & Kroupa              & 6.47  & 6.48  & 6.58  & 6.88  & 4.67  & 5.24  & 4.52  & 6.73  \\
      & Chabrier            & 6.53  & 6.53  & 6.64  & 6.93  & 4.72  & 5.30  & 4.58  & 6.79  \\ 
  \hline
      & Salpeter            & 7.00  & 6.98  & 7.05  & 7.38  & 5.22  & 5.75  & 5.10  & 7.21  \\
IP13  & Kroupa              & 6.70  & 6.68  & 6.75  & 7.08  & 4.92  & 5.45  & 4.80  & 6.91  \\
      & Chabrier            & 6.76  & 6.73  & 6.81  & 7.14  & 4.98  & 5.50  & 4.86  & 6.97  \\
  \hline   
      & Salpeter            & 7.16  & 6.98  & 6.90  & 7.39  & 5.54  & 5.75  & 5.51  & 7.12  \\
H16   & Kroupa              & 6.86  & 6.68  & 6.60  & 7.09  & 5.24  & 5.45  & 5.21  & 6.82  \\
      & Chabrier            & 6.92  & 6.74  & 6.65  & 7.15  & 5.29  & 5.51  & 5.27  & 6.88  \\
  \hline   
      & Salpeter            & 6.92  & 6.81  & 6.78  & 7.21  & 5.23  & 5.57  & 5.17  & 6.98  \\
D20   & Kroupa              & 6.62  & 6.51  & 6.48  & 6.91  & 4.93  & 5.27  & 4.87  & 6.68  \\
      & Chabrier            & 6.68  & 6.56  & 6.54  & 6.97  & 4.99  & 5.33  & 4.92  & 6.74  \\
  \hline\hline      
  \end{tabular*}
  \tablecomments{All parameters listed here correspond to the values of $\log_{10}(M_{*}/M_\odot)$.}
\end{table*}

\begin{table*}[!t]
\centering
\caption{Comparison Of UV Photometric Magnitude And Exposure Time}
\label{tab:04}
\setlength{\leftskip}{-35pt}
\begin{tabular}{llllllllllllll}
\hline \hline
&AGC                & 123216  & 215415 & 219841 & 225880  & 229197  & 229361 & 258690 & & GALEX&AIS & MIS & DIS \\
\hline
Exp. time (FUV, s) && 105     & 121    & 111    & 122     & 1654    & 109    & 110   &  & &100        & 1500       & 30000      \\
mag (FUV)          && 21.66   & 20.25  & 21.48  & 21.56   & 22.85   & 21.26  & 22.33 &  & &19.9       & 22.6       & 24.8       \\
\hline
Exp. time (NUV, s) && 105     & 1637   & 111    & 1501    & 1698    & 109    & 1058  &  & &100        & 1500       & 30000      \\
mag (NUV)          && 21.39   & 18.93  & 19.98  & 20.49   & 21.58   & 20.66  & 20.76 &  & &20.8       & 22.7       & 24.4      \\
\hline\hline
    \end{tabular}
\end{table*}

\begin{table*}[!ht]
    \centering
    \caption{Surrounding Of AGC\,123216}
    \label{tab:05}
    \begin{tabular}{llllll}
    \hline
       num   & Name        &  RA      & Dec      & D (Mpc)    & Separation \\ 
        (1)  &(2)  &(3) &(4)  &(5) &(6)  \\\hline
        1    &AGC\,123216  & 31.17792 & 28.80583 & 71.51±5.01 & ~          \\ 
        2    &KUG 0202+283 & 31.38837 & 28.61056 & 74.56±5.40 & 16.12208   \\ 
        3    &NGC\,807     & 31.23192 & 28.98744 & 66.48±4.66 & 11.25976   \\ 
        4    &NGC\,805     & 31.12321 & 28.81233 & 63.26±4.46 & 2.90263    \\ 
        5    &MRK\,365     &31.07729  & 28.65592 & 72.17±5.06 & 10.43716   \\ \hline
    \end{tabular}
    \tablecomments{Galaxies with the same redshift with AGC\,123216 within the projection area of 0.6\,Mpc.\\Col.(1): The number of the source on the diagram; \\ Col.(2), (3)$\&$(4): Name, right ascension, and declination of the sources; \\Col.(5): The CMB distance and error of the sources from NASA/IPAC Extragalactic Database (NED); \\Col.(6): The projected distance between the galaxy and our source, in the unit of arcmins.}
\end{table*}

\begin{table*}[!ht]
    \centering
    \caption{Surrounding Of AGC\,215415}
    \label{tab:06}
    \begin{tabular}{llllll}
    \hline
      num  & Name       &  RA       & Dec      & D (Mpc)    & Separation  \\ 
      (1)  &(2)  &(3) &(4)  &(5) &(6)  \\\hline  
        1  & AGC\,215415 &171.141249 & 12.67778   & 19.95±1.45  & ~        \\ 
        2  & NGC\,3628   &170.07071  & 13.589684  & 17.60±1.28  & 1.38504  \\ 
        3  & NGC\,3627   &170.06419  & 12.98406   & 15.79±1.16  & 1.09390  \\ 
        4  & NGC\,3623   &169.74113  & 13.08342   & 16.98±1.24  & 1.42385  \\ 
        5  & AGC\,215303 &172.787042 & 13.570389  & 20.14±1.46  & 1.83450  \\ 
        6  & NGC\,3666   &171.10862  & 11.34222   & 20.82±1.50  & 1.33591  \\ 
        7  & IC\,2787    &170.82949  & 13.62978   & 18.14±1.32  & 0.99922  \\ 
        8  & AGC\,215414 &170.809    & 13.7139    & 18.06±1.32  & 1.08542  \\ 
        9  & IC\,2782    &170.73066  & 13.44129   & 17.69±1.29  & 0.86192  \\ 
        10 & IC\,2767    &170.59662  & 13.07781   & 21.04±1.52  & 0.66476  \\ 
        11 & SDSS\,J111921.39+140431.6&169.8393   & 14.0755     & 17.86±1.30  & 1.88615  \\ 
        12 & IC\,2715    &169.809958 & 11.952167  & 17.87±1.30  & 1.48932  \\ 
        13 & AGC\,215286 &169.802917 & 14.327778  & 19.82±1.44  & 2.10127  \\ 
        14 & H\textsc{i}PASS\,J1122+13&170.63833  & 13.67556   & 18.34±1.33  & 1.11144   \\ 
        15 & AGC\,215413 &170.59625  & 13.648611  & 18.46±1.35  & 1.10639  \\ 
        16 & AGC\,215412 &170.4475   & 13.621389  & 18.51±1.35  & 1.16048  \\ 
        17 & AGC\,215406 &169.889583 & 13.862222  & 19.62±1.42  & 1.69905  \\ 
        18 & AGC\,215401 &169.799167 & 13.595278  & 17.42±1.28  & 1.59679  \\ 
        19 & AGC\,215397 &169.726667 & 14.218611  & 18.51±1.35  & 2.06552  \\ 
        20 & AGC\,215393 &169.71833  & 13.40917   & 17.84±1.30  & 1.56726  \\ 
        21 & AGC\,215389 &169.6175   & 14.303611  & 18.62±1.37  & 2.19955  \\ 
        22 & AGC\,215386 &169.46083  & 13.985     & 17.95±1.31  & 2.09329  \\ \hline
    \end{tabular}
    \tablecomments{Same with Table \ref{tab:05}, except for the unit of Col.(6), which is in degrees here.}
\end{table*}

\begin{table*}[!ht]
    \centering
    \caption{Surrounding Of AGC\,219841}
    \label{tab:07}
    \begin{tabular}{lllllll}
    \hline
        num  & Name       &  RA       & Dec      & D (Mpc)   & Separation  \\ 
        (1)  &(2)  &(3) &(4)  &(5) &(6)  \\\hline
       1 &AGC 219841 &179.28708 & 32.25444     & 50.36±3.62  & ~ \\ 
       2 &SDSS J115754.24+322358.6 &179.47639  & 32.39977    &  51.48±3.63 & 12.9676 \\ 
       3 &NGC 3995   &179.43374 & 32.29407     &  52.51±3.69 & 7.81081 \\ 
       4 &NGC 3994   &179.40361 & 32.27762     &  49.60±3.48 & 6.0735 \\ 
       5 &NGC 3991   &179.3795  & 32.33778     &  51.22±3.60 & 6.85385 \\ 
       6 &IC 2979    &179.226   & 32.15881     &  49.42±3.47 & 6.52211 \\ 
       7 &NGC 3966   &179.18412 & 32.02178     &  52.05±3.66 & 14.9075 \\ 
       8 &WISEA J115634.07+320740.8 &179.14219 & 32.12814    &  50.63±3.56 & 10.56179\\ 
       9 &WISEA J115632.48+320749.5 &179.13554 & 32.13036    &  49.65±3.49 & 10.7066 \\ 
      10 &IC 2978    &179.09679 & 32.03872     & 51.71±3.63  & 16.15478 \\ 
      11 &WISEA J115618.80+320203.2 &179.07851 & 32.03423 &  50.74±3.75 & 16.93652 \\ \hline
    \end{tabular}
    \tablecomments{Same with Table \ref{tab:05}.}
\end{table*}

\begin{table*}[!ht]
    \centering
    \caption{Surrounding of AGC\,258609}
    \label{tab:08}
    \begin{tabular}{lllllll}
    \hline
       num & Name         &  RA      & Dec     & D (Mpc)     & Separation  \\ 
       (1)  &(2)  &(3) &(4)  &(5) &(6)  \\\hline
        1  &AGC\,258609   &236.49125 & 2.5     &  58.63±4.13 & ~  \\ 
        2  &NGC\,5990     &236.56816 & 2.41542 &  58.43±4.09 & 6.85631 \\ 
        3  &CGCG\,050-098 &236.556   & 2.45294 &  59.33±4.16 & 4.79977 \\ 
        4  &CGCG\,050-095 &236.50546 & 2.71423 &  57.45±4.02 & 12.88199 \\ 
        5  &CGCG\,050-093 &236.4413  & 2.40984 &  61.92±4.34 & 6.18298 \\ 
        6  &CGCG\,050-091 &236.25146 & 2.468   &  58.53±4.10 & 14.50155 \\ \hline
    \end{tabular}
    \tablecomments{Same with Table \ref{tab:05}.}
\end{table*}


\begin{table*}[!ht]
    \centering
    \caption{Surrounding Of AGC\,225880}
    \label{tab:09}
    \begin{tabular}{lllllll}
    \hline
       num  & Name       &  RA       & Dec      & D (Mpc)    & Separation \\ 
       (1)  &(2)  &(3) &(4)  &(5) &(6)  \\\hline
       1 & AGC\,225880   & 182.1896  & 11.92139 & 23.16±1.66      & ~     \\ 
       2 & [K2010]Cloud 1 South &182.1975 & 11.91333 & 23.08±1.65 & 0.671 \\
       3 & [K2010]Cloud 1 North &182.1983 & 11.9325  & 23.21±1.66 & 0.842 \\ 
       4 & IC\,3028      & 182.6487  & 11.76082 & 24.12±1.72      & 28.63 \\
       5 & LEDA\,1402698 & 181.6695  & 12.03454 & 26.09±1.92      & 31.269 \\\hline
    \end{tabular}
    \tablecomments{Same with Table \ref{tab:05}.}
\end{table*}

\begin{table*}[!ht]
    \centering
    \caption{Surrounding of AGC\,229361}
    \label{tab:10}
    \begin{tabular}{lllllll}
    \hline
        num  & Name        &  RA        & Dec      & D (Mpc)    & Separation\\ 
        (1)  &(2)  &(3) &(4)  &(5) &(6)  \\\hline
        1    & AGC\,229361 & 186.67083  & 19.76139 & 29.11±2.07 & ~         \\ 
        2    & VIRGOH\textsc{i}\,27 & 186.6875   & 19.74389 & 28.98±2.05 & 1.41019   \\ 
        3    & AGC\,229360 & 186.66542  & 19.85306 & 27.82±2.03 & 5.50867   \\ \hline
    \end{tabular}
    \tablecomments{Same with Table \ref{tab:05}.}
\end{table*}
=


\addtocounter{figure}{-1}
\begin{figure*}
    \centering
    \includegraphics[width=0.24\textwidth]{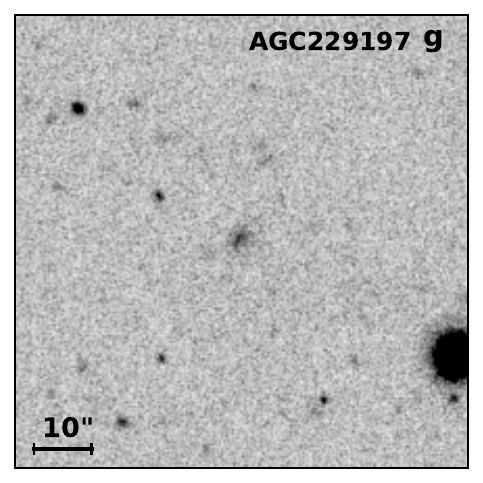}
    \includegraphics[width=0.24\textwidth]{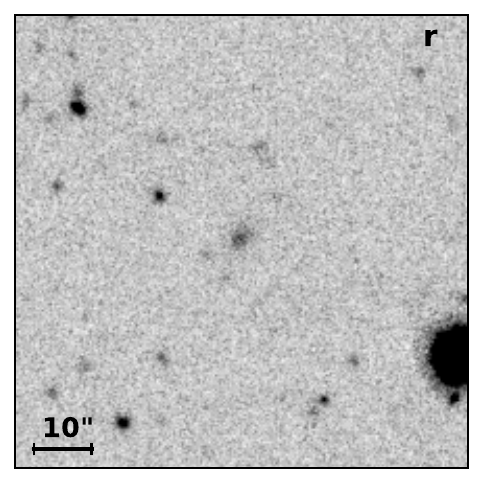}
    \includegraphics[width=0.24\textwidth]{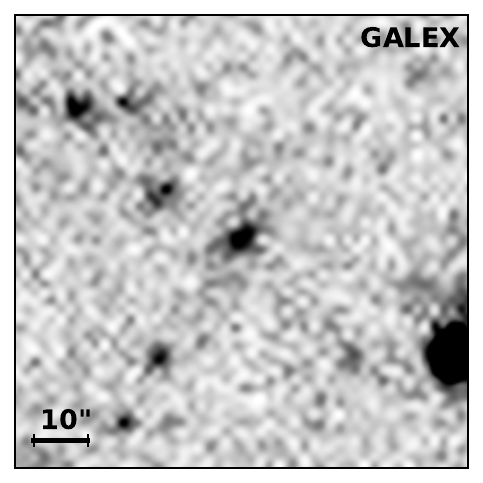}
    \\
    \includegraphics[width=0.24\textwidth]{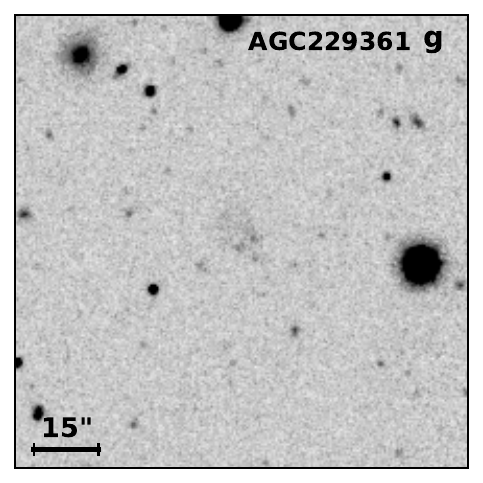}
    \includegraphics[width=0.24\textwidth]{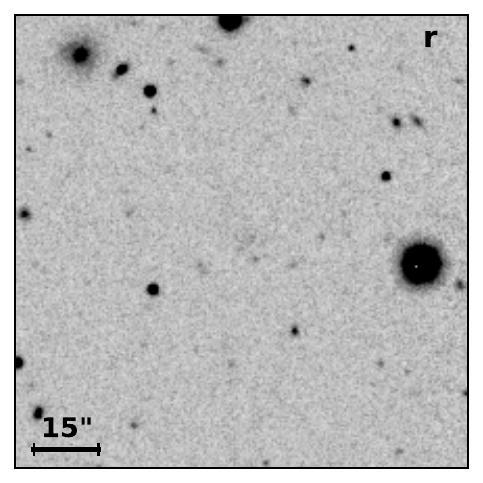}
    \includegraphics[width=0.24\textwidth]{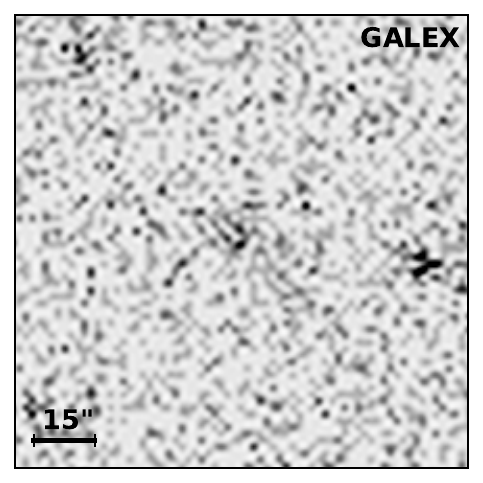} 
    \\ \newpage
    \includegraphics[width=0.24\textwidth]{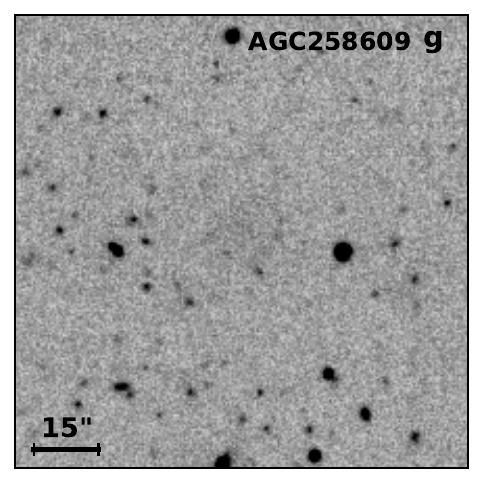}
    \includegraphics[width=0.24\textwidth]{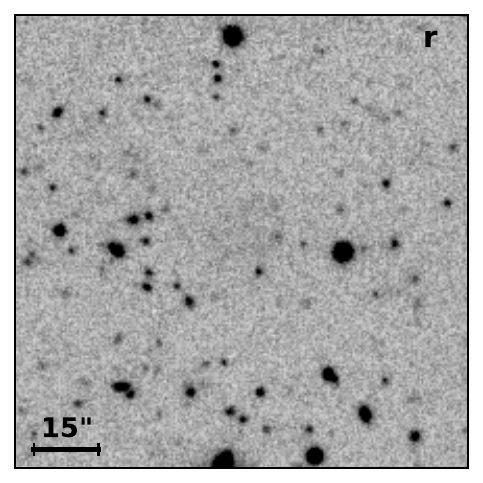}
    \includegraphics[width=0.24\textwidth]{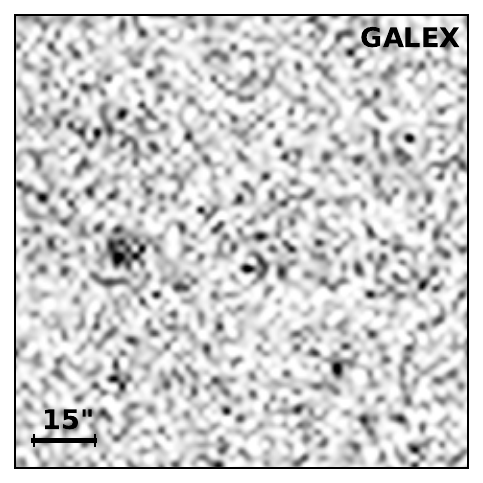}
    \caption{Continue of Fig. \ref{fig:01}.}
\end{figure*}

\begin{figure*}
    \centering
\begin{minipage}{.24\textwidth}
    \centering
    \includegraphics[width=1\textwidth]{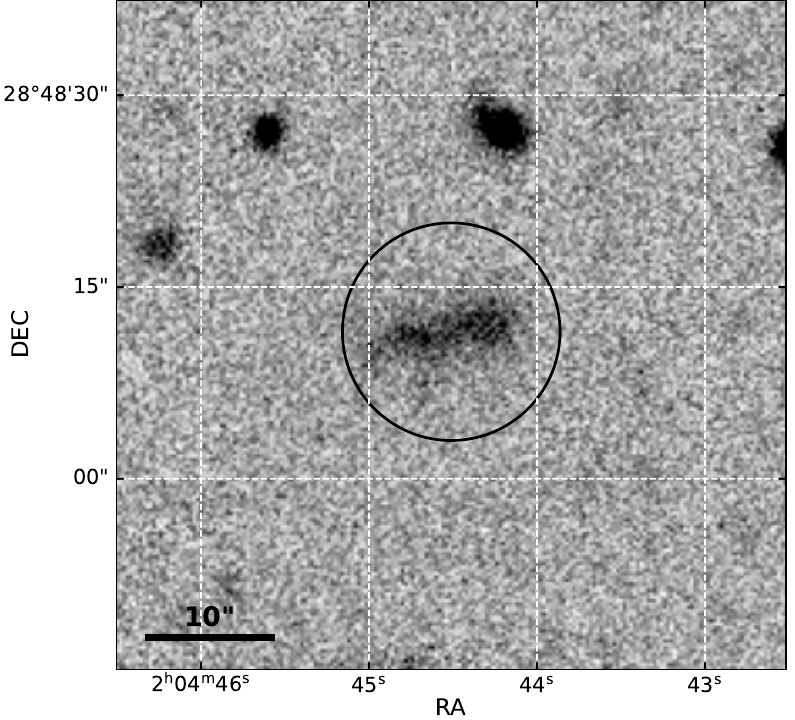}
\end{minipage}
\begin{minipage}{.24\textwidth}
    \raisebox{0.08\height}{\includegraphics[width=0.845\textwidth]{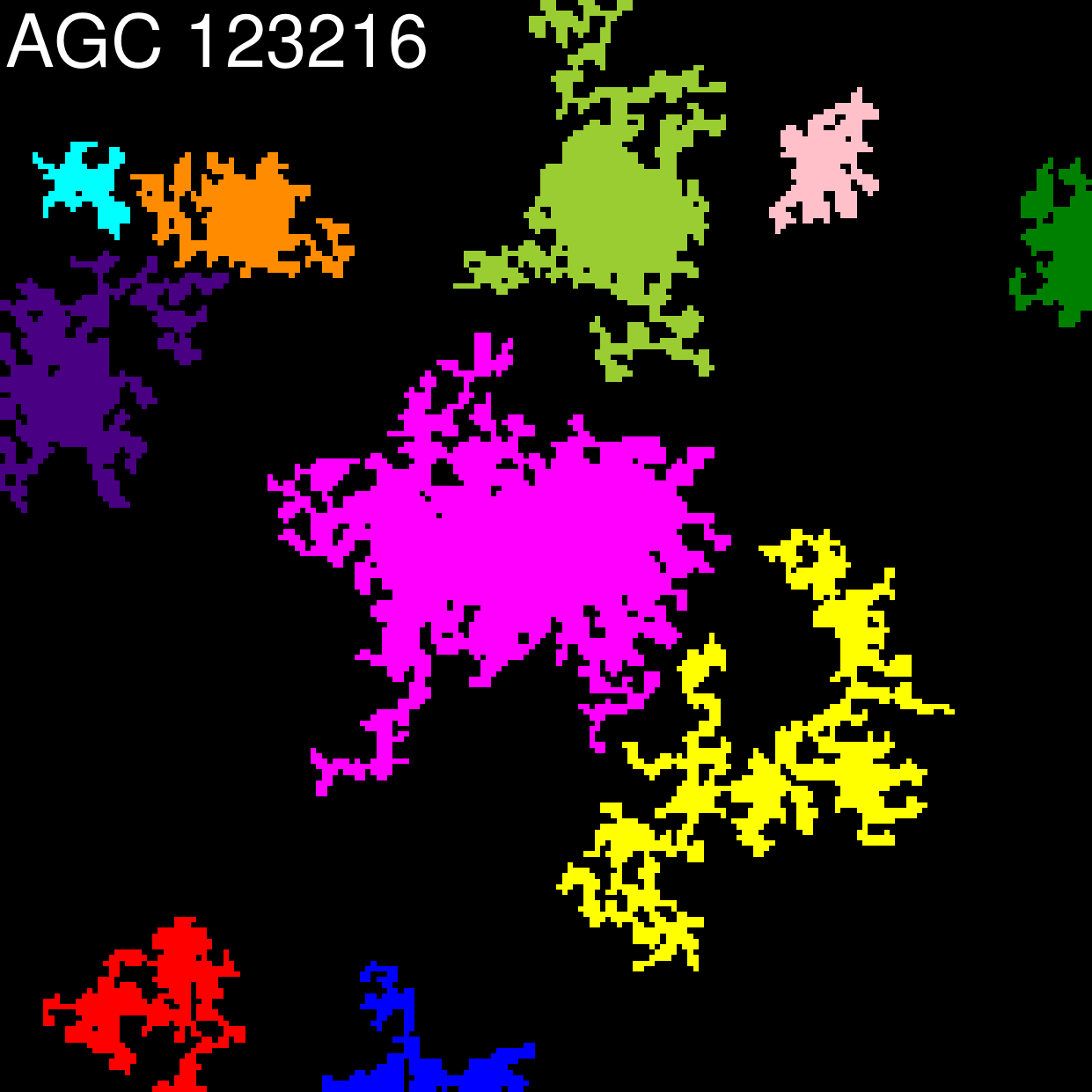}}\qquad
\end{minipage}
\begin{minipage}{.24\textwidth}
    \centering
    \includegraphics[width=1\textwidth]{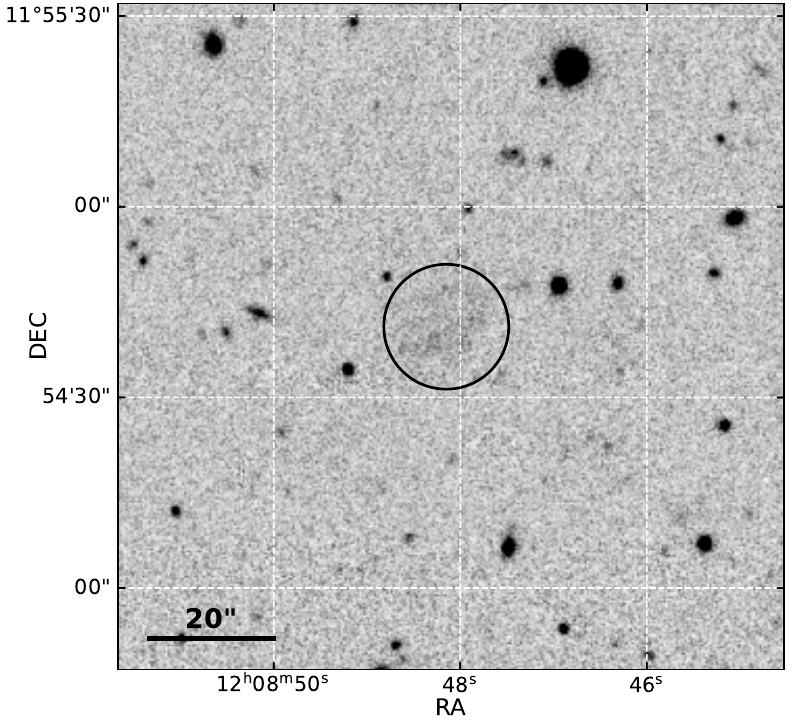}
\end{minipage}
\begin{minipage}{.24\textwidth}
    \raisebox{0.08\height}{\includegraphics[width=0.845\textwidth]{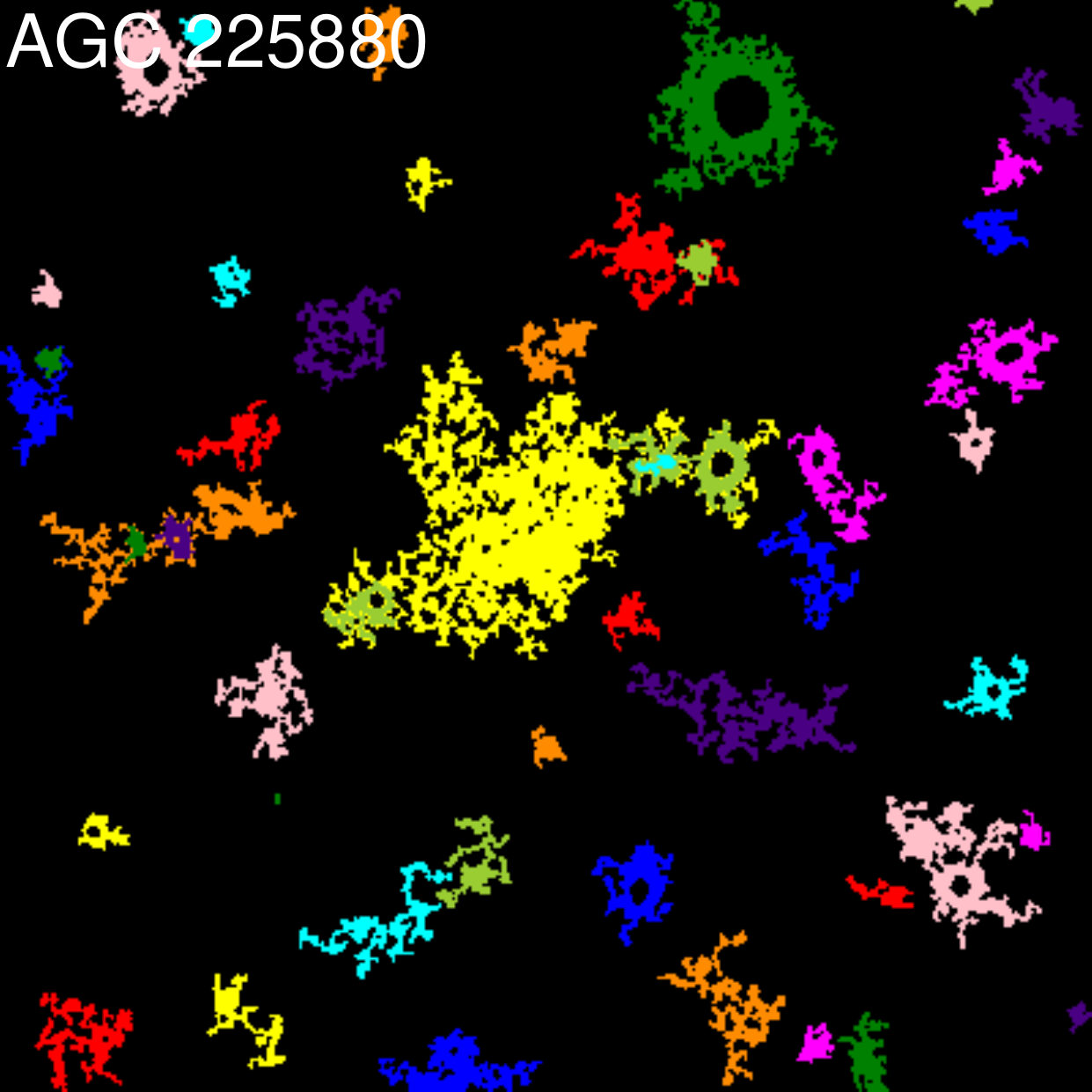}}\qquad
\end{minipage}
\begin{minipage}{.24\textwidth}
    \includegraphics[width=1\textwidth]{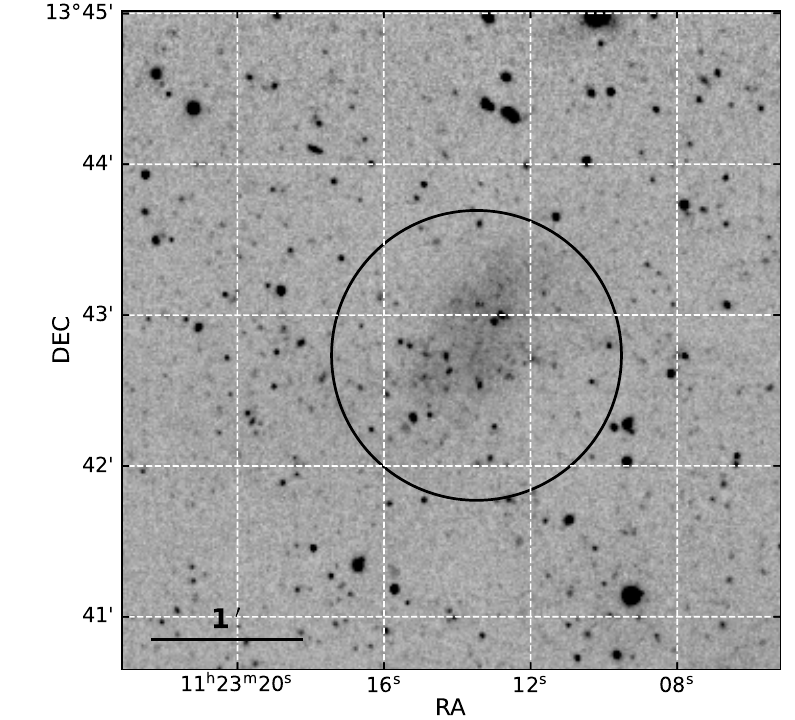}
\end{minipage}
\begin{minipage}{.24\textwidth}
    \raisebox{0.08\height}{\includegraphics[width=0.845\textwidth]{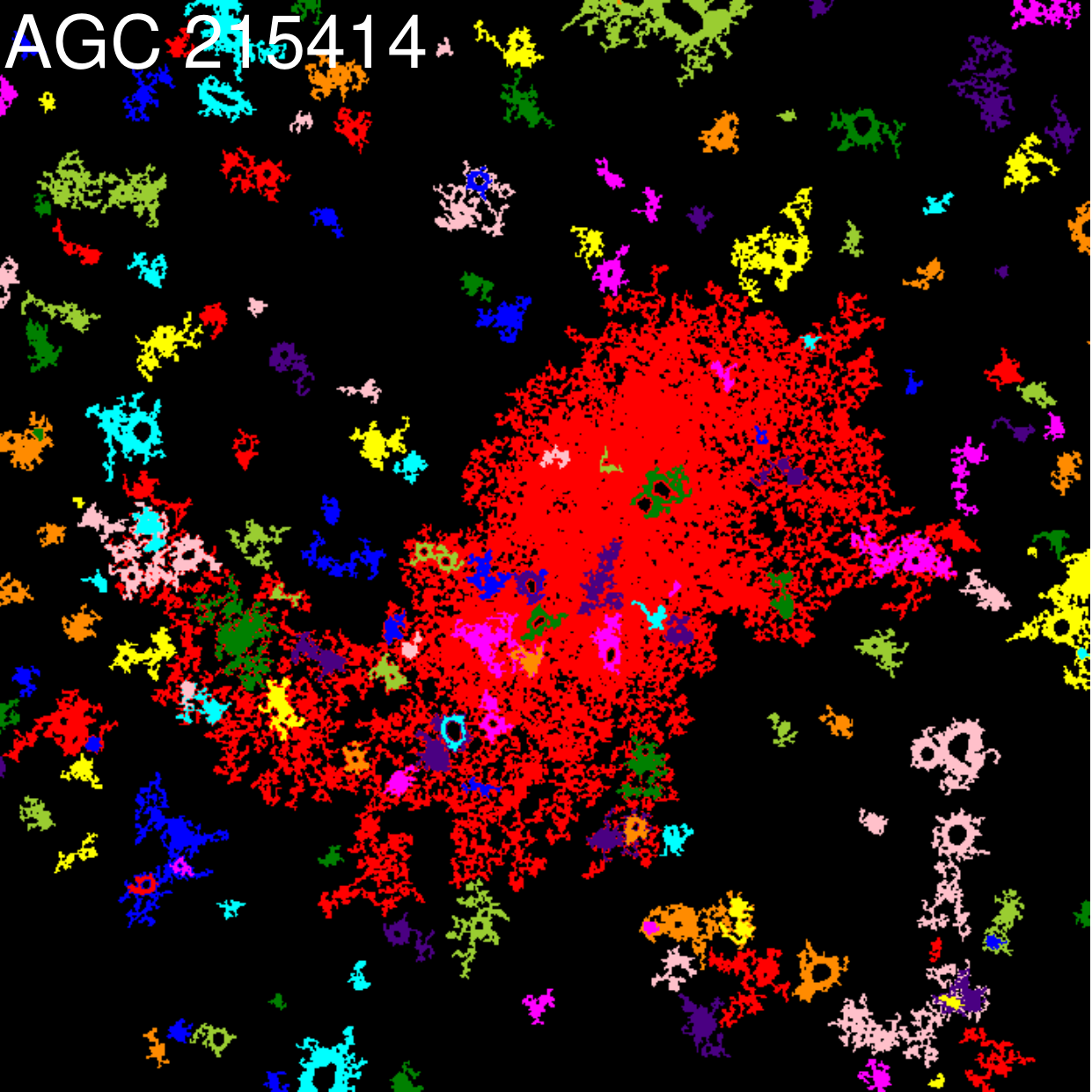}}\qquad
\end{minipage}
\begin{minipage}{.24\textwidth}
    \includegraphics[width=1\textwidth]{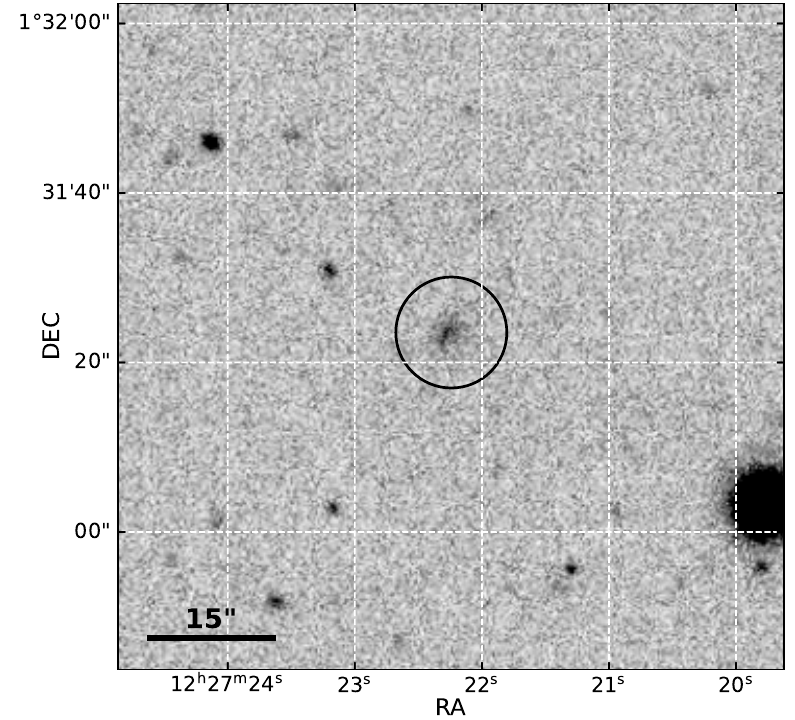}
\end{minipage}
\begin{minipage}{.24\textwidth}
    \raisebox{0.08\height}{\includegraphics[width=0.845\textwidth]{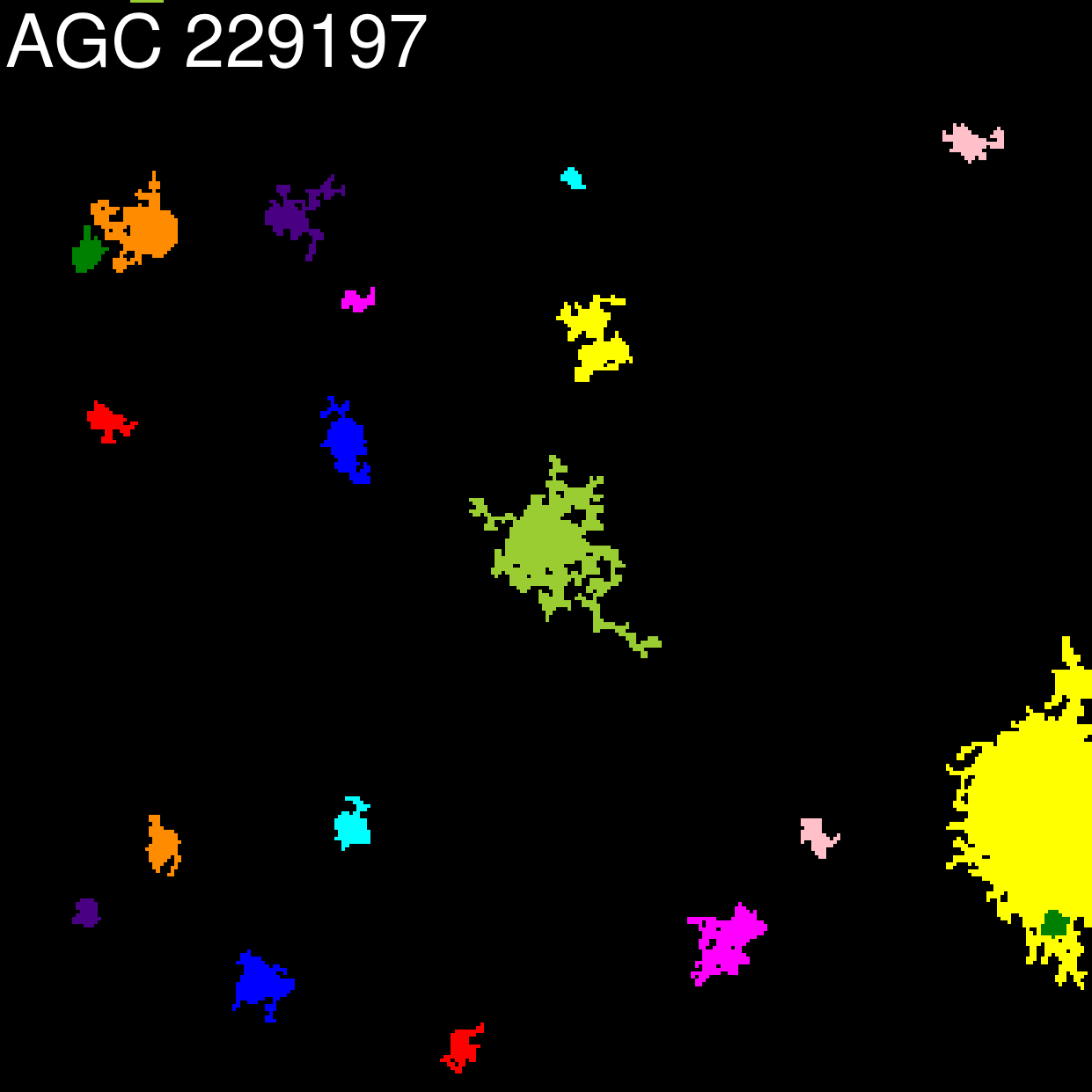}}\qquad
\end{minipage}
\begin{minipage}{.24\textwidth}
    \centering
    \includegraphics[width=1\textwidth]{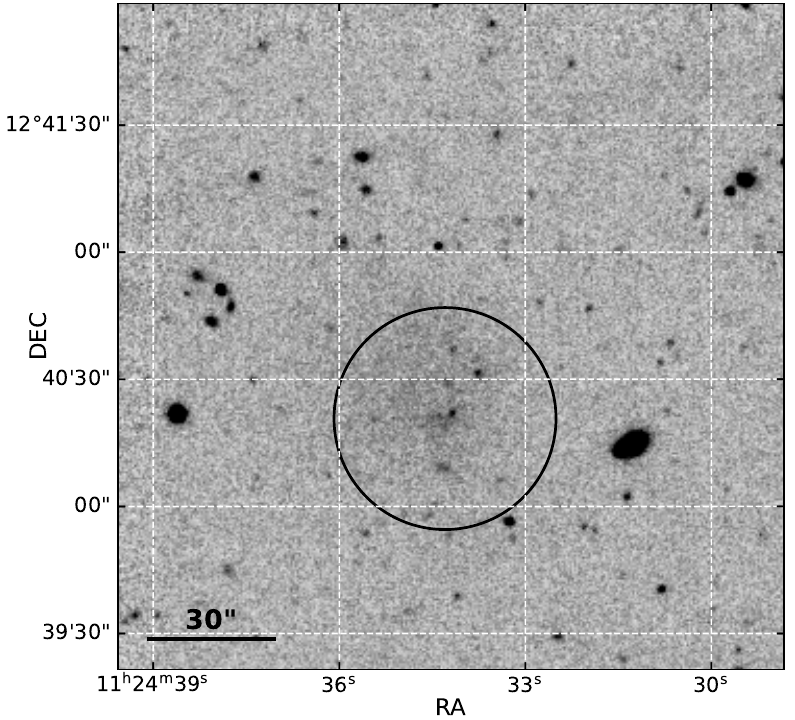}
\end{minipage}
\begin{minipage}{.24\textwidth}
    \raisebox{0.08\height}{\includegraphics[width=0.845\textwidth]{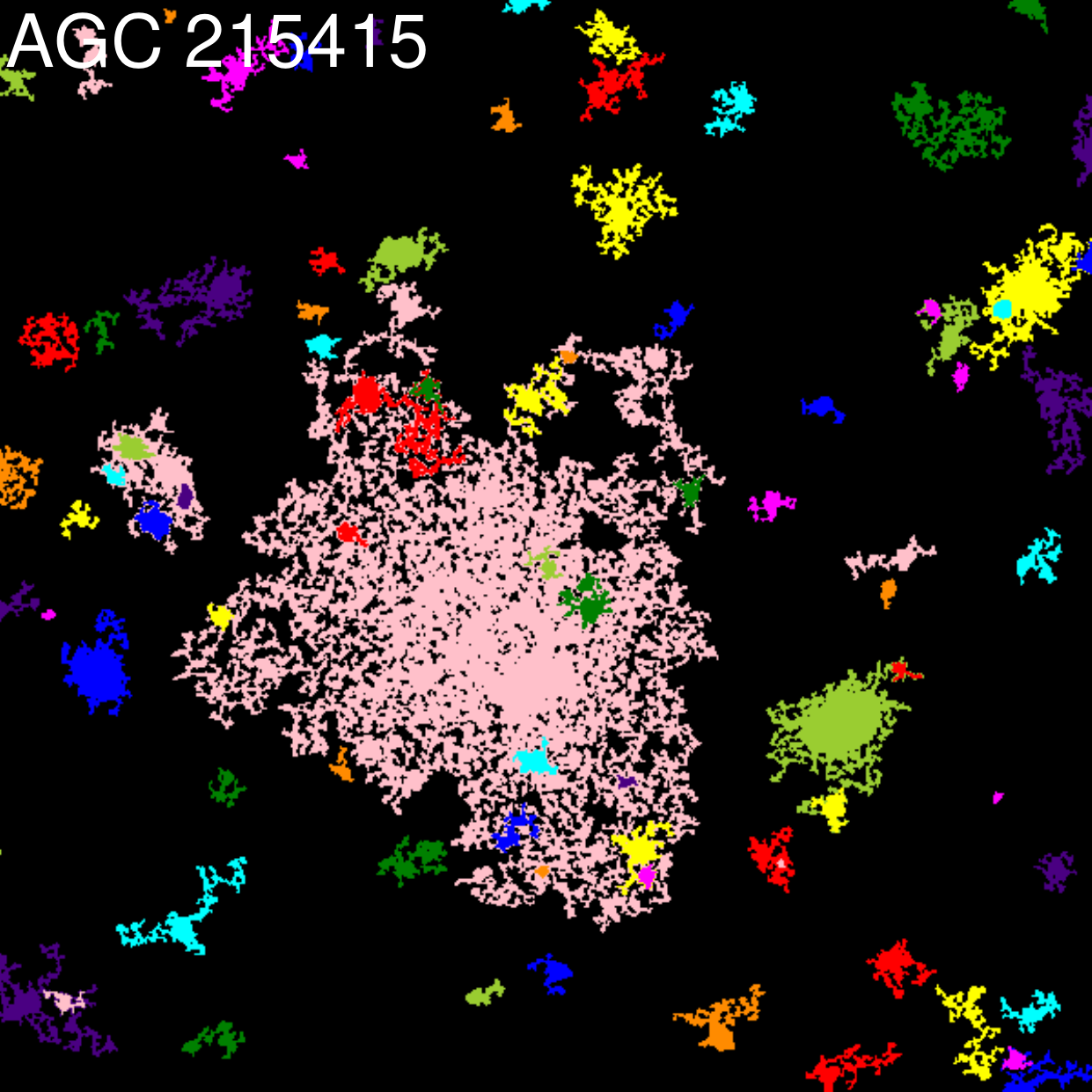}}\qquad
\end{minipage}
\begin{minipage}{.24\textwidth}
    \centering
    \includegraphics[width=1\textwidth]{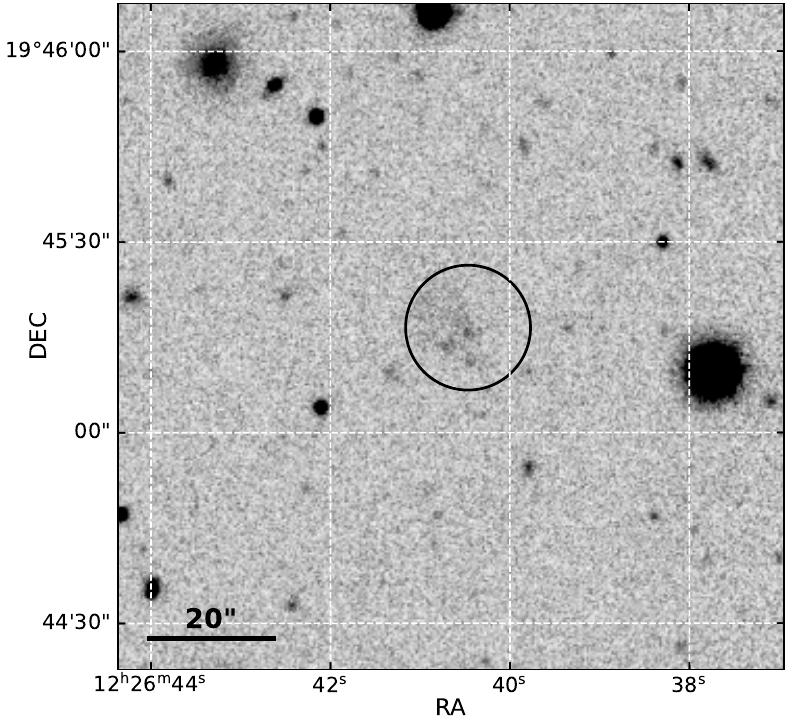}
\end{minipage}
\begin{minipage}{.24\textwidth}
    \raisebox{0.08\height}{\includegraphics[width=0.845\textwidth]{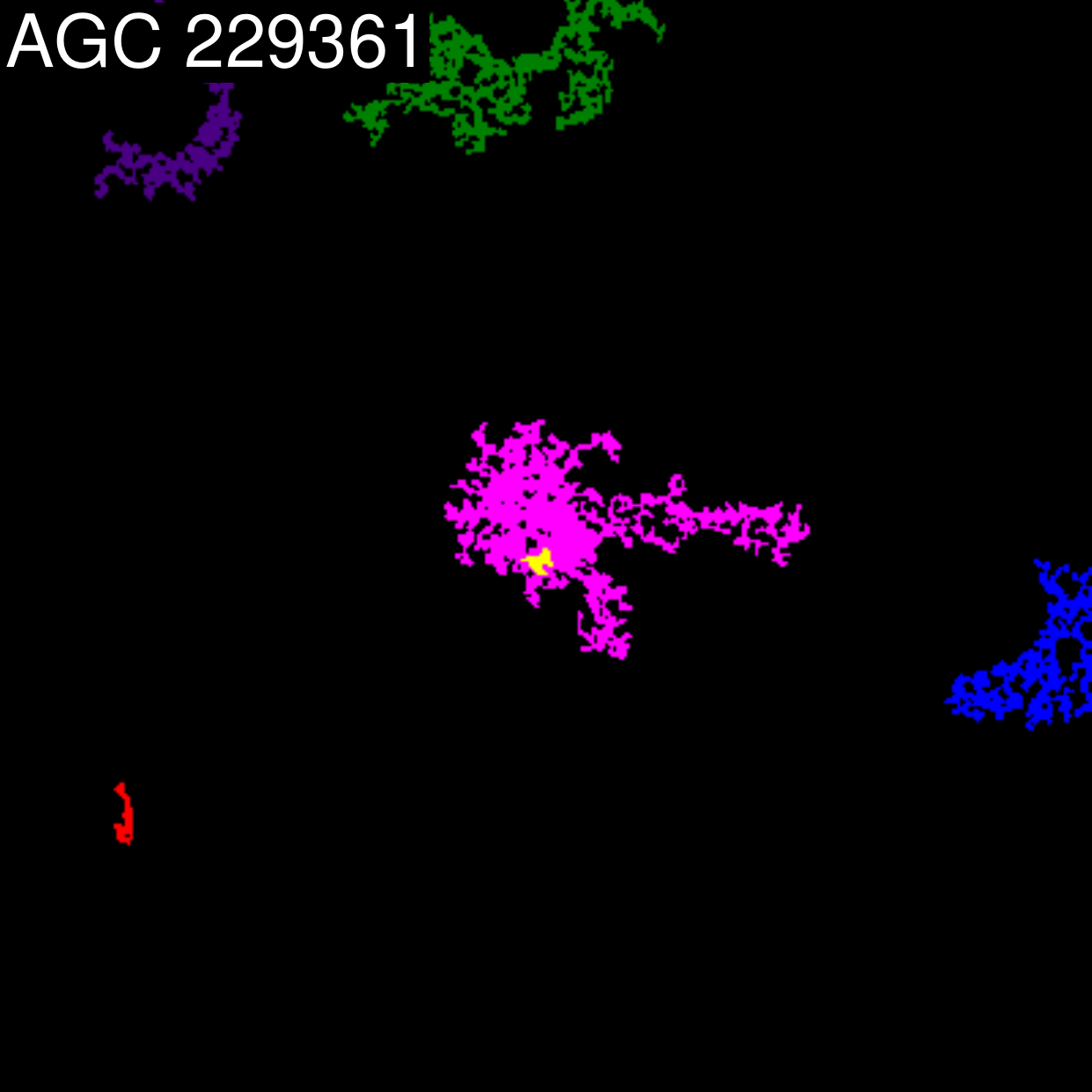}}\qquad
\end{minipage}
\begin{minipage}{.24\textwidth}
    \includegraphics[width=1\textwidth]{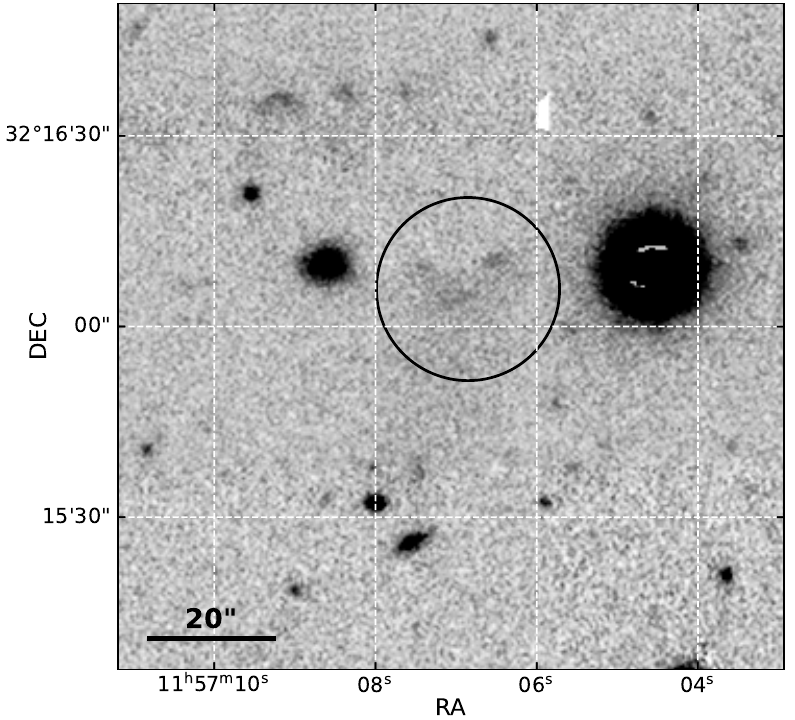}
\end{minipage}
\begin{minipage}{.24\textwidth}
    \raisebox{0.08\height}{\includegraphics[width=0.845\textwidth]{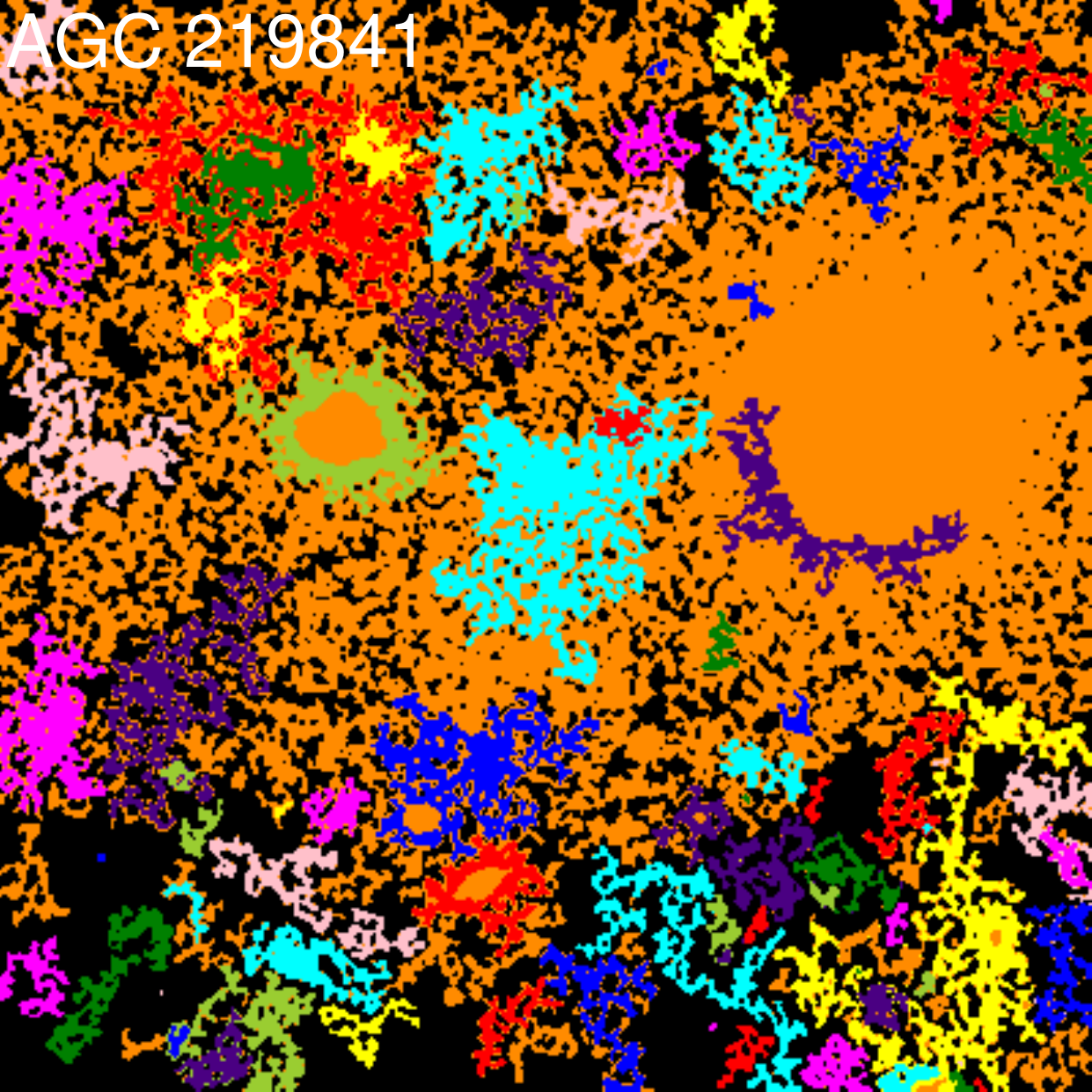}}\qquad
\end{minipage}
\begin{minipage}{.24\textwidth}
    \includegraphics[width=1\textwidth]{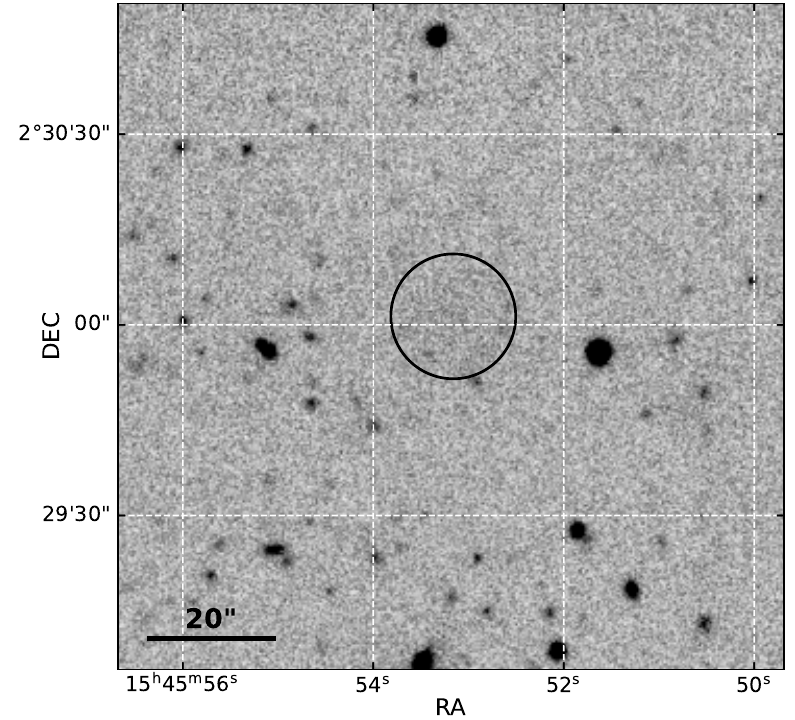}
\end{minipage}
\begin{minipage}{.24\textwidth}
    \raisebox{0.08\height}{\includegraphics[width=0.845\textwidth]{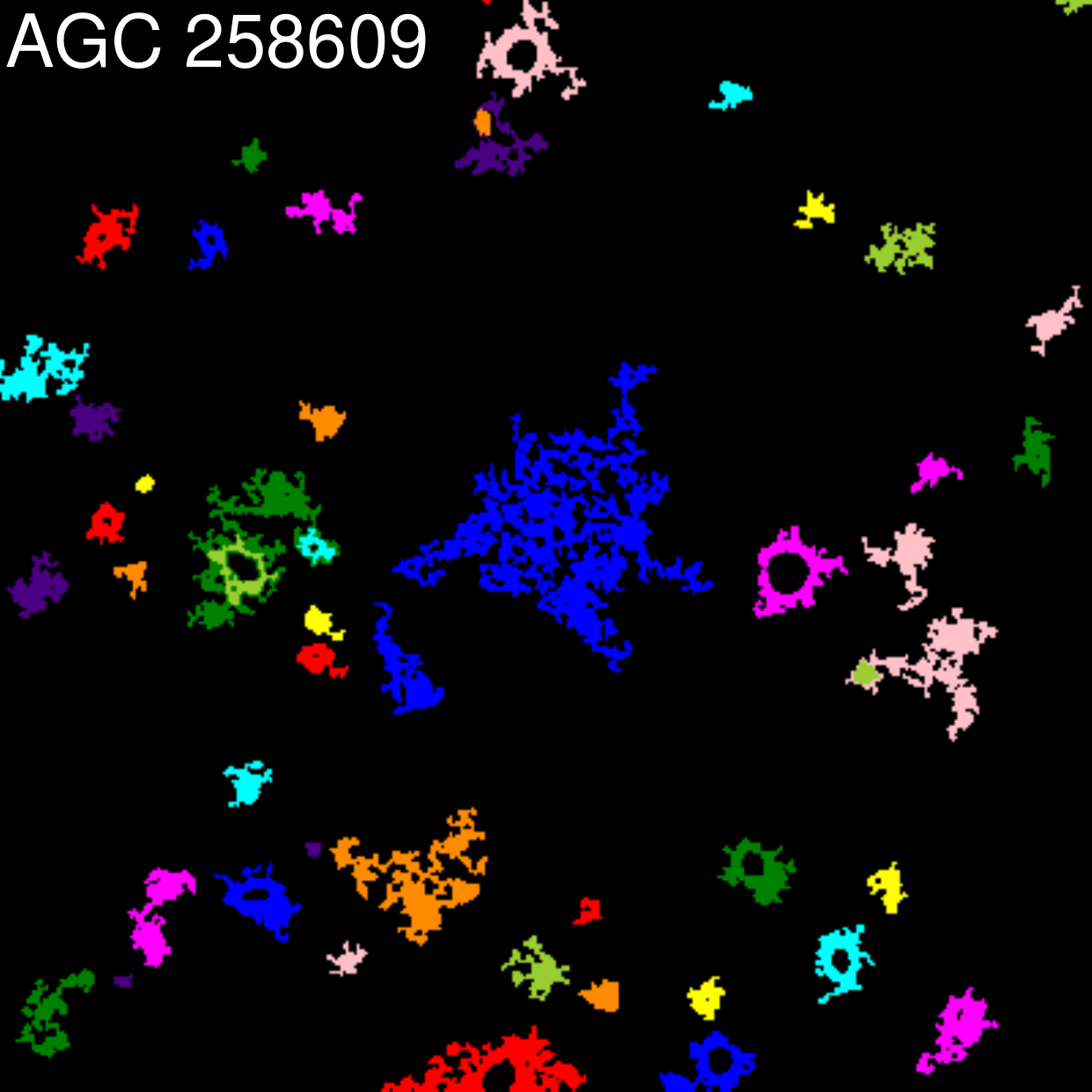}}\qquad
\end{minipage}

\caption{The black photometric apertures overlaid on the $g$-band images (left), and the color-segmentation images by \texttt{MTObjects} detection (right), arranged in two columns from top to bottom, following the same order as Table \ref{tab:01}. The apertures are determined by \texttt{SExtractor} with the initial input parameters based on \texttt{MTObjects} with minor adjustments, and they appear slightly smaller than the sources detected in segmentation images. It is due to the possibility that \texttt{MTObjects} may over-detect additional pixels from the periphery of the sources. We have empirically removed several such pixels that are unlikely to physically be part of the sources. \textbf{See more details about the photometric aperture in Section \ref{sec:03-01}.}}
\label{fig:02}
\end{figure*}


\begin{figure*}
    \centering
    \includegraphics[width=0.23\textwidth]{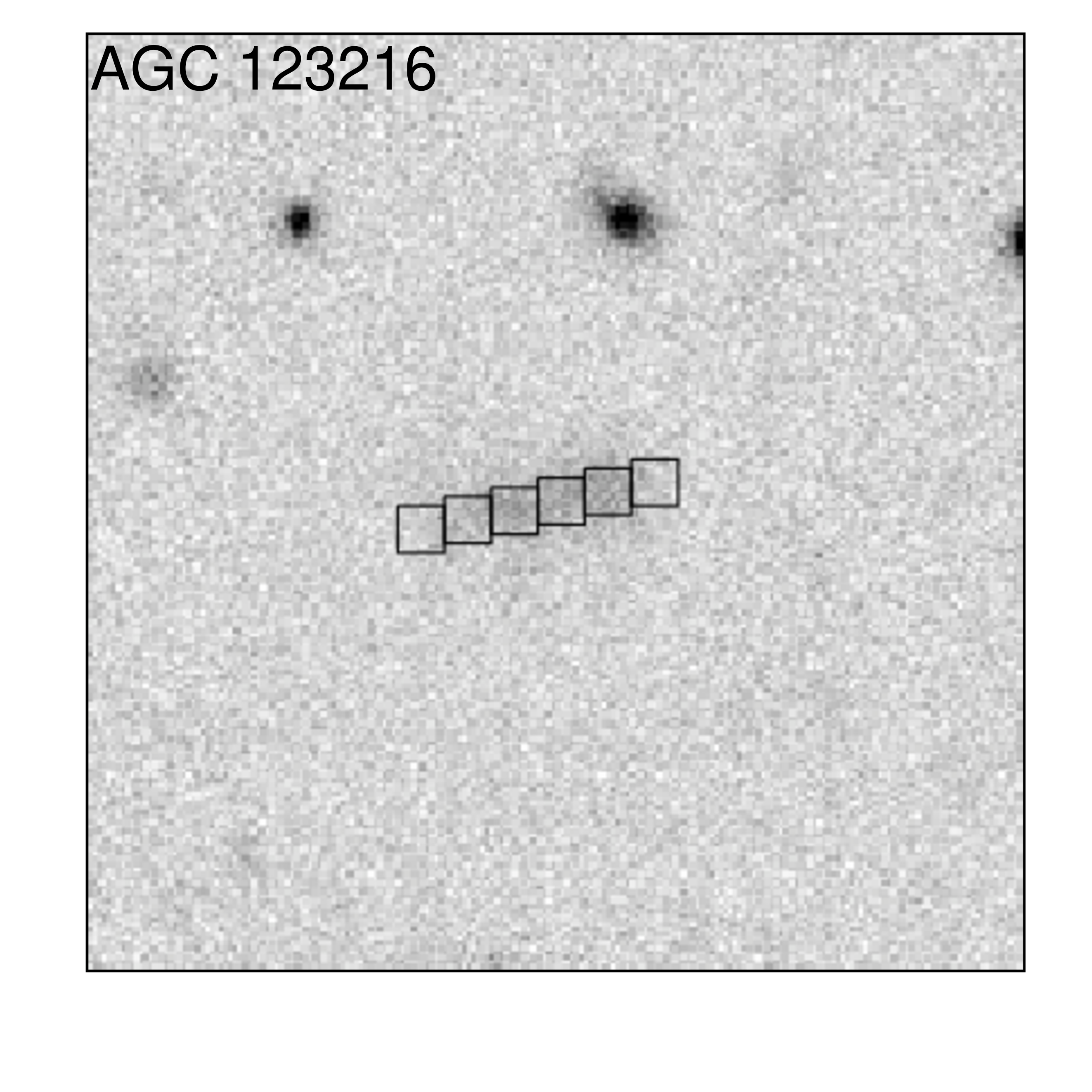}
    \includegraphics[width=0.229\textwidth]{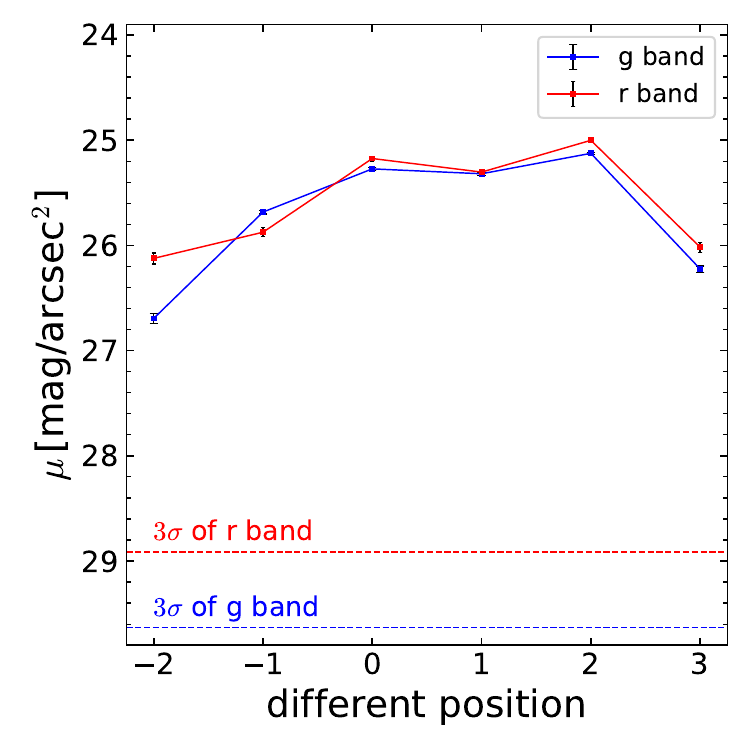}\qquad
    \includegraphics[width=0.23\textwidth]{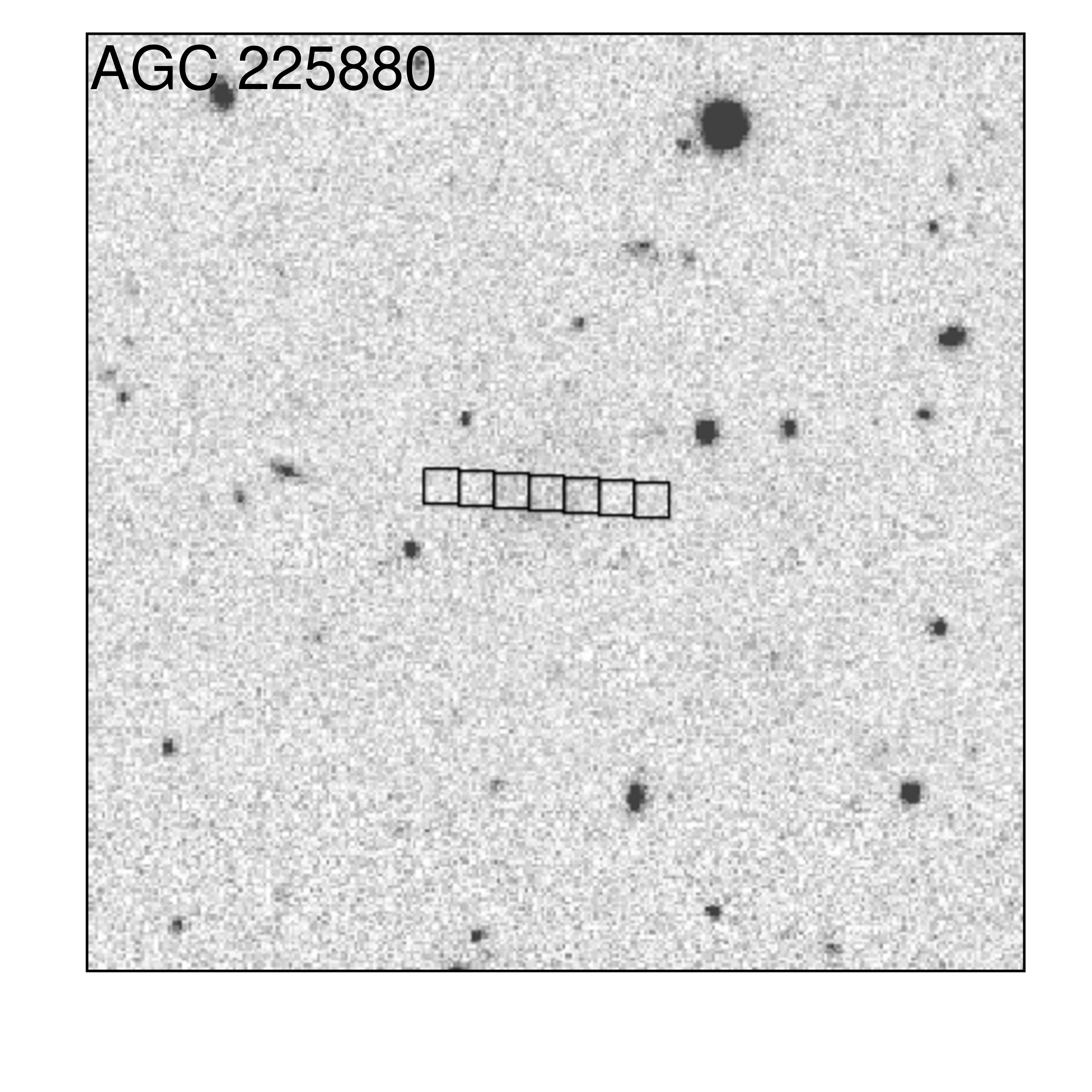}
    \includegraphics[width=0.229\textwidth]{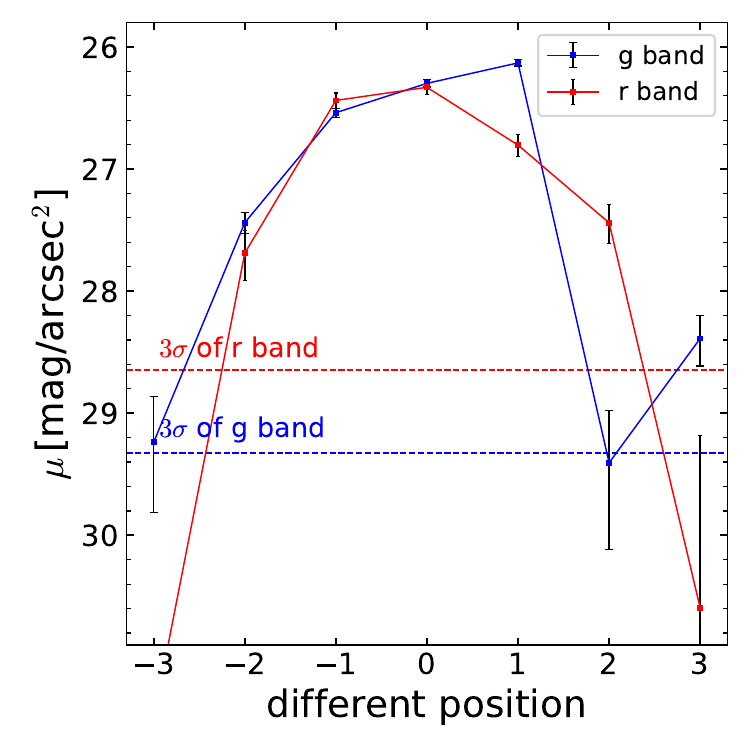}
     \\
    \includegraphics[width=0.23\textwidth]{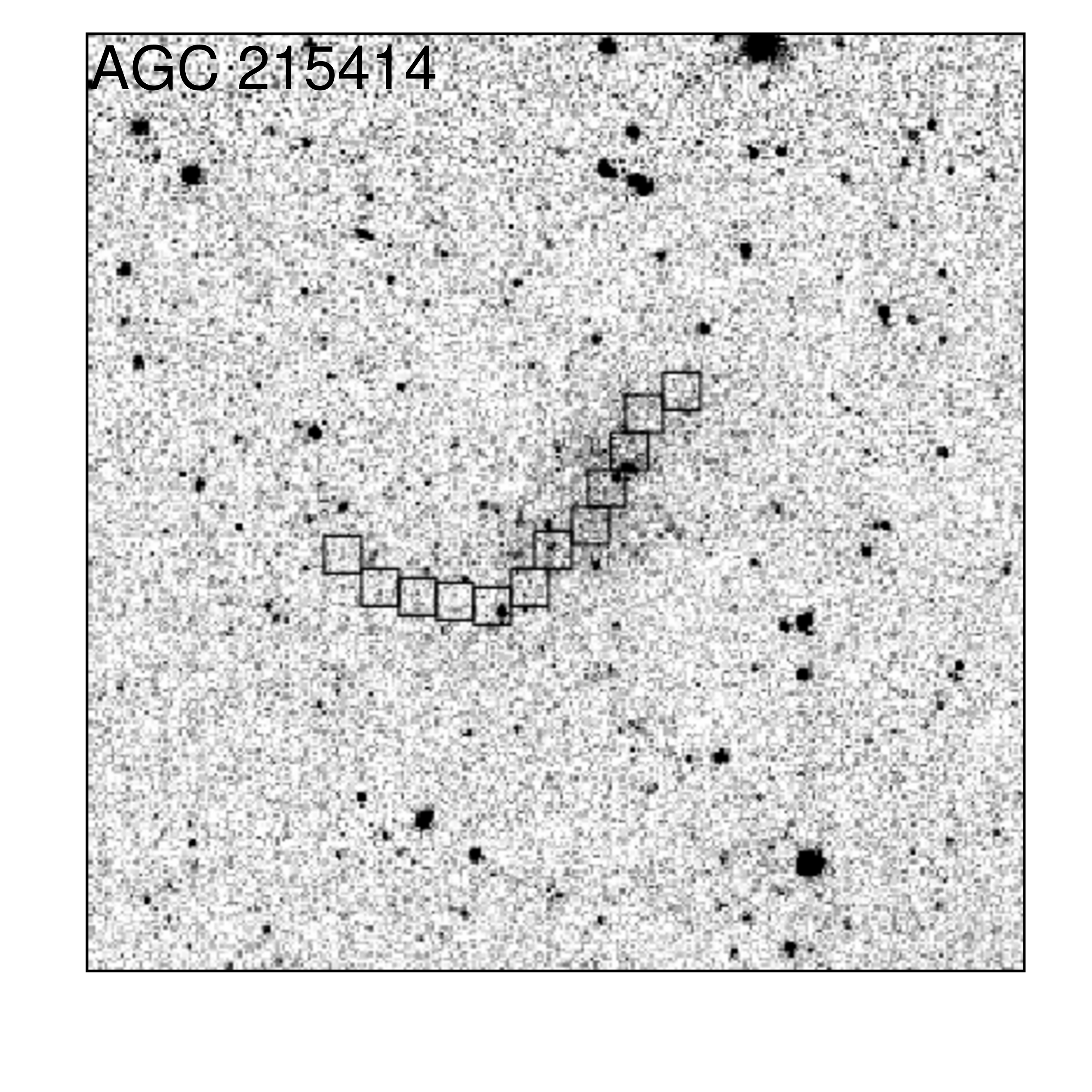}
    \includegraphics[width=0.229\textwidth]{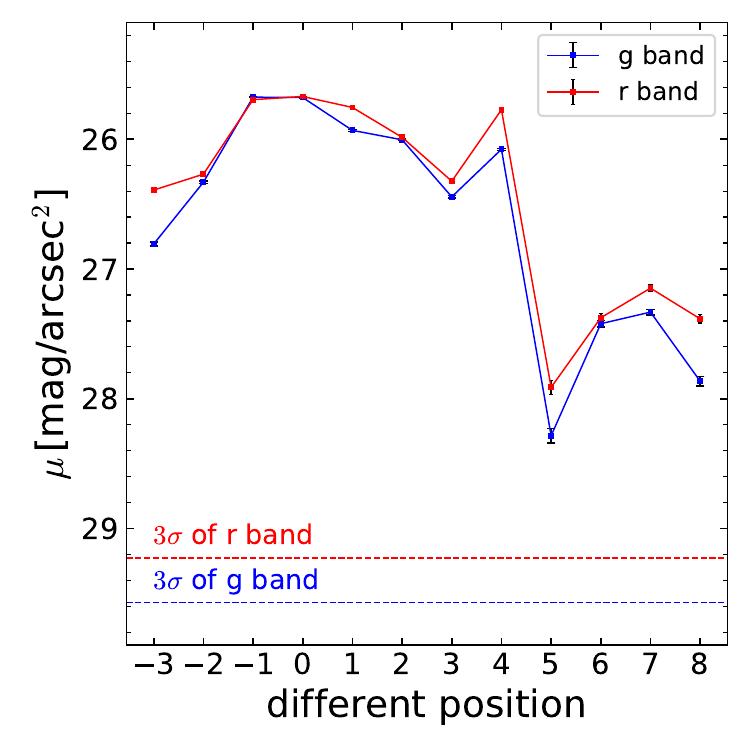}\qquad
    \includegraphics[width=0.23\textwidth]{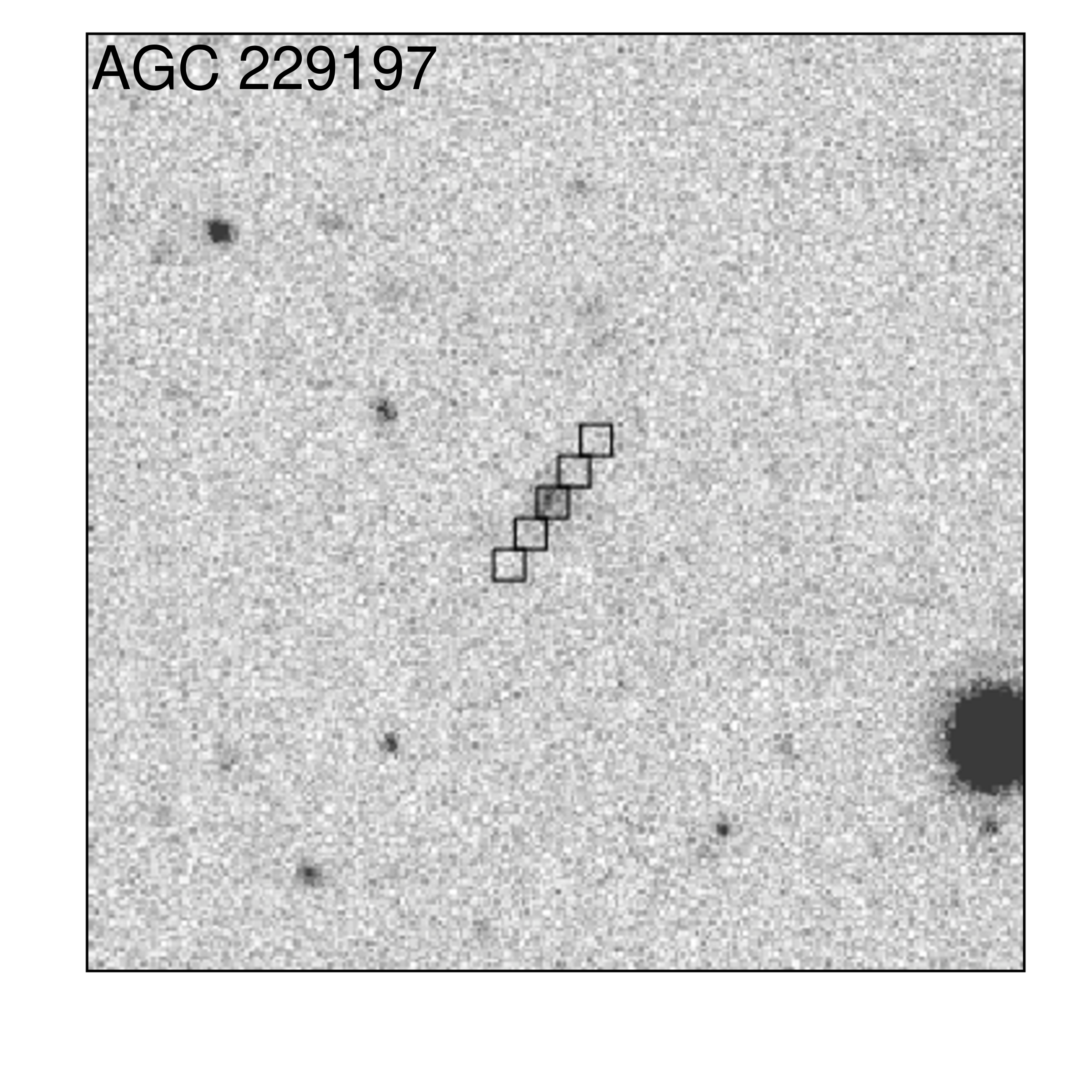}
    \includegraphics[width=0.229\textwidth]{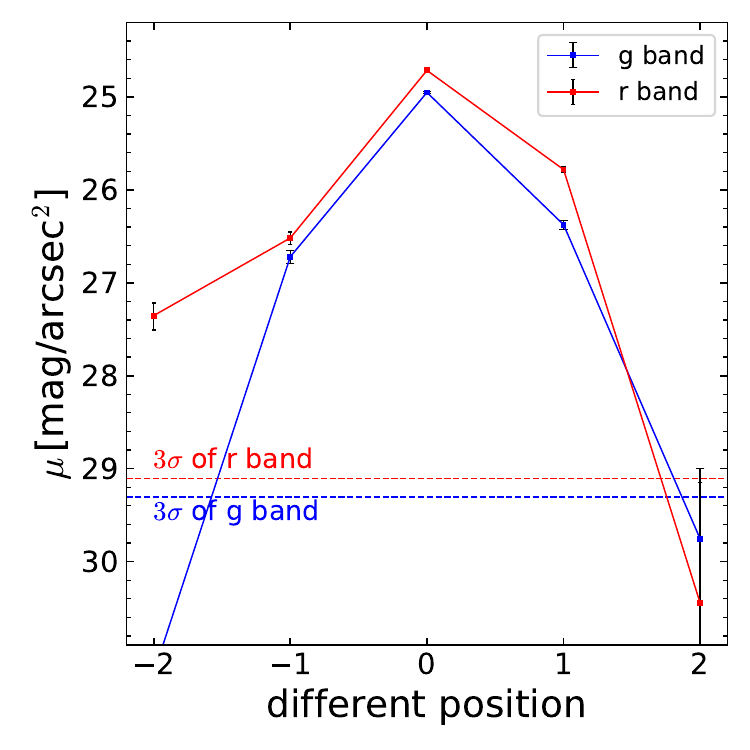}
    \\
    \includegraphics[width=0.23\textwidth]{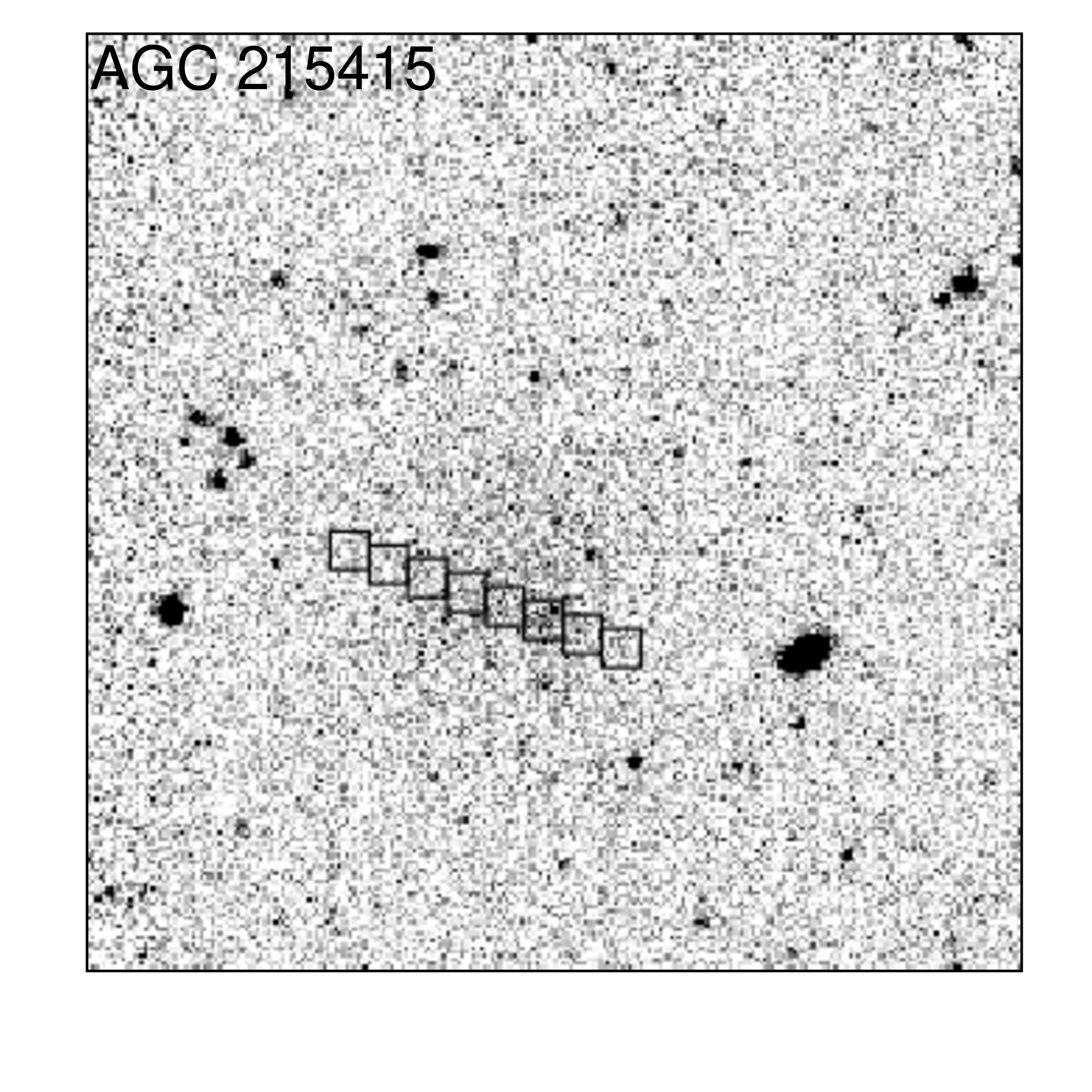}
    \includegraphics[width=0.229\textwidth]{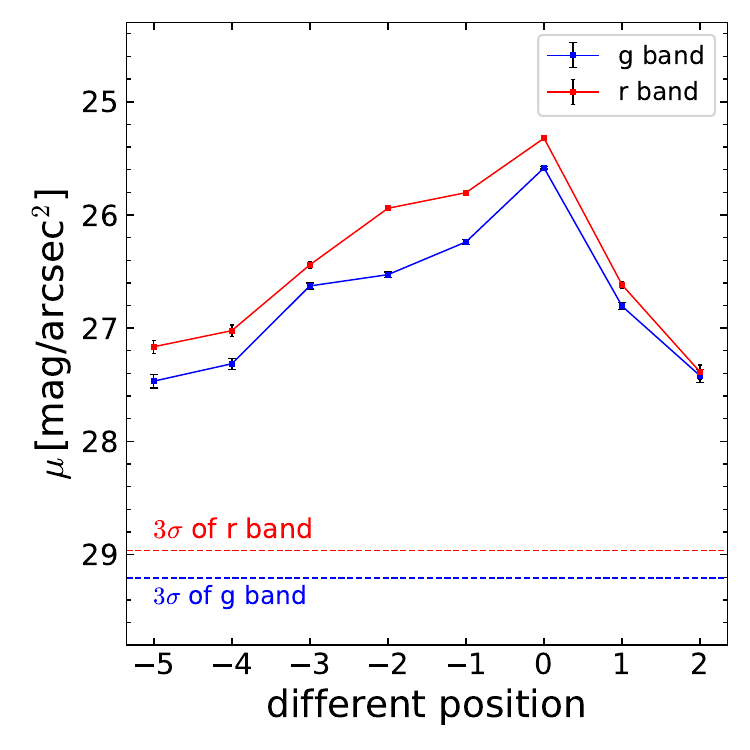}\qquad
    \includegraphics[width=0.23\textwidth]{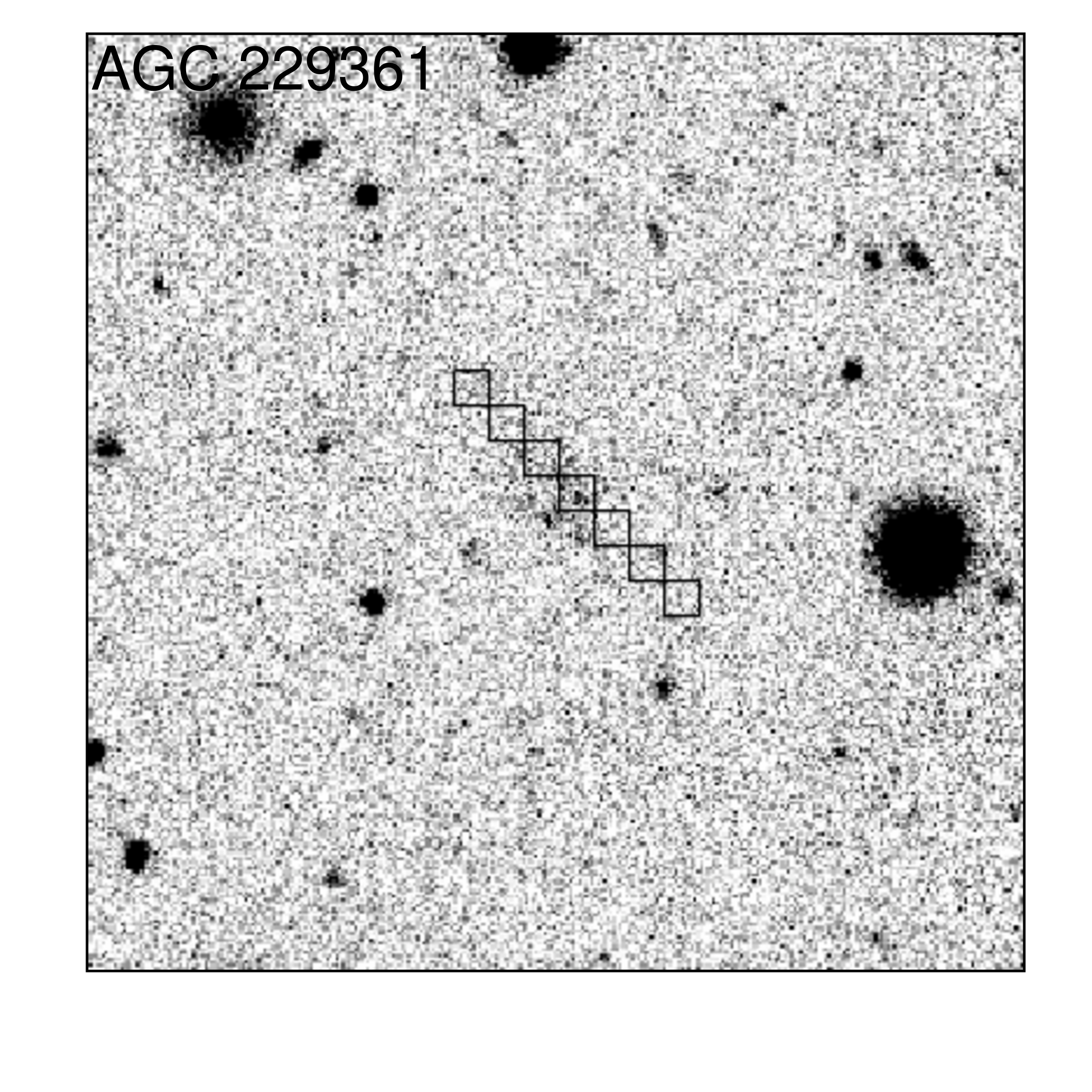}
    \includegraphics[width=0.229\textwidth]{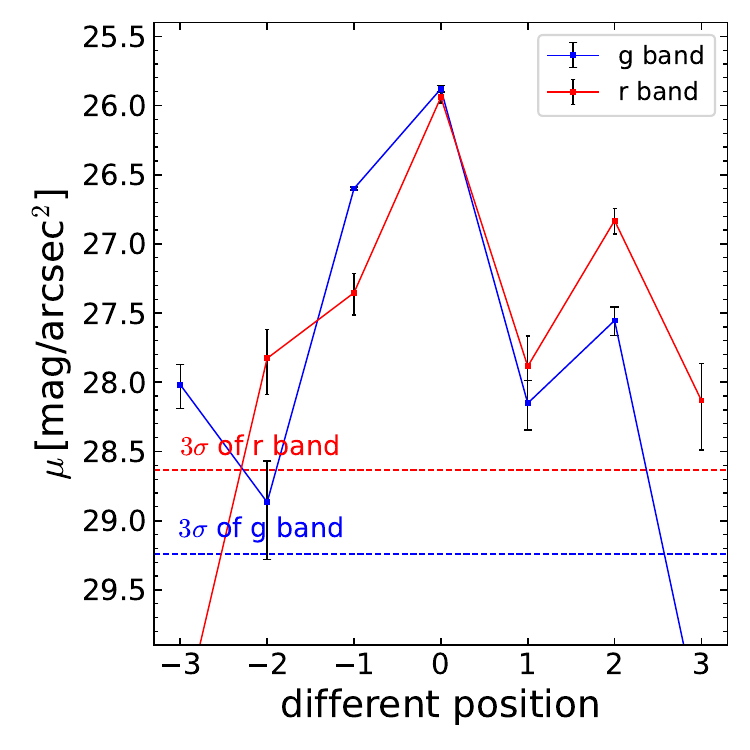}
    \\
    \includegraphics[width=0.23\textwidth]{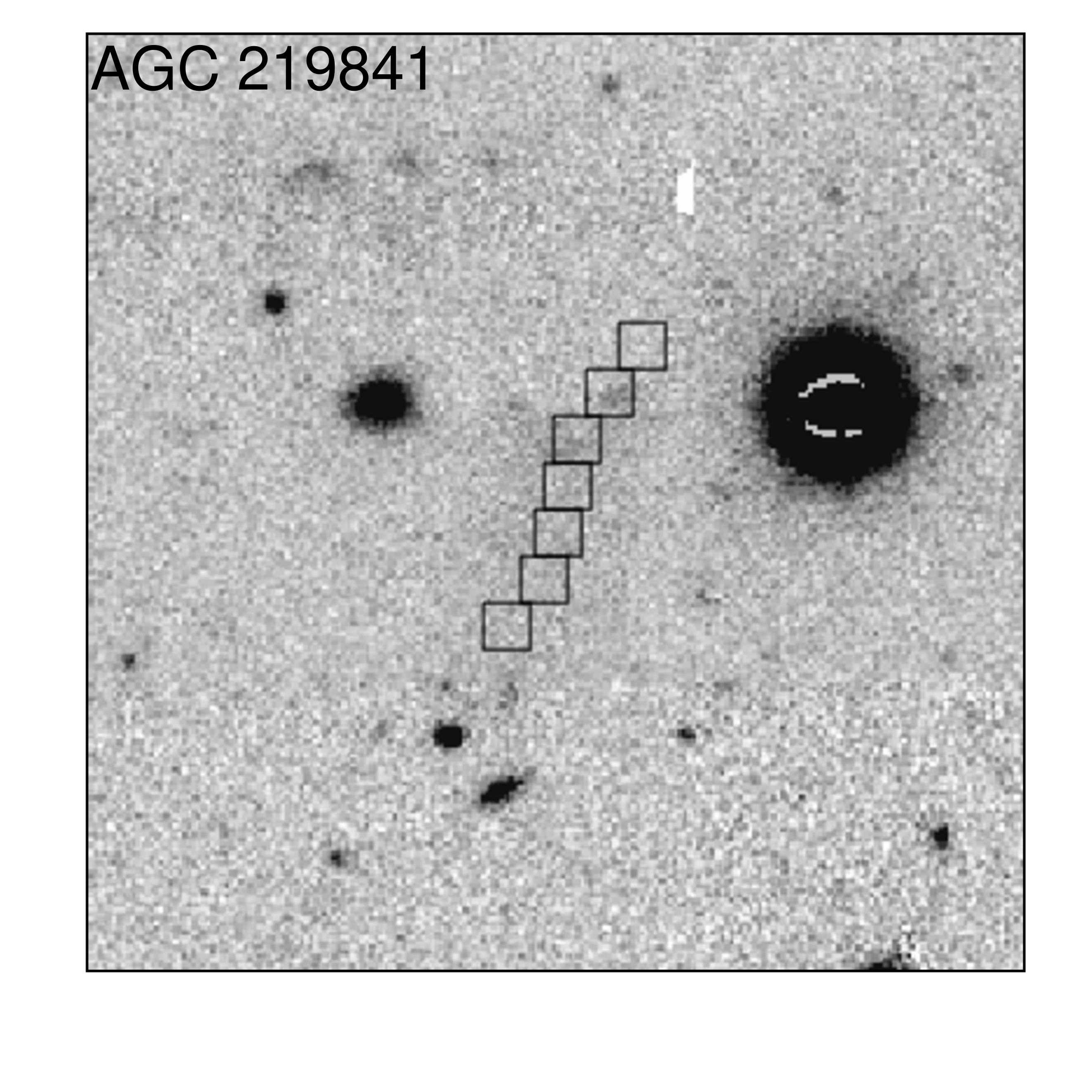}
    \includegraphics[width=0.229\textwidth]{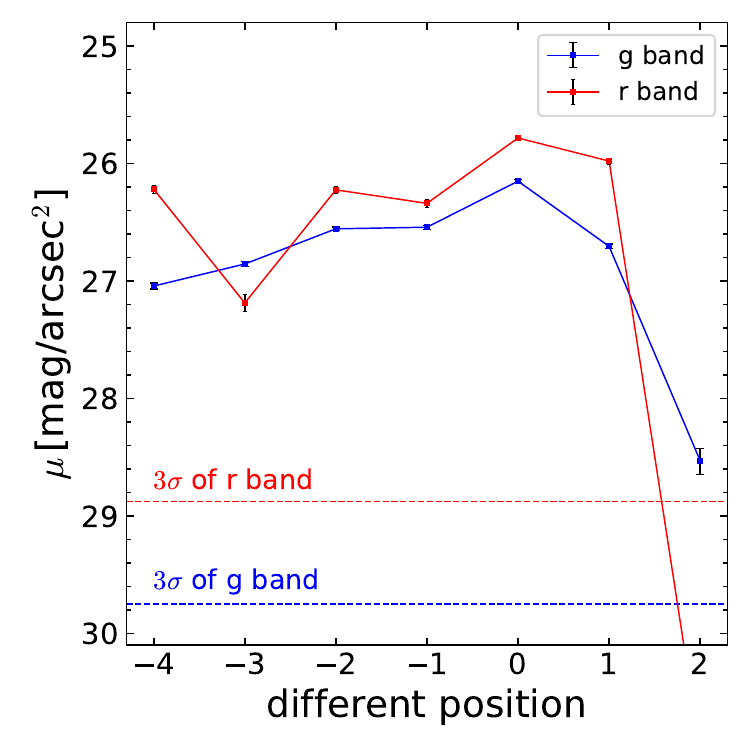}\qquad
    \includegraphics[width=0.23\textwidth]{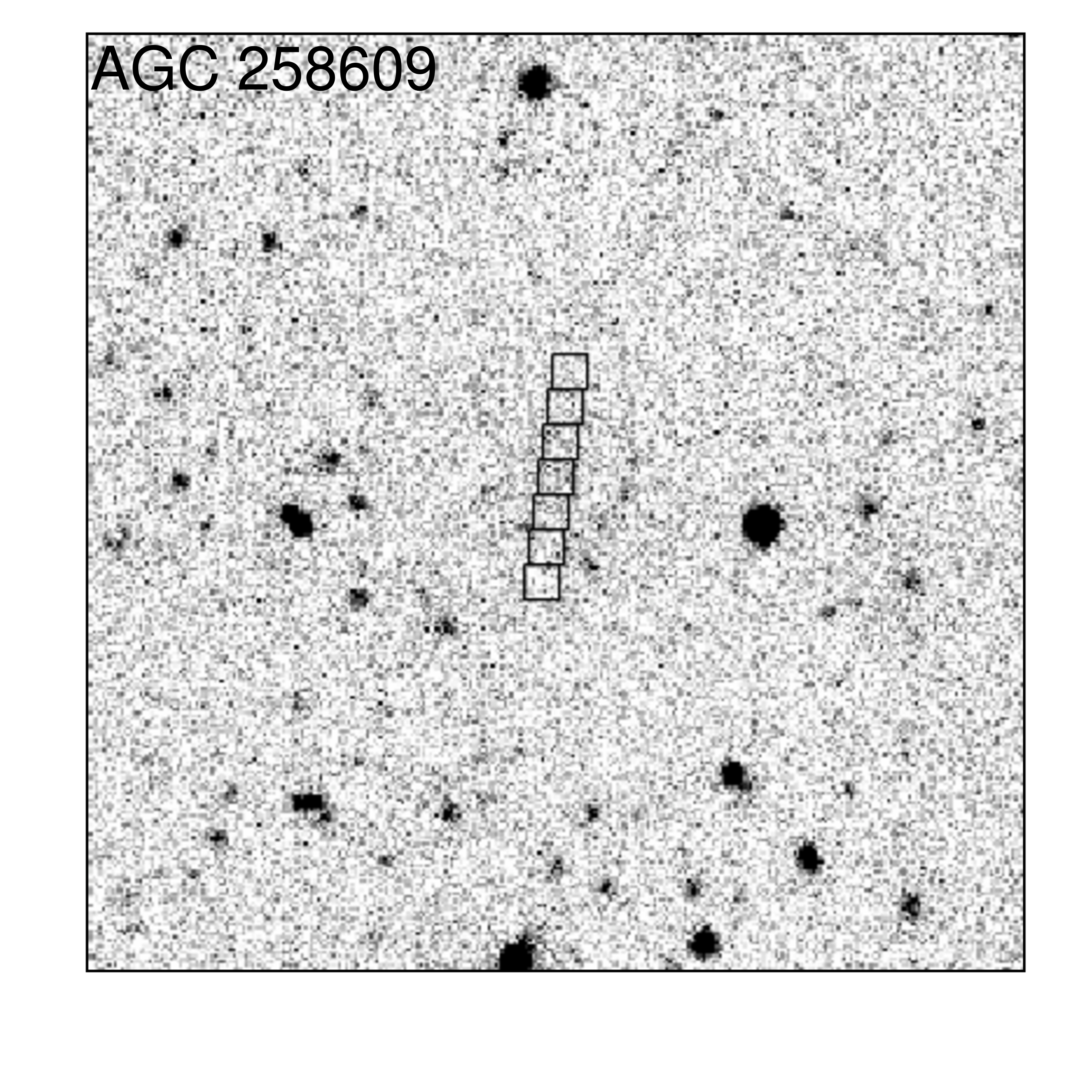}
    \includegraphics[width=0.229\textwidth]{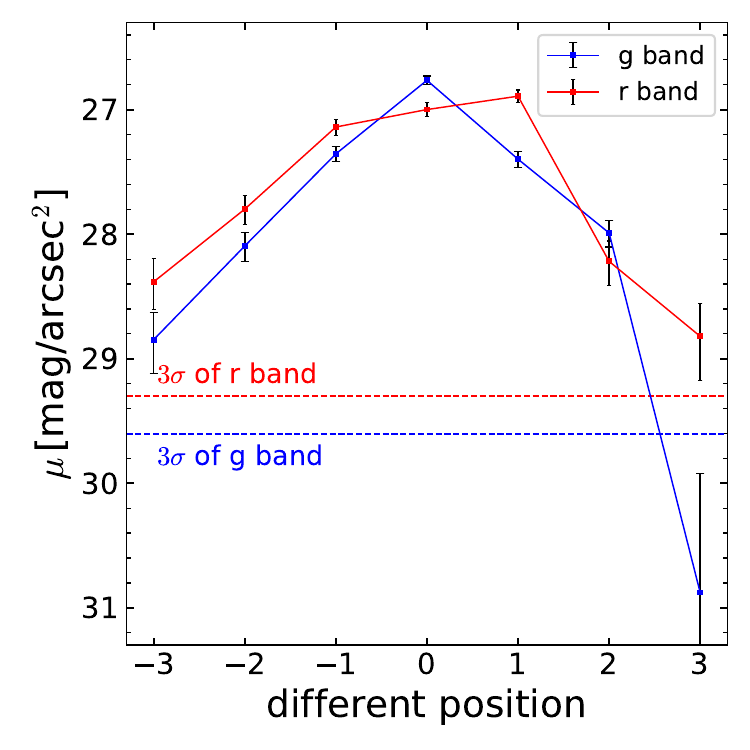}
    \caption{The positions of boxes for the surface brightness measurements (left), and the variation of surface brightness in these boxes (right), arranged in two columns from top to bottom, in the same order as Table \ref{tab:01}. For each source, the left image shows a zoomed-in view of the DECaLS $g$-band image with black boxes indicating where the surface brightness was measured; the right image displays the surface brightness of the $g$- and $r$-band images within these boxes, from left to right. The photometric results of the $g$- and $r$-band images are depicted with points and lines in blue and red, respectively. The error bars are shown in black lines, which are derived from the inverse variance images. The dotted lines correspond to the $3\sigma$ detection limits, determined by the background variance.}
    \label{fig:03}
\end{figure*}

\begin{figure*}
    \plotone{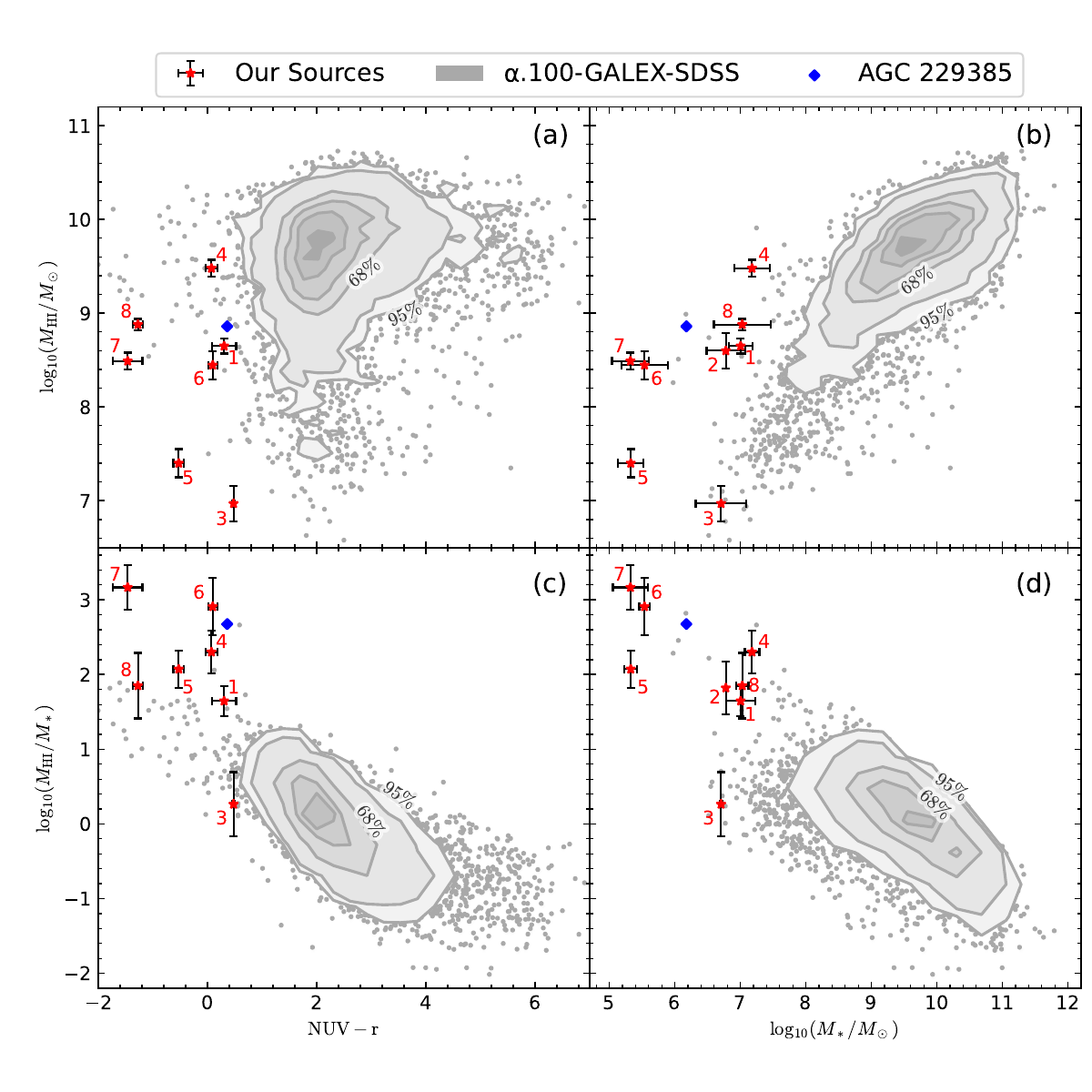}
    \centering
    \caption{The scaling relations between $\log_{10}M_{\rm H\textsc{i}}$, $\log_{10}M_{*}$, $\log_{10}(M_{\rm H\textsc{i}}/M_{*})$ and the (NUV-$r$) color. Our sources are shown as red pentagrams, with numbers corresponding to the ordinal numbers in Table \ref{tab:01}. The $\alpha.100$-GALEX-SDSS sample forms the background, shown as gray contours and dots. Contours in each panel from inside to outside represent 10$\%$, 30$\%$, 50$\%$, 68$\%$, 90$\%$ and 95$\%$ of the sample based on the concentration, with 1$\sigma$ and 2$\sigma$ lines specially marked. Outliers are shown as dots. Panels (a) and (c) show the relations between (NUV-$r$) color with $M_{\rm H\textsc{i}}$ and $\log_{10}(M_{\rm H\textsc{i}}/M_{*})$. Lacking NUV data of AGC\,215414, only seven other sources appear on the two panels. Panels (b) and (d) show relations of $M_{\rm H\textsc{i}}$ and $\log_{10}(M_{\rm H\textsc{i}}/M_{*})$ with $M_{*}$. Notably, except for AGC\,215415 (number 3), which is near the 95$\%$ range of $\alpha.100$ sample on panel (c), all other sources significantly deviated from the scaling relations of $\alpha.100$ sample.}
    \label{fig:04}
\end{figure*}

\begin{figure*}
    \centering
    \includegraphics[scale=0.65]{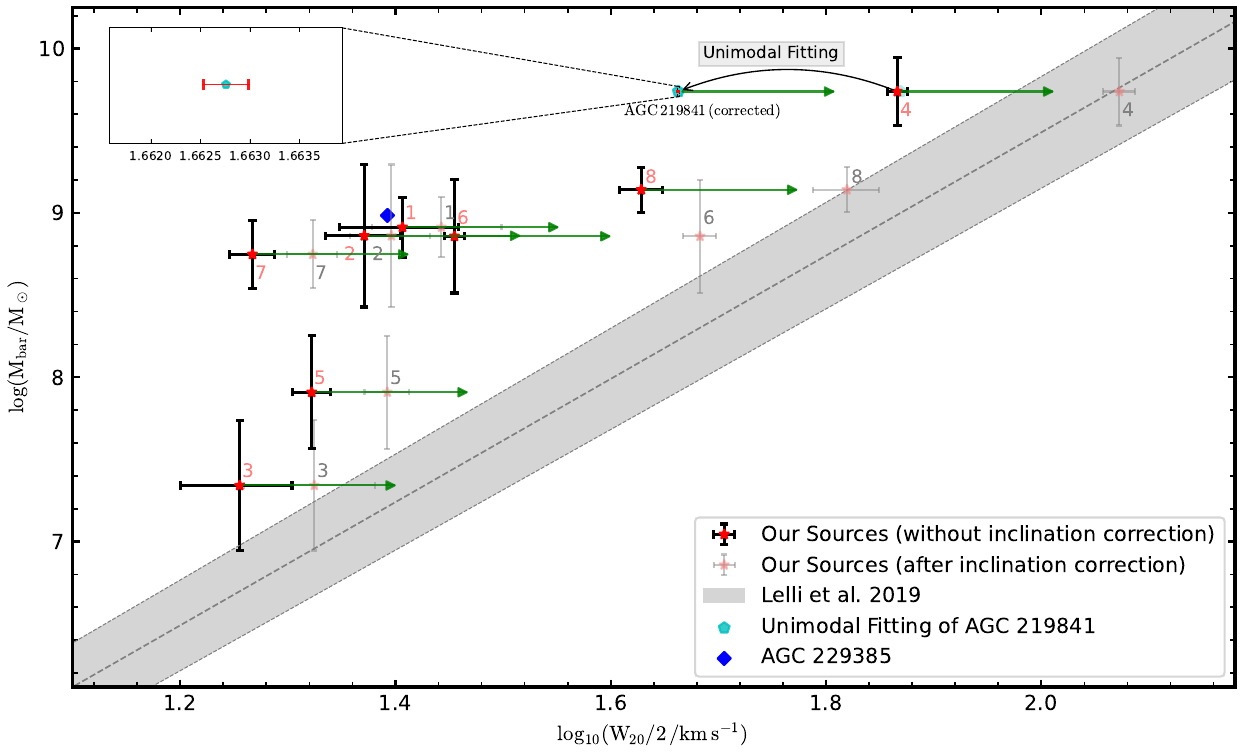}
    \caption{Baryonic Tully-Fisher Relation of our sources using  $W_{20}$ as velocity estimators. Our sources are represented by red pentagrams with black error bars. The gray stars and error bars indicate the position of sources after optical inclination corrections. The green arrows indicate where our sources may appear after accounting for a large error in optical inclination. BTFrs for bright galaxies from \citet{2019MNRAS.484.3267L} and the corresponding scatters are shown as black dotted lines and gray areas. The labeling of the eight sources as 1 to 8 corresponds to that in Table \ref{tab:01}. Moreover, the result of the single-peak refitting for the H\textsc{i} profile of AGC\,219841 is indicated by a green star symbol. The \textit{almost dark} galaxy AGC\,229385 \textbf{\citep{2015ApJ...801...96J}} is represented by a blue diamond. We found that three of our sources exhibit deviations from classical BTFrs towards the heavier baryonic mass end, which is similar to AGC\,229385.}
    \label{fig:05}
\end{figure*}

\begin{figure*}
    \centering
    \includegraphics[width=0.498\textwidth]{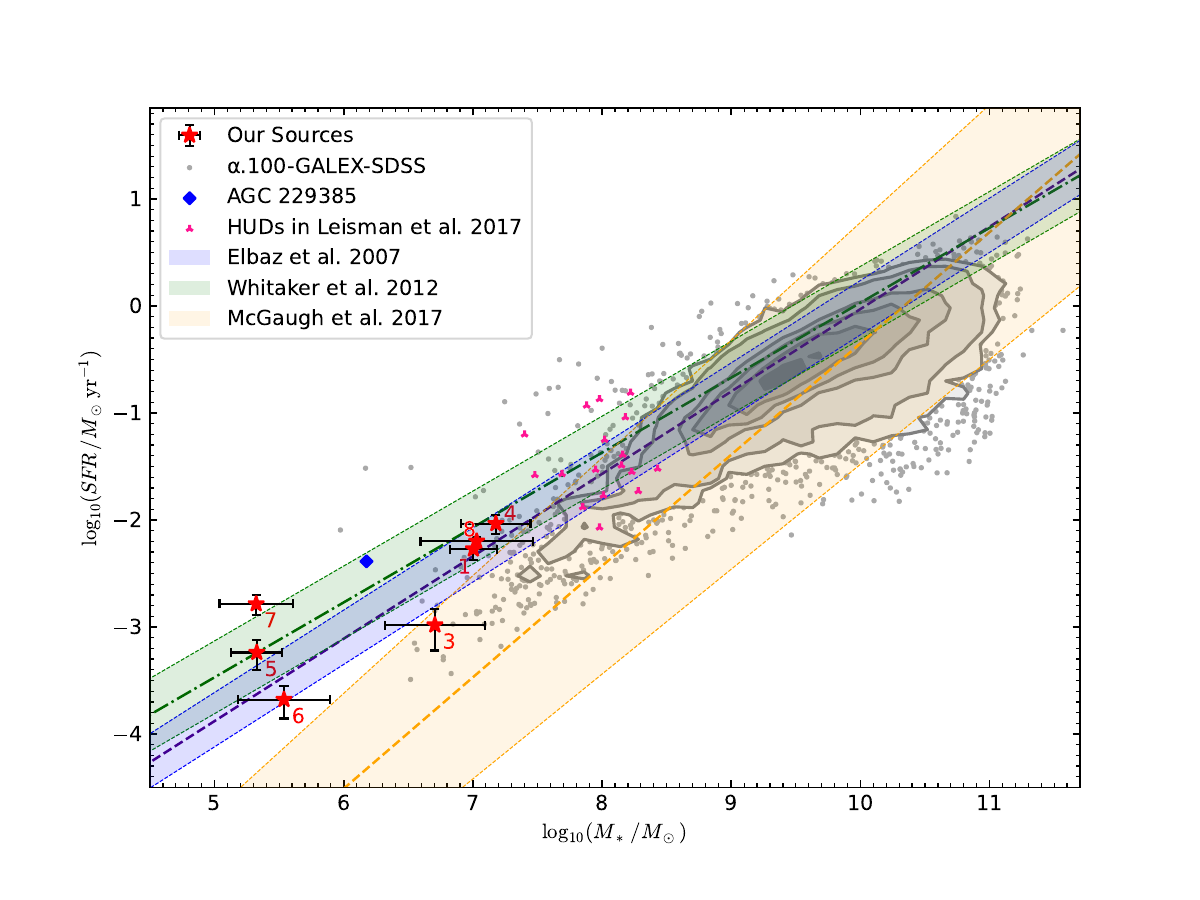}
    \includegraphics[width=0.498\textwidth]{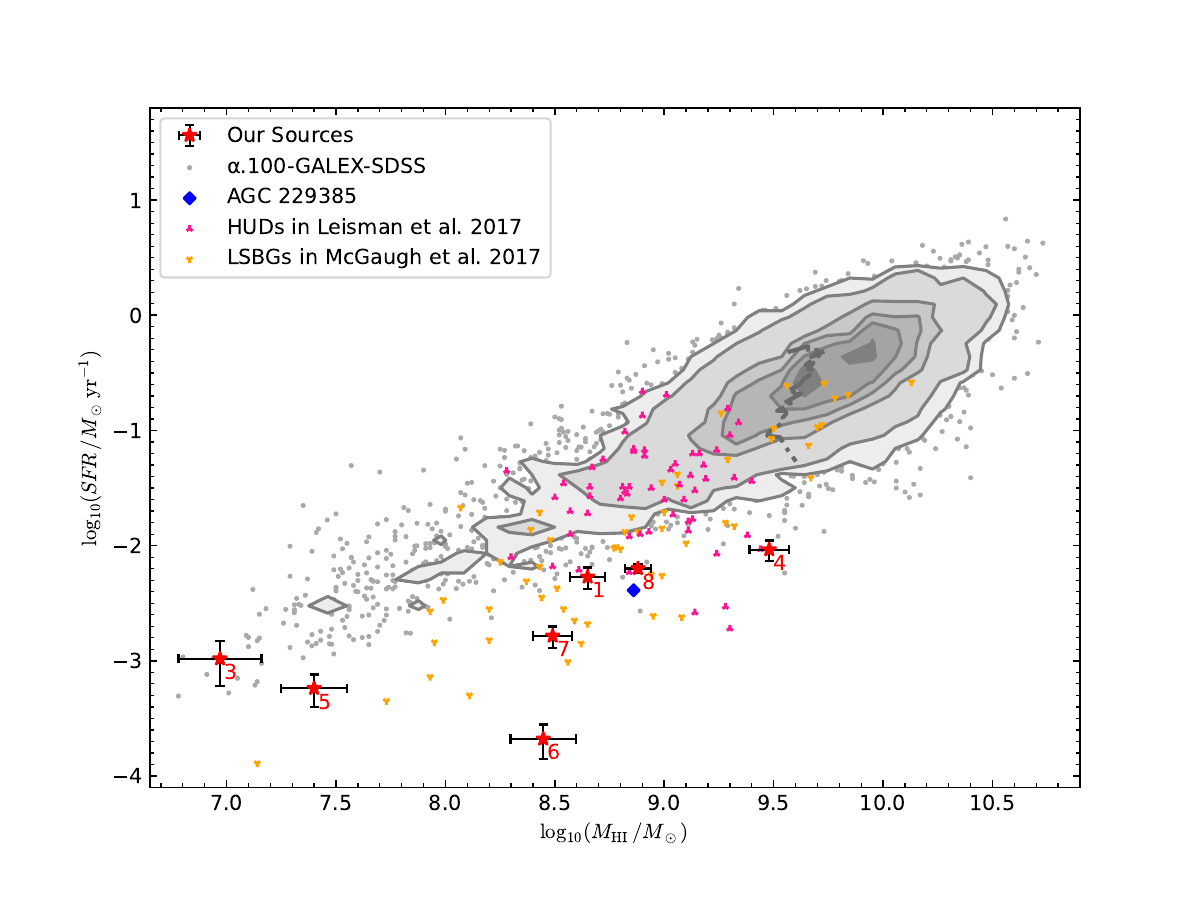}
    \caption{Left panel: The SFR-$M_\ast$ relation of our sources and comparisons with $\alpha.100$ sample, LSB objects, and the empirical MSs. Red pentagrams with black error bars represent our sources, and their numbers correspond to the ordinal numbers in Table \ref{tab:01}. The $\alpha.100$-GALEX-SDSS sample is shown in the form of grey contours and dots, and the contours in each panel from inside to outside are the same as in Fig. \ref{fig:04}. The sample of 115 HUDs in \citet{2017ApJ...842..133L} and the  \textit{almost dark} galaxy AGC\,229385 \textbf{\citep{2015ApJ...801...96J}} are represented by pink triangular symbols and a blue diamond, respectively. Three empirical MSs and their scatters, which are from \citet{2007A&A...468...33E}, \citet{2012ApJ...754L..29W}, and \citet{2017ApJ...851...22M}, are shown as dotted lines and shadows in blue, green, and orange, respectively. Right panel: The SFR-$M_{\rm H\textsc{i}}$ relation of our sources and comparisons with $\alpha.100$ sample and LSB objects. The symbols \textbf{and data references} of our sources, $\alpha.100$ sample, AGC\,229385, and HUD sample are the same with the left panel. The grey dot-dashed line donates the median values of the $\alpha.100$ sample. The LSBGs in \citet{2017ApJ...851...22M} are marked as orange triangular symbols.}
    \label{fig:06}
\end{figure*}


\begin{figure*}
    \centering
    \includegraphics[scale=0.45]{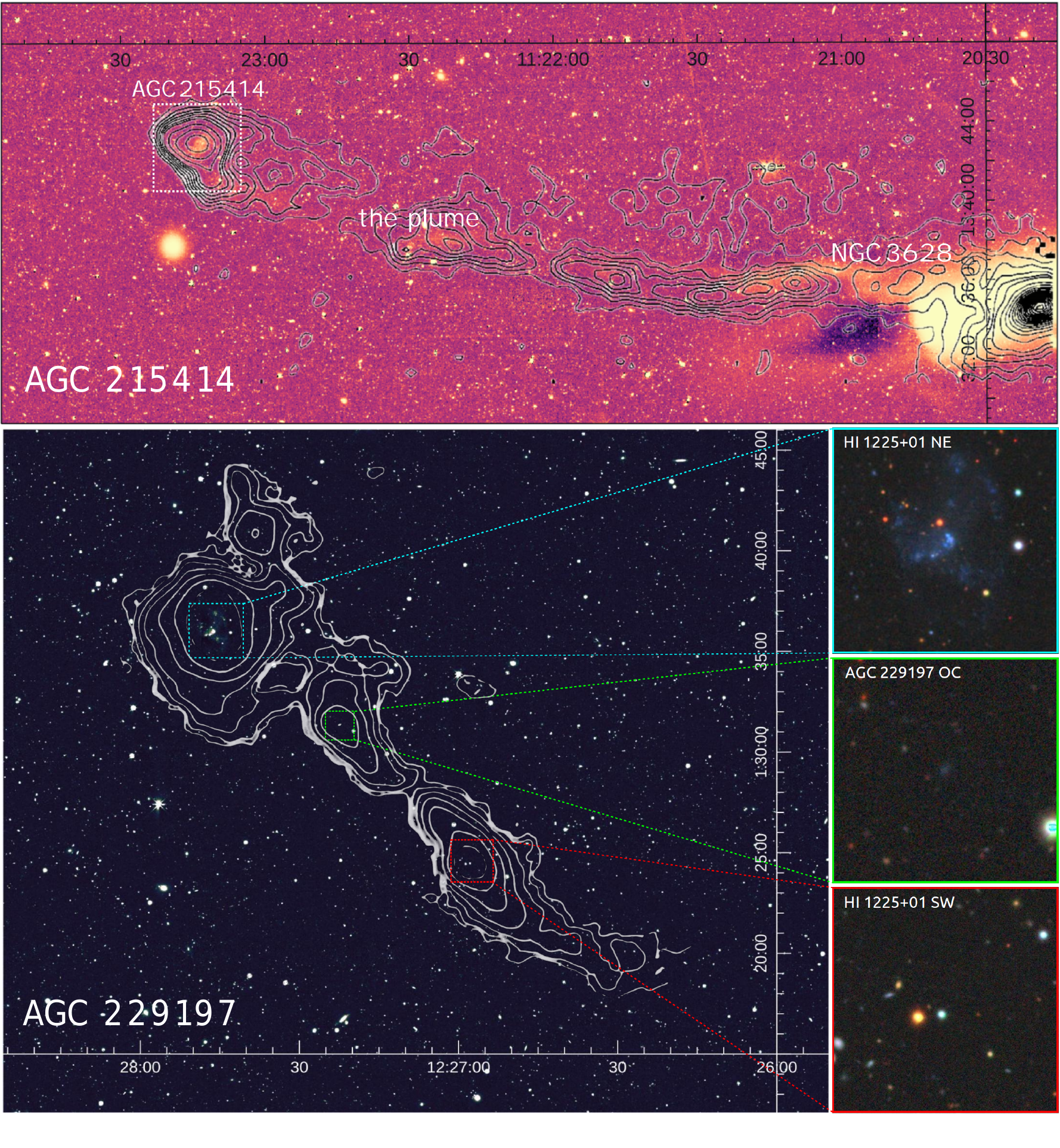}
    \caption{The environment of the sources with tidal origins, which are AGC\,215414 and AGC\,229197, as discussed in Section \ref{sec:05-01-01}. Both the upper and lower images are pseudo-color $g$-band images of DECaLS, with H\textsc{i} distribution contours from VLA overlaid. The H\textsc{i} contours are from Fig. 4 of \citet{2022A&A...658A..25W} and Fig. 2 of \citet{1995AJ....109.2415C}, respectively. The upper image illustrates the surroundings of AGC\,215414, displaying an optical-visible plume trailing from NGC\,3628, and AGC\,215414 is located at the tip. All of these features are marked by white dashed boxes or white-colored text. The lower image shows the environment of AGC\,229197, characterized by a large gas structure. Three areas on the lower image, circled by the cyan, green, and red dashed boxes, are zoomed in on the right, from top to bottom being H\textsc{i}\,1225+01\,NE, AGC\,229197\,OC, and H\textsc{i}\,1225+01\,SW, with FOV size of $2.62^{\prime}$, $1.31^{\prime}$ and $2^{\prime}$, respectively.}
    \label{fig:07}
\end{figure*}

\begin{figure*}
    \centering
    \includegraphics[scale=0.1]{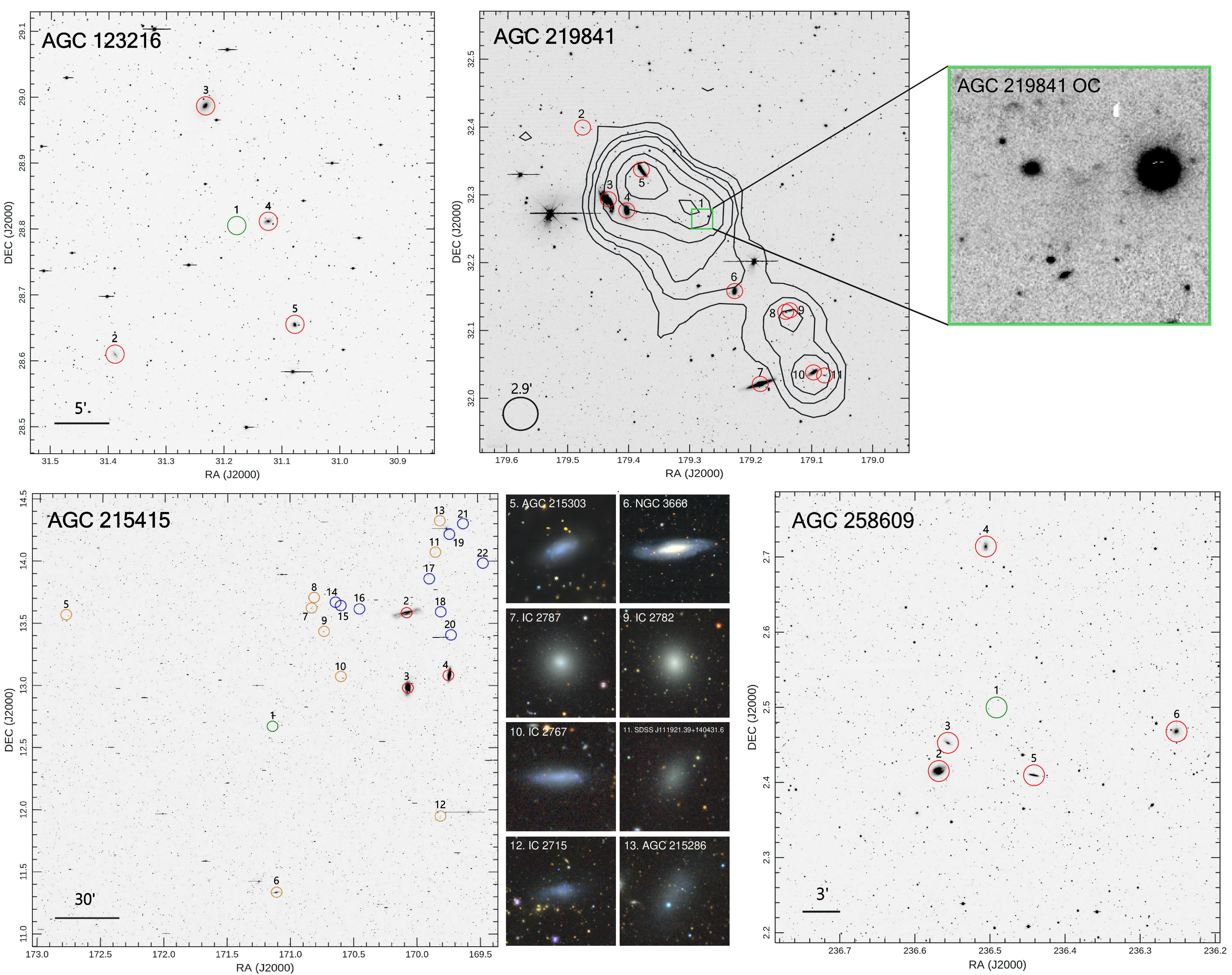}
    \caption{The environment of sources the in groups, as discussed in Section \ref{sec:05-01-02}, from upper left to lower right are AGC\,123216, AGC\,219841, AGC\,215415, and AGC\,258609. These images show the positions of nearby galaxies with the same redshift of the sources. Our sources are marked with green circles or boxes, while others are marked with red. The numbers labeled on the images correspond to those in Table \ref{tab:05}, \ref{tab:07}, \ref{tab:06}$\&$\ref{tab:08}, respectively. In the upper right image (AGC\,219841), the contour overlaid is the H\textsc{i} distribution observed by FAST. The green box highlights the region where our sources are located which is magnified on the right. In the lower left image (AGC\,215415), we further divided the other sources into the Leo Triplet, other optically visible galaxies, and H\textsc{i} sources without optical counterparts, which are marked in red, orange, and blue, respectively; we show the details of the orange-marked sources on the right, all of which show LSB characteristics.}
    \label{fig:08}
\end{figure*}

\begin{figure*}
    \centering
    \includegraphics[scale=0.128]{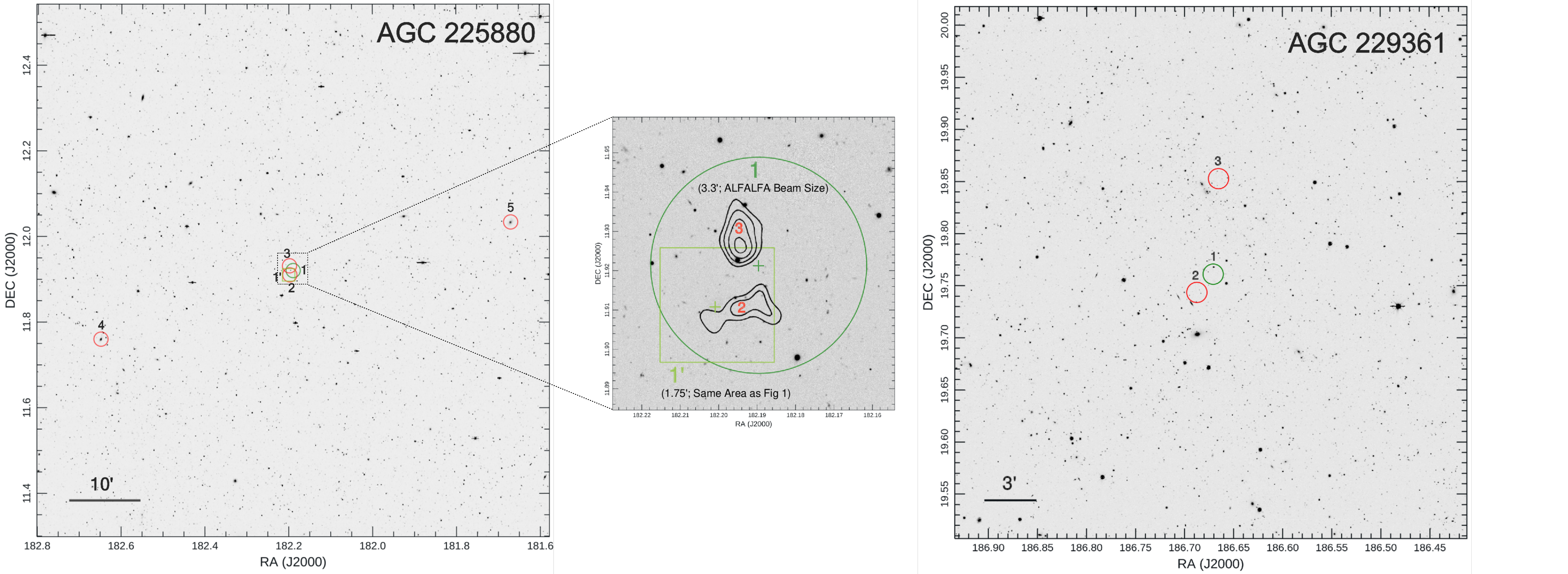}
    \caption{The environment of the sources with potentially isolated origins, AGC\,225880 (left) and AGC\,229361 (right), discussed in Section \ref{sec:05-01-03}. The colors of circles and numbers marked on the images are the same as Fig. \ref{fig:08}. Besides, in the left image (AGC\,225880), we marked the area with a dashed box where our source is located and enlarged it on the right side. In this partial image, the deep green circle indicates the observation of AGC\,225880 by ALFALFA, with the beam size of $3.3^{\prime}$; the light green square marks the location of the optical counterparts, with the size the same as the area shown in \ref{fig:01}; and the light green cross indicates the center position of AGC\,225880\,OC. Contours display the H\textsc{i} distribution of Cloud 1 North and Cloud 1 South, observed by VLA from \citet{2010ApJ...725.2333K}.}
    \label{fig:09}
\end{figure*}

\begin{figure*}
    \centering
    \includegraphics[scale=0.3]{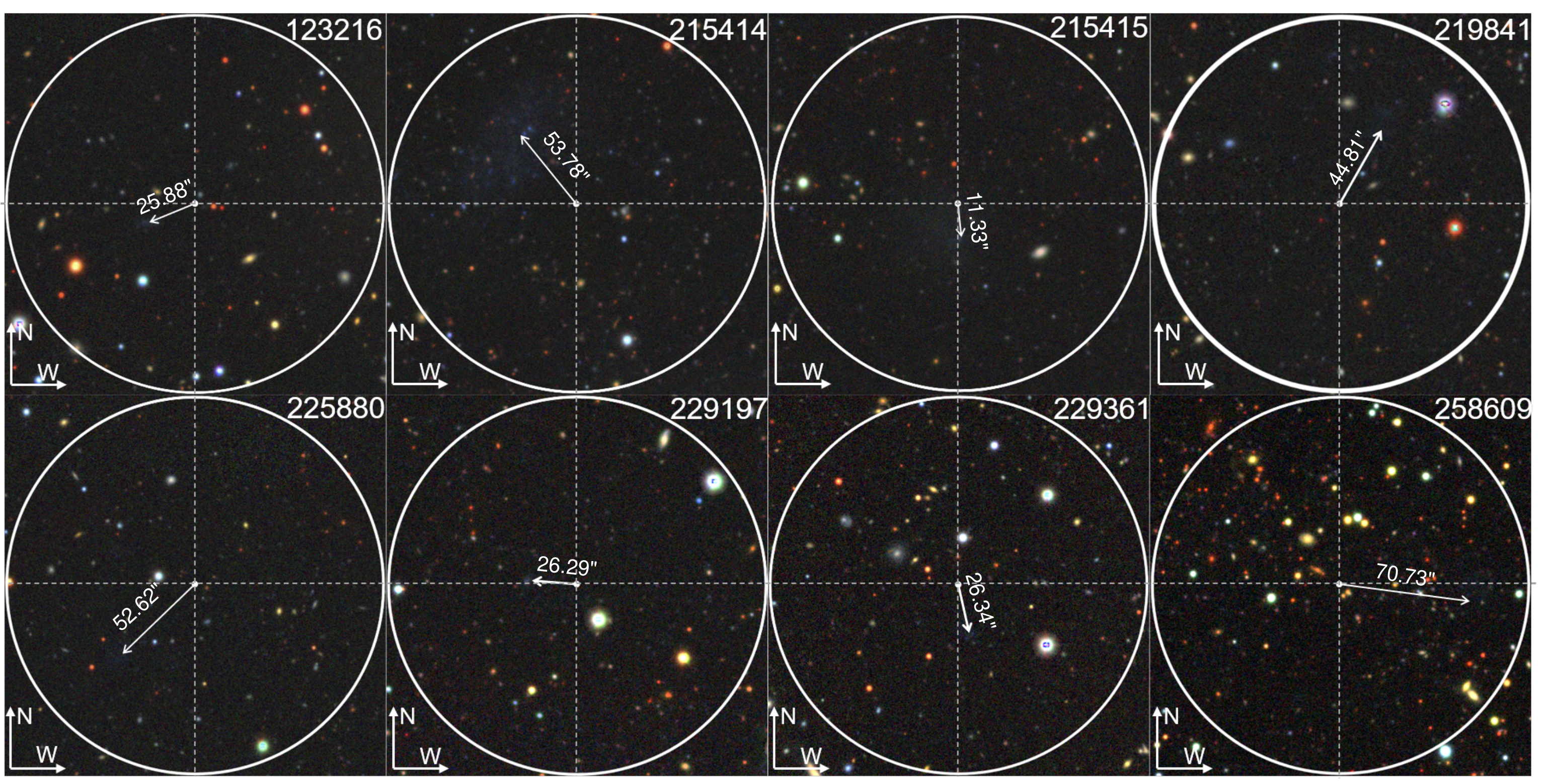}
    \caption{The offsets of optical and H\textsc{i} centers shown on the $grz$ pseudo-color images of DECaLS, arranged from upper left to lower right in two rows, in the same order as Table \ref{tab:01}. Each subplot is centered on the H\textsc{i} position detected by ALFAFALA, and the white circles correspond to the area covered by a beam of ALFALFA ($3.3^{\prime}$ in diameter). The position indicated by the arrow is the center of the optical counterpart in our work. The specific values of the offsets are marked in the diagram.}
    \label{fig:10}
\end{figure*}


\begin{figure*}[htbp]
  \centering
  \includegraphics[width=0.9\linewidth]{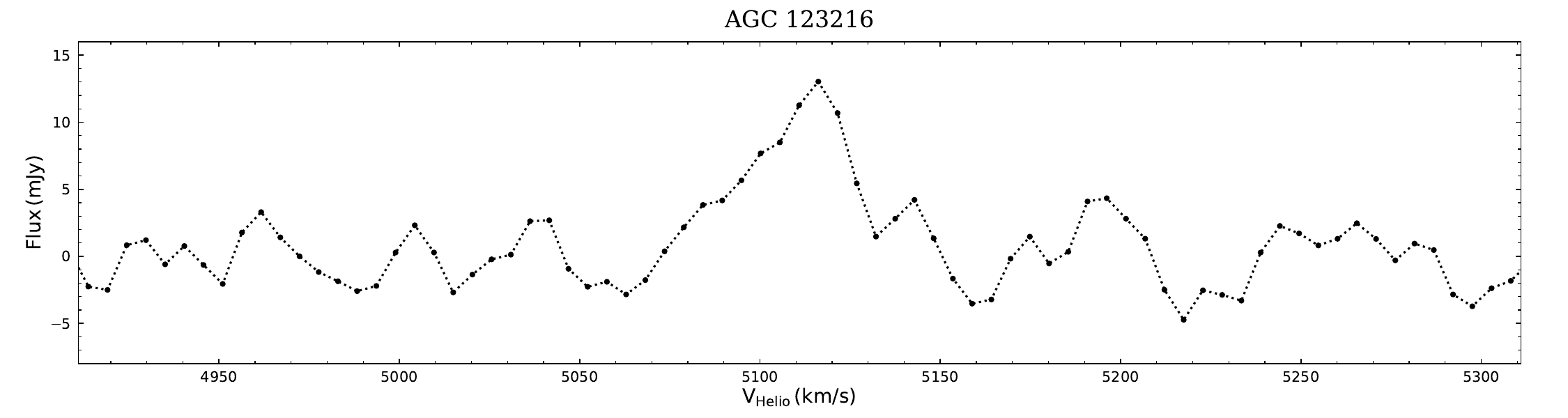}\\
  \includegraphics[width=0.9\linewidth]{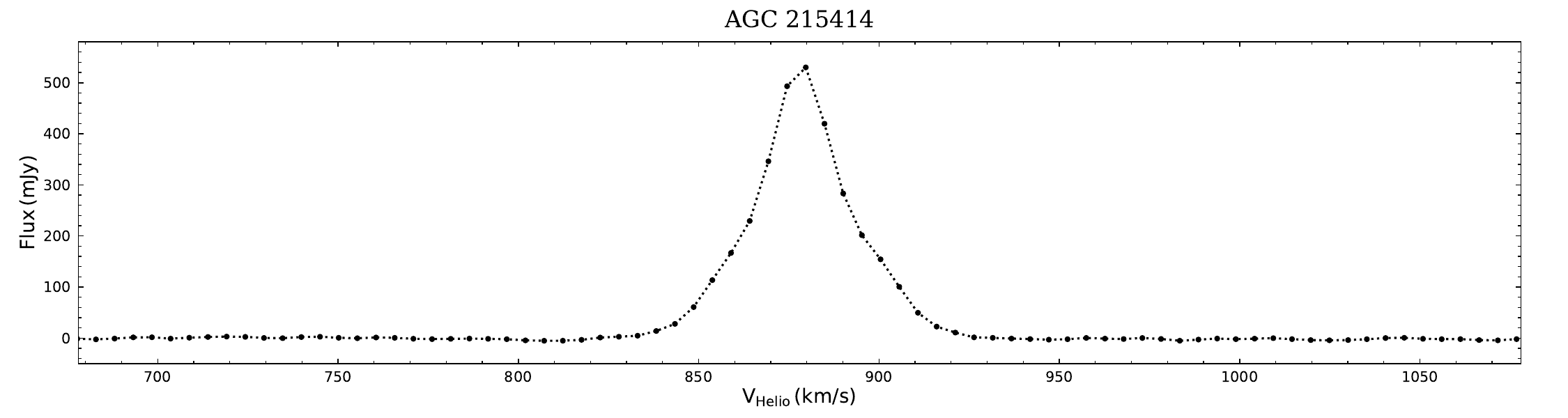}\\
  \includegraphics[width=0.9\linewidth]{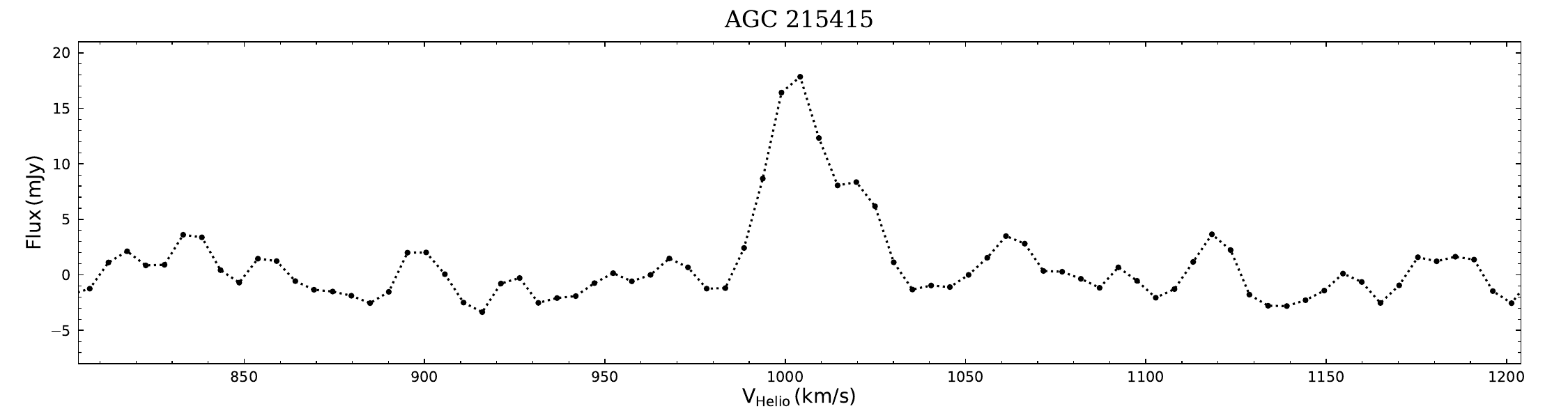}\\
  \includegraphics[width=0.9\linewidth]{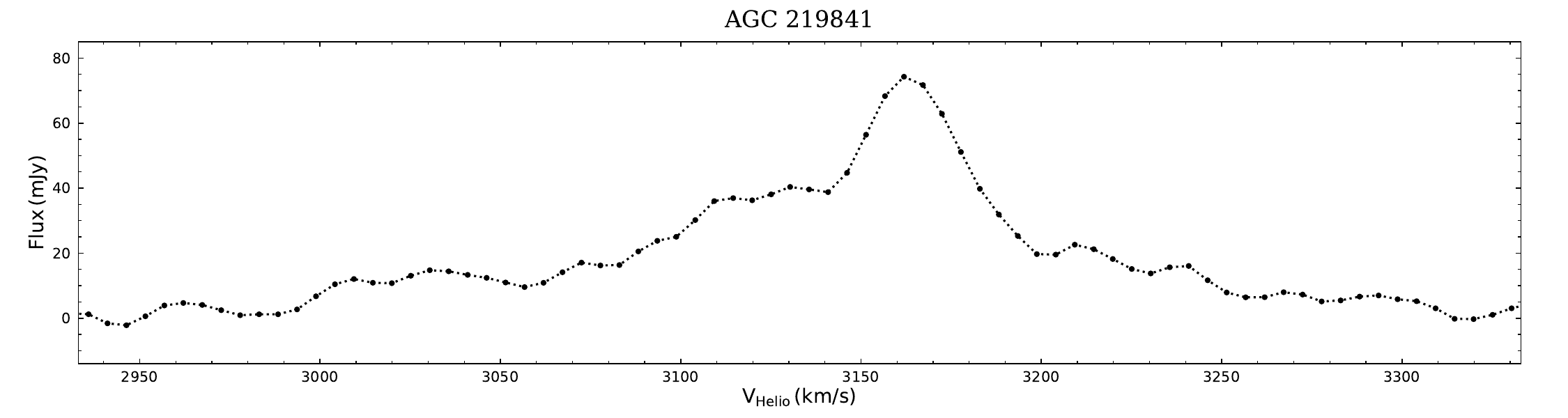}
  \caption{The H\textsc{i} spectral line profiles of the eight sources obtained from ALFALFA, in the same order as Table \ref{tab:01}. The range of the x-axis ($V_{\rm Helio}$) is set to $\pm200\,\kms$ near the heliocentric velocity of each H\textsc{i} source, while the range of the y-axis (Flux) is determined by the H\textsc{i} emission intensity of each source.}
  \label{fig:overall}
\end{figure*}
\addtocounter{figure}{-1}
\begin{figure*}[htbp]
  \centering
  \includegraphics[width=0.9\linewidth]{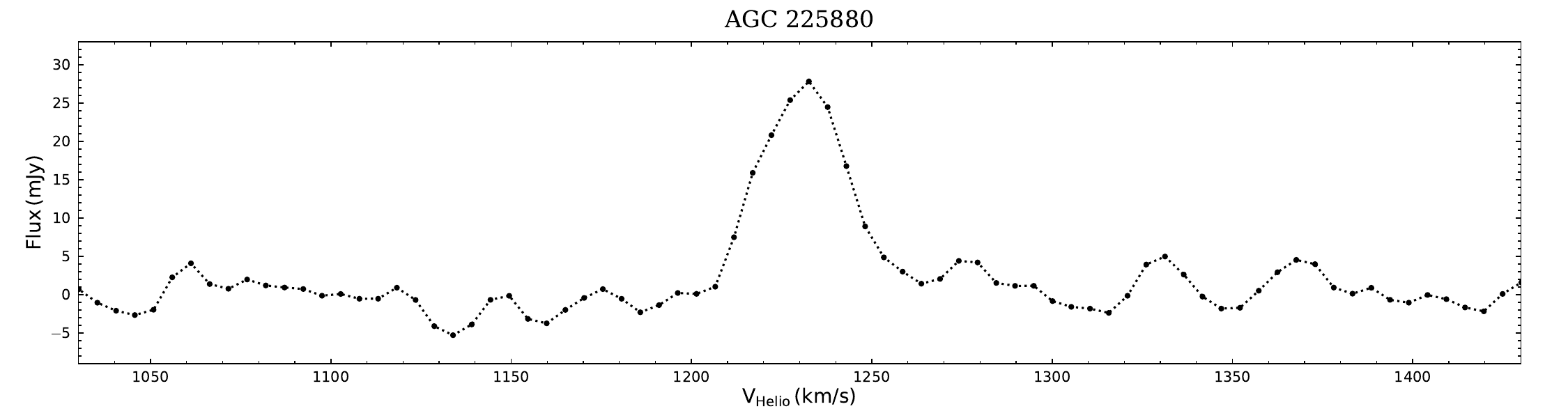}\\
  \includegraphics[width=0.9\linewidth]{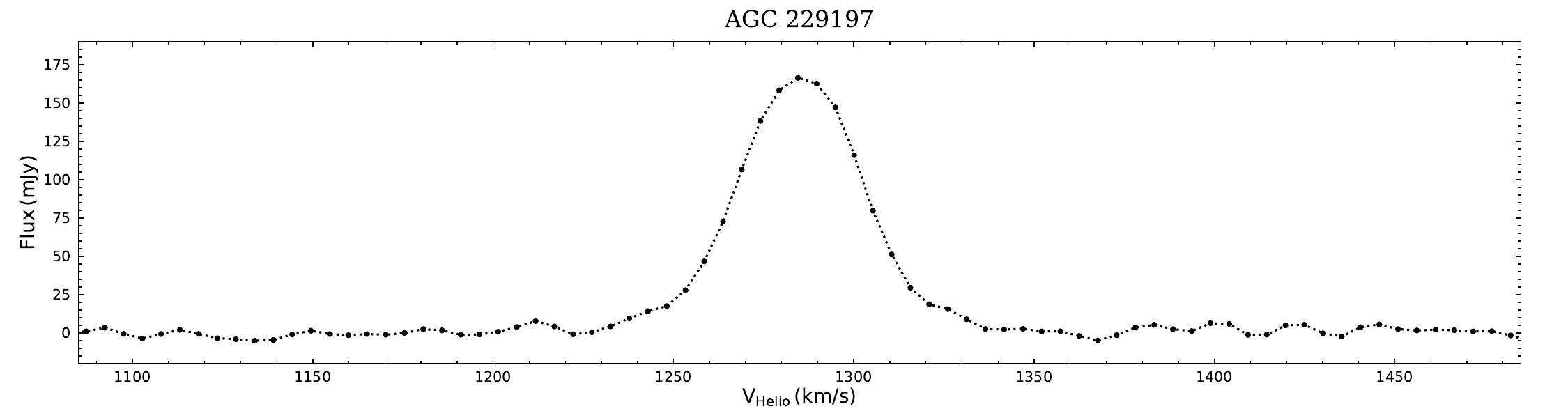}\\
  \includegraphics[width=0.9\linewidth]{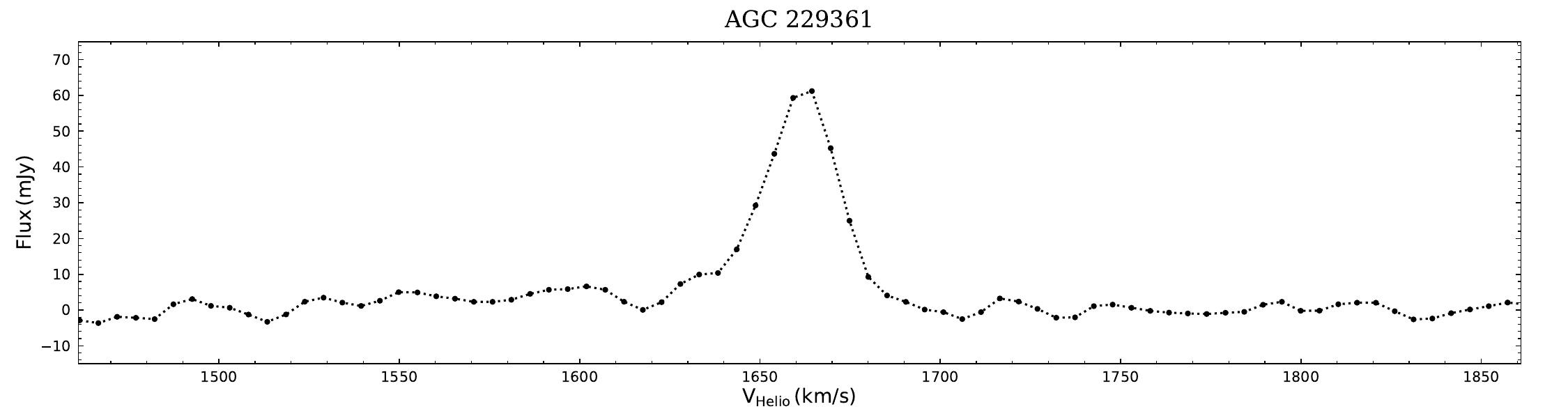}\\
  \includegraphics[width=0.9\linewidth]{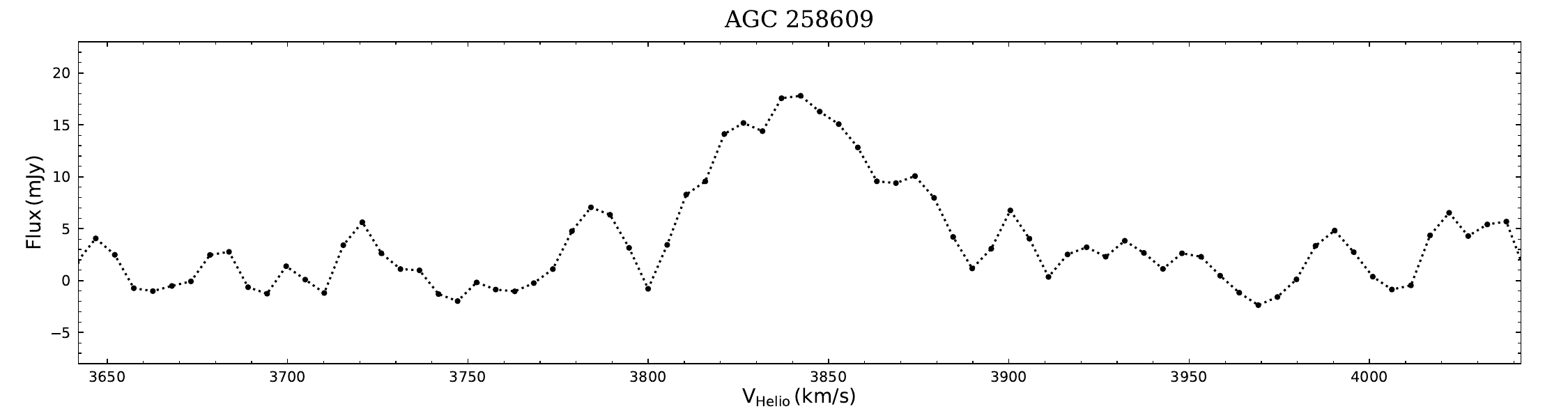}
  \caption{Continue of Fig. \ref{fig:overall}.}
\end{figure*}

\end{CJK*}
\end{document}